\documentclass[useAMS,usenatbib,usegraphicx]{mn2e}

\newcommand{\pin}{\sc pinocchio}
\newcommand{\gal}{\sc morgana}
\newcommand{\zm}{$z_{\rm merge}$}
\newcommand{\be}{\begin{equation}}
\newcommand{\ee}{\end{equation}}
\newcommand{\bea}{\begin{eqnarray}}
\newcommand{\eea}{\end{eqnarray}}
\newcommand{\msun}{M$_\odot$}
\newcommand{\msunyr}{M$_\odot$ yr$^{-1}$}
\newcommand{\kms}{km s$^{-1}$}

\newcommand{\rcool}{$r_{\rm cool}$}
\newcommand{\rh}{$r_{\rm H}$}
\newcommand{\gammap}{${\gamma_{\rm p}}$}
\newcommand{\cnfw}{$c_{\rm nfw}$}
\newcommand{\rs}{$r_{\rm s}$}
\newcommand{\heff}{$H_{\rm eff}$}
\newcommand{\surf}{M$_\odot$ pc$^{-2}$}

\title[\gal]
{The {\gal} model for the rise of galaxies and active nuclei}
\author[Monaco, Fontanot \& Taffoni]
{Pierluigi Monaco$^{1,2}$, Fabio Fontanot$^{1,3}$ \& Giuliano Taffoni$^2$\\
$^1$Dipartimento di Astronomia, Universit\`a di Trieste, 
via Tiepolo 11, 34131 Trieste, Italy \\
$^2$INAF-Osservatorio Astronomico di Trieste, via Tiepolo 11, 34131 Trieste, Italy\\
$^3$Max Planck Institute for Astronomy, K\"onigstuhl 17, D-69117 Heidelberg, Germany\\
email: monaco, taffoni@oats.inaf.it; fontanot@mpia-hd.mpg.de}

\begin{document}

\date{Accepted ... Received ...}

\pagerange{\pageref{firstpage}--\pageref{lastpage}} \pubyear{}

\maketitle

\label{firstpage}

\begin{abstract}

We present the MOdel for the Rise of GAlaxies aNd Active nuclei
({\gal}), a new code for the formation and evolution of galaxies and
AGNs.  Starting from the merger trees of dark matter halos and a model
for the evolution of substructure within the halos, the complex
physics of baryons is modeled with a set of state-of-the-art models
that describe the mass, metal and energy flows between the various
components (baryonic halo, bulge, disc) and phases (cold and hot gas,
stars) of a galaxy.  These flows are then numerically integrated to
produce predictions for the evolution of galaxies.  The processes
of shock-heating and cooling, star formation, feedback, galactic winds
and super-winds, accretion onto BHs and AGN feedback are described by
new models. In particular, the evolution of the halo gas explicitly
follows the thermal and kinetic energies of the hot and cold phases,
while star formation and feedback follow the results of the
multi-phase model by Monaco (2004a).  The increased level of
sophistication of these models allows to move from a phenomenological
description of gas physics, based on simple scalings with the depth of
the DM halo potential, toward a fully physically motivated one.  We
deem that this is fully justified by the level of maturity and rough
convergence reached by the latest versions of numerical and
semi-analytic models of galaxy formation.  The comparison of the
predictions of {\gal} with a basic set of galactic data reveals from
the one hand an overall rough agreement, and from the other hand
highlights a number of well- or less-known problems: (i) producing the
cutoff of the luminosity function requires to force the quenching of
the late cooling flows by AGN feedback, (ii) the normalization of the
Tully-Fisher relation of local spirals cannot be recovered unless the
dark matter halos are assumed to have a very low concentration, (iii)
the mass function of HI gas is not easily fitted at small masses,
unless a similarly low concentration is assumed, (iv) there is an
excess of small elliptical galaxies at $z=0$.  These discrepancies,
more than the points of agreement with data, give important clues on
the missing ingredients of galaxy formation.

\end{abstract}



\section{Introduction}
\label{section:intro}

Several datasets spanning a large range of distances and luminosities
in most of our past-light cone are now constraining the background
cosmology and the properties of primordial fluctuations with a
remarkable precision.  Temperature fluctuations of the CMB (Spergel et
al. 2003, 2006), distant Supernovae (SN, Perlmutter et al, 1999; Knop
et al., 2003), the large-scale structure traced by galaxies (2dF,
Percival et al. 2002; SDSS, Eisenstein et al. 2005), the statistics of
microlensing (Refregier 2003), the abundance and clustering of galaxy
clusters (Rosati, Borgani \& Norman 2002) and the statistics of the
Lyman$\alpha$ forest transmission (Viel, Haehnelt \& Springel 2004)
are now giving a consistent picture of our Universe, well described by
the $\Lambda$CDM model with parameters $(\Omega_0, \Omega_\Lambda,
\Omega_b, h, \sigma_8) \simeq (0.3, 0.7, 0.04, 0.7, 0.8)$ (with the
last quantity ranging from 0.75 to 09).

This ``concordance'' model provides the initial conditions for the
evolution of perturbations, responsible for the gathering of dark
matter (DM) into bound and relaxed halos.  These DM halos host
most of the astro-physical processes that rule the formation of stars
and galaxies from the primordial gas.  However, while the initial
conditions are specified and the basic physical laws are known, the
formation of galaxies is still an open problem, due to the high level
of non-linearity of the processes involved and to the wide coupling of
scales, from the km scale of imploding cores of SNe to the Mpc scale
of galaxy super-winds.  Galaxy formation is thus a problem of
complexity.

Numerical simulations aimed at resolving galaxies (see, e.g., Pearce
et al. 2001; Weinberg, Hernquist \& Katz 2002; Steinmetz \& Navarro
2002; Mathis et al. 2002; Lia, Portinari \& Carraro 2002; Recchi et
al. 2002; Toft et al. 2002; Springel \& Hernquist 2003; Tornatore et
al. 2003; Governato et al. 2004) are still struggling both to tame the
full complexity of the problem and to reach a sufficient mass and
spatial resolution to resolve the ``microphysics'' of the injection of
energy by SNe and accreting BHs.  As a matter of fact, most
simulations of single galaxies are limited to mass resolutions of
$\sim 10^5$ {\msun} and space resolutions of $\sim1$ kpc, so the
microphysical level is still treated as ``sub-grid physics'' and
inserted in the codes with the aid of simple effective models.  The
problem is even more severe when trying to introduce accreting BHs
within a galaxy (Di Matteo et al. 2003; Kazantzidis et al. 2005),
something that many authors regard as the missing ingredient of galaxy
formation.  Considering then that most constraints on galaxies and
AGNs are of a statistical nature, so that thousands if not
hundreds of thousands of galaxies need to be produced for each model,
it is clear that, at variance with what has happened with DM, a
straighforward solution of the galaxy formation problem with N-body
simulations is beyond hope for years to come.

A quicker approach is given by the so-called ``semi-analytic'' galaxy
formation models (White \& Frenk 1991; Kauffmann et al. 1999;
Somerville \& Primack 1999; Cole et al. 2000; Wu, Fabian \& Nulsen
2000; Granato et al. 2001; Hatton et al. 2003; Menci et al. 2004; Kang
et al. 2005; Nagashima et al. 2005; Cattaneo et al. 2006; De Lucia et
al. 2006; Bower et al. 2006), where all the processes that take place
in the formation of galaxies are taken into account with simple
approximated recipes.  The main advantage of these models resides in
the possibility of predicting the properties of whole galaxy
populations in a short amount of computing time, thus making it
possible to achieve a good sampling of the parameter space.  However,
it has been remarked that the (mostly phenomenological) recipes
used in these models have often a weak physical motivation and require
the inclusion of free parameters that are difficult to constrain
otherwise.  As a consequence, the agreement between model and data,
when achieved, may not shed much light on the process of galaxy
formation.  Moreover, given the intrinsic complexity of the problem,
the models have often struggled to reproduce longly known pieces of
evidence, like the shape of the luminosity function of galaxies,
without convincingly showing their predictive power.  This is
highlighted by the difficulties of specific versions of semi-analytic
models to reproduce some pieces of evidence, like the high-mass cutoff
of the luminosity function (Benson et al. 2003), the normalization of
the Tully-Fisher relation (Kauffmann et al. 1999), the sub-mm counts
(Baugh et al. 2005), the level of $\alpha$-enhancement in ellipticals
(Thomas et al. 2005; Nagashima et al. 2005), the redshift distribution of
K-band sources (Cimatti et al. 2002).  From this point of view, it is
incorrect to claim that these models have too many free parameters, as
it is clearly possible to falsify them.

\begin{figure}
\centerline{
\includegraphics[width=8cm]{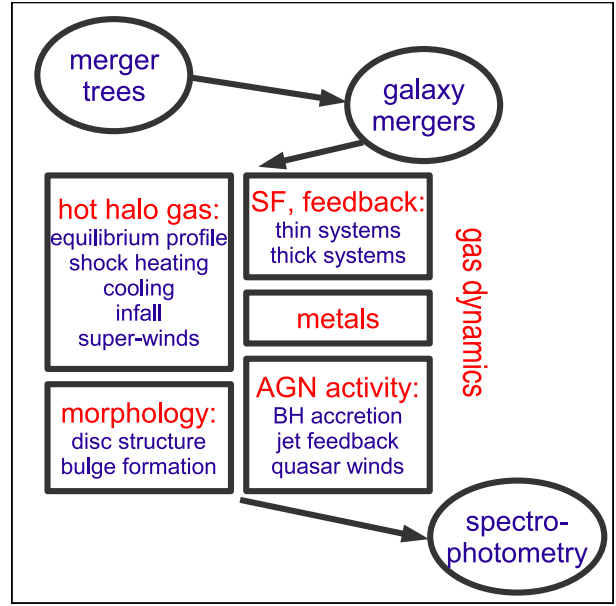}}
\caption{Outline of the main ingredients in the {\gal} code.
Merger trees are taken from the {\pin} tool (Monaco et al. 2002a);
galaxy mergers are treated following Taffoni et al. (2003);
spectrophotometry is performed with {\sc grasil} (Silva et al. 1998).
The treatment of gas dynamics of baryons within the DM halos is the
main argument of the present paper.}
\label{fig:outline}
\end{figure}

This paper is the first of a series devoted to presenting {\gal}, a
new galaxy formation model which, in comparison to the ones listed
above, treats many of the physical processes at an increased level of
sophistication.  Figure~\ref{fig:outline} shows a general outline of
this model, that could be applied to most similar models.  The
most relevant features of {\gal} are the following: 

\begin{itemize}
\item the evolution of the various components and phases of a galaxy
is followed by integrating a differential system of equations along
each branch of a merger tree, thus allowing for the most general (and
non-linear) set of equations for mass, energy and metal flows;
\item the halo gas is treated as a multi-phase medium (hot gas,
cooling flow, halo stars) and its evolution is described by a
model that takes into account the thermal and kinetic energies of
the hot and cold phases; this model treats cooling and infall as
separated processes, takes into account the mass and energy injection
from the galaxy to the halo (galactic winds) and from the halo to the
Inter-Galactic Medium (IGM; galactic super-winds);
\item feedback and star formation are inserted following the results
of the multi-phase model by Monaco (2004a, hereafter M04; 2004b), plus
an additional prescription for kinetic feedback;
\item accretion onto BHs and its feedback onto the galaxy 
(through radio jets and quasar-triggered galaxy winds) are built-in.
\end{itemize}
The increased level of sophistication allows to move from a
phenomenological description of gas physics,
based on simple scalings with the depth of the DM halo potential,
toward a fully physically motivated one.

{\gal} has not been developed to construct a ``theory of everything''
for galaxies; we consider this model simply as a precious tool (i) to
understand the interplay and relative importance of the various
physical processes that take place in galaxy formation, (ii) to
incorporate easily more processes that are thought to influence
galaxies, so as to test their effects, (iii) to produce mock galaxy
catalogues which reproduce particular selection criteria, in order to
investigate the properties of galaxy surveys like sample variance and
completeness level.  The ultimate aim is to increase the predictive
power of such galaxy formation models.

On the other hand, the increase in the level of sophistication can
only lead to an increase in the number of parameters of the
model. Many of these parameters are fixed either by observations (like
the cosmological parameters) or purely gravitational N-body
simulations (like the parameters related to galaxy mergings), while
others can be fixed in principle with the aid of hydro simulations
(like the parameters related to gas cooling and super-winds).
Besides, fundamental quantities like the energy of a single SN, the
Initial Mass Function (IMF) of stars, the chemical yields or the
parameters connected to the feedback and accretion onto BHs are very
uncertain.  This is again a problem of complexity, which will not be
solved by neglecting important degrees of freedom.  So, instead of
decreasing the number of parameters we aim to test our model versus a
large number of observational constraints, so as to fix all the
parameters and propose predictions for upcoming and future
observational campaigns.

This paper presents the algorithm in full detail and some results that
show the main potentialities, merits and problems of the model.  The
stellar mass function produced by this model has already been
presented in Fontana et al. (2006).  Other upcoming papers will
address specific cosmogonical topics, like the statistics of the
AGN/quasar population (Fontanot et al. 2006a), the assembly of the
stellar mass of bright galaxies (Fontanot et al. 2006b) and the
construction of the Stellar Diffuse Component in galaxy clusters
(Monaco et al. 2006).  The paper is organized as follows:
Section~\ref{section:outline} presents an outline of the model,
describing its structure and the mass, energy and metal flows.  The
next sections, from \ref{section:halos} to \ref{section:post},
describe in detail the treatment of all the physical processes
inserted in the model; the reader willing to skip the details may go
directly to section~\ref{section:results}.
Section~\ref{section:halos} describes the merger trees of DM halos,
section~\ref{section:mergers} the merging of galaxies,
section~\ref{section:halogas} the physics of the halo gas phases,
section~\ref{section:structure} the formation of bulges and discs,
section~\ref{section:feedback} the modeling of stellar feedback,
section~\ref{section:metals} the production of metals,
section~\ref{section:agn} the accretion onto BHs.
Section~\ref{section:parameters} discusses the parameters involved in
the model, while section~\ref{section:post} outlines the main
post-processing phases.  Section~\ref{section:results} gives some
basic results of the model and compares them to available
observations, while section~\ref{section:conclusions} gives the
conclusions.  Two appendices report details on the merging and
destruction times of satellites and a very preliminary analysis of
stability with mass resolution.



In this paper we use a ``concordance'' $\Lambda$CDM cosmology with
parameters $\Omega_0=0.3$, $\Omega_\Lambda=0.7$, $\Omega_b=0.044$,
$\sigma_8=0.9$, $H_0=70$ \kms\ Mpc$^{-1}$.  All physical quantities
are scaled to this $H_0$ value.  The new WMAP results (Spergel et
al. 2006), published when this paper was at a very advanced state of
preparation, suggest $\sigma_8\simeq0.75$, and this could lead to some
modest though significant changes in the predictions given in this
paper.  


\section{Outline of the model}
\label{section:outline}

In this section we outline the model, describing the structure of the
code and the system of equations for the mass, energy and metal flows
that is integrated by the code.

%


\subsection{DM halos and galaxies}
\label{section:dmhalos}

The merger trees of DM halos are obtained using the {\pin} code (the
details are given later in section~\ref{section:trees}).  This is not
considered as an integrated part of the galaxy formation code; any
other code for generating merger trees can be used in its place.  With
respect to the Millennium Simulation (Springel et al. 2005), the
{\pin} trees do not give any information on the substructure of DM
halos.

Like N-body simulations, {\pin} is based on realizations of Gaussian
initial conditions on a cosmological grid, so the mass resolution of merger
trees is determined by the grid particle mass.  Each DM halo that gets
as massive as at least ten particles\footnote{
As shown in Monaco et al. (2002b), ten particles are not enough to
reconstruct robustly a DM halo.  On the other hand, in {\pin} the
halos gather around the peaks of the {\em inverse collapse time
field}, so the natural choice for the appearance time of a halo would
be the collapse time of its first (peak) particle, corresponding to
the creation of a 1-particle object.  We deem that 10 particles is a
good compromise between the need for mass resolution and robustness.
} constitutes a starting branch of the tree; the time at which this
happens is named {\it appearance time}.  A galaxy is associated to
each starting branch.  When a DM halo merges with a larger one, it
disappears as an individual entity and becomes a substructure (or
satellite) of the larger DM halo.  The fate of substructures is
followed using the model of Taffoni et al. (2003;
section~\ref{section:dynfriction}): while the external regions of the
satellite DM halo are tidally stripped, its core, which contains its
associated galaxy, survives for some time.  A galaxy that is
associated with an existent DM halo is named {\em central}; in general
(though not always) the central galaxy is the largest in a DM halo.
Galaxies associated with substructures are named {\em satellites}.
Dynamical friction brings the orbit of the substructure toward the
centre, making it eventually merge with the main DM halo; when this
happens the satellite galaxy merges with the central one.  In some
cases, the substructure and its associated galaxy can be destroyed by
tides.  Substructures can also merge between themselves, but this
process is neglected in the present code.

In general, DM halos will contain one central galaxy and several
satellites, each associated with a substructure. When two DM halos
merge, the central galaxy of the more massive DM halo will become the
central galaxy of the merger, while the one of the smaller halo will
become a satellite like the others.  The bound between satellites
belonging to the same substructure is assumed to be lost after the
merger, i.e. we do not allow for substructure of substructures.  As a
consequence, an existing substructure will always be associated with a
single galaxy.


\subsection{Algorithm}
\label{section:algorithm}

\begin{figure*}
\centerline{
\includegraphics[width=17.7cm]{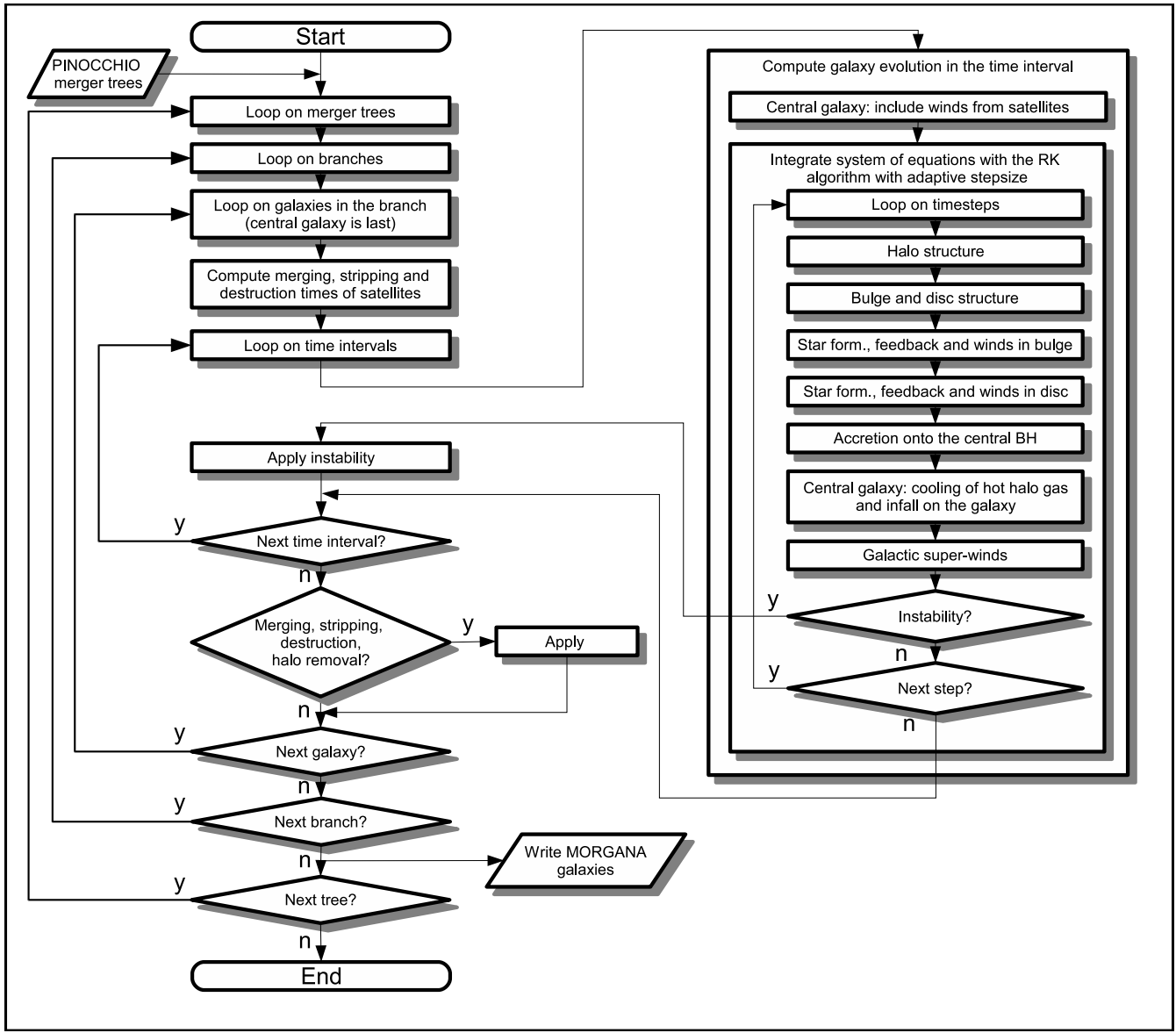}}
\caption{ Flow chart of the {\gal} code; see text for all details.}
\label{fig:flowchart}
\end{figure*}

The flow chart of the algorithm is given in
figure~\ref{fig:flowchart}.  The algorithm can be ideally divided into
two main parts, represented by the blocks on the left or right of the
figure; the first part (in the left half of the flow chart) handles
the merger trees and calls the numerical integrator for all the
galaxies, the second part (in the right half) performs the
integration.  The algorithm proceeds as follows:

\begin{itemize}
\item The merger trees are read from the {\pin} output file.  Then a
loop on all the trees is started.
\item Each merger tree is subdivided into branches.  A branch is
defined as the evolution of a DM halo between two consecutive
mergings, be them major or minor; in other words, it corresponds to a
time interval in which no new substructures are added to the DM halo.
\item The galaxies contained in the DM halo during the branch are
looped on. The central galaxy is always addressed as the last one, so
that the evolution of the hot halo gas associated with it can take
into account the energetic input of all the satellites.
\item For each satellite galaxy the dynamical friction, tidal
stripping and tidal destruction times are computed
(section~\ref{section:dynfriction}).
\item The time interval corresponding to the branch is subdivided into
smaller intervals, starting or ending at times $i_{\rm
sample}\Delta_t$, where $i_{\rm sample}$ is a sampling index and
$\Delta_t$ is usually set to 0.1 Gyr. This is done to allow for a
regular time sampling of baryonic variables.  In general the branch
ends are not integer multiples of $\Delta_t$; for instance, if a
galaxy is to be evolved from $t_1=4.386$ to $t_2=4.728$ Gyr, the
branch will be subdivided into the intervals $[4.386,4.4]$,
$[4.4,4.5]$, $[4.5,4.6]$, $[4.6,4.7]$ and $[4.7,4.728]$.
\item The evolution of the galaxy, the second main part of the code,
is performed in the time interval by integrating a system of
differential equations with a Runge-Kutta integrator with adaptive
time-steps (Press et al. 1992).  The time-steps are chosen so as to
have an accurate results to within $10^{-4}$ on the mass and energy
flows described below.  Of course, the integration of the galaxy stops
at its merger or destruction time whenever these events occur.
\item The system of equations integrated by the code and the
implemented physical processes are described below
(section~\ref{section:physproc} and tables~1, 2 and 3).
\item The integration is interrupted whenever an instability is found;
presently we take into account disc instabilities and
quasars-triggered galaxy winds.
\item If relevant, the effect of such instabilities is applied at the
end of the integration, and the loop on time intervals is closed.
\item At the end of the evolution of the galaxy in the branch, the
predicted events of galaxy mergers, tidal stripping and tidal
destruction, and the removal of the baryonic halo component of new
satellites (described in section~\ref{section:satellite}) are applied.
\item The loops on galaxies and on branches are closed, and the
results for all the galaxies in the tree are written on the output
file.
\item The loop on trees is closed.
\end{itemize}


\subsection{Baryons}
\label{section:baryons}

\begin{figure}
\centerline{
\includegraphics[width=8cm]{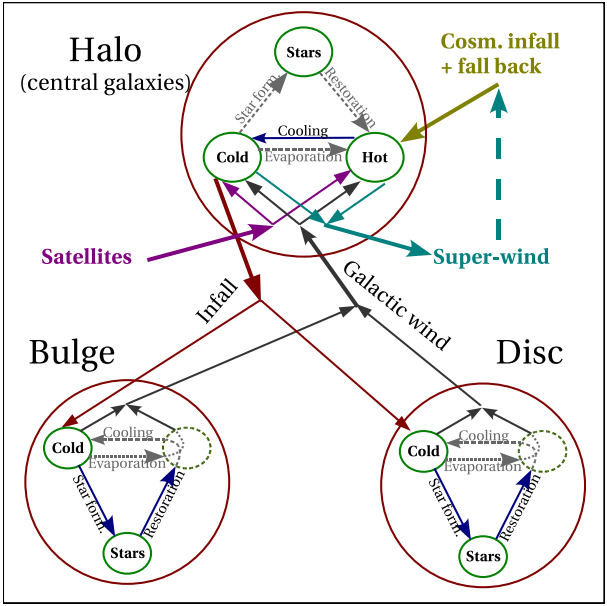}
}
\caption{General scheme of mass flows in a model galaxy.  The
flows highlighted by dashed lines and gray colours are not used in the
present version of the model. In particular, star formation (and then
restoration and evaporation) in the halo is not active, while in bulge
and disc components the evaporation and hot wind flows, as well as the
restoration and cooling flows, are equated so as to leave the hot
phase empty.}
\label{fig:flows}
\end{figure}

The baryonic content of each galaxy is divided into three components,
namely a halo, a bulge and a disc (figure~\ref{fig:flows}).  Each
component is made up by three phases, i.e. cold gas, hot gas and
stars.  The halo component of central galaxies contains the virialized
gas pervading the DM halo, cold gas associated to the cooling flow and
halo stars.  Satellite galaxies do not have any mass (or energy or
metals) in their halo component, and this is justified by the fact
that tidal stripping is very efficient in unbinding the halo component
from satellites.  (From the computational point of view, it is
straightforward to relax this assumption, but we do not implement this
feature here.)  For the halo component, the code follows the mass and
metal content of the three phases (cold gas, hot gas, stars), plus the
thermal energy of the hot gas (that determines its temperature) and
the kinetic energy of the cold gas (that determines its velocity
dispersion).  In total, 8 variables are associated to this component.

Bulge and disc components are formally described in the code by the
same variables.  This approach was indeed undertaken as a follow-up of
the M04 multi-phase model of a star-forming ISM, which is presented in
some detail in section~\ref{section:feedback}.  However, a
straightforward implementation of the M04 model, attempted in one of
the first versions of the code, led to a number of unwelcome numerical
complications.  We then decided to collect the main results of the M04
model and insert them as a set of simple recipes.  This allows also to
have a more transparent grasp of the effect of feedback in galaxy
formation without losing the physical motivation of the multi-phase
approach; this is a welcome feature in account of the lack of a widely
accepted model for the evolution of a star-forming ISM.  In summary,
the bulge and disc components are described, in an effectively
single-phase formalism, by a more limited set of 4 variables, namely
the masses of stars and gas, and their metal masses.

Together with the 16 variables (8+4+4) associated to the three
components, 9 more variables are integrated by the code.
The complete list of the 25 variables is the following:

\begin{itemize}
\item mass, kinetic energy and metals of cold halo gas;
\item mass, thermal energy and metals of hot halo gas;
\item mass and metals of halo stars;
\item mass and metals of gas in bulge and disc;
\item mass and metals of stars in bulge and disc;
\item cooling radius of the hot halo gas;
\item mass, metals and kinetic energy of the cold gas ejected from DM halos as a super-wind;
\item mass, metals and thermal energy of the hot gas  ejected from DM halos as a super-wind;
\item black hole mass;
\item black hole reservoir.
\end{itemize}

The sampling of variables in the time grid defined above
(section~\ref{section:algorithm}) is performed as follows.  At times
$i_{\rm sample}\Delta_t$ the values of the 16 variables relative to
the galaxy components are stored together with bulge and disc radii
and velocities, black hole masses, punctual values for the star
formation rates of bulge and disc and accretion rate onto the black
hole.  Cooling radii, ejected matter and black-hole reservoirs are
not sampled.  With our choice of $\Delta t =0.1$ Gyr the time sampling
is adequate to describe old stars, but is too poor for young stars.
For this reason we store the punctual values of star formation in
bulge and disc; in this way we can reconstruct, with a minimal amount
of information, the recent star formation which gives rise to the
massive stars.  This procedure is presented and tested in Fontanot et
al. (2006b) (see also section~\ref{section:post}).

\begin{table}
\label{table:system}
\begin{tabular}{l}
\hline
 Mass flows \\
\hline
$\dot{M}_{\rm c,H} = \dot{M}_{\rm co,H} - \dot{M}_{\rm in,H} - \dot{M}_{\rm csw} + \dot{M}_{\rm cw,H} + \dot{M}_{\rm csat}$ \\
$\dot{M}_{\rm h,H} = \dot{M}_{\rm cosm} - \dot{M}_{\rm co,H} - \dot{M}_{\rm hsw} + \dot{M}_{\rm hw,H} + \dot{M}_{\rm hsat}$ \\
$\dot{M}_{\rm s,H} = 0$ \\
$\dot{M}_{\rm c,B} = \dot{M}_{\rm in,B} - \dot{M}_{\rm sf,B} + \dot{M}_{\rm rs,B} - \dot{M}_{\rm hw,B} - \dot{M}_{\rm cw,B}$ \\ 
$\dot{M}_{\rm s,B} = \dot{M}_{\rm sf,B} - \dot{M}_{\rm rs,B}$ \\
$\dot{M}_{\rm c,D} = \dot{M}_{\rm in,D} - \dot{M}_{\rm sf,D} + \dot{M}_{\rm rs,D} - \dot{M}_{\rm hw,D} - \dot{M}_{\rm cw,D}$ \\ 
$\dot{M}_{\rm s,D} = \dot{M}_{\rm sf,D} - \dot{M}_{\rm rs,D}$ \\
\hline
Energy Flows \\
\hline
$\dot{K}_{\rm H} = \dot{K}_{\rm co,H} - \dot{K}_{\rm in,H} - \dot{K}_{\rm csw} + \dot{K}_{\rm w,H} + \dot{K}_{\rm csat} - \dot{K}_{\rm ad}$ \\
$\dot{E}_{\rm H} = \dot{E}_{\rm cosm} - \dot{E}_{\rm co,H} - \dot{E}_{\rm hsw} + \dot{E}_{\rm hw,H} + \dot{E}_{\rm hsat} - \dot{E}_{\rm ad}$ \\
\hline
Metal Flows \\
\hline
$\dot{M}^Z_{\rm c,H} = \dot{M}^Z_{\rm co,H} - \dot{M}^Z_{\rm in,H} - \dot{M}^Z_{\rm csw} + \dot{M}^Z_{\rm cw,H} + \dot{M}^Z_{\rm csat}$ \\
$\dot{M}^Z_{\rm h,H} = \dot{M}^Z_{\rm cosm} - \dot{M}^Z_{\rm co,H} - \dot{M}^Z_{\rm hsw} + \dot{M}^Z_{\rm hw,H} + \dot{M}^Z_{\rm hsat}$ \\
$\dot{M}^Z_{\rm s,H} = 0 $\\
$\dot{M}^Z_{\rm c,B} = \dot{M}^Z_{\rm in,B} - \dot{M}^Z_{\rm sf,B} + \dot{M}^Z_{\rm rs,B} - \dot{M}^Z_{\rm hw,B} - \dot{M}^Z_{\rm cw,B} + \dot{M}^Z_{\rm yi,B}$ \\
$\dot{M}^Z_{\rm s,B} = \dot{M}^Z_{\rm sf,B} - \dot{M}^Z_{\rm rs,B}$ \\
$\dot{M}^Z_{\rm c,D} = \dot{M}^Z_{\rm in,D} - \dot{M}^Z_{\rm sf,D} + \dot{M}^Z_{\rm rs,D} - \dot{M}^Z_{\rm hw,D} - \dot{M}^Z_{\rm cw,D} + \dot{M}^Z_{\rm yi,D}$ \\
$\dot{M}^Z_{\rm s,D} = \dot{M}^Z_{\rm sf,D} - \dot{M}^Z_{\rm rs,D}$\\
\hline
Ejected matter \\
\hline
$\dot{M}_{\rm c,E} = \dot{M}_{\rm csw}$ \\
$\dot{K}_{\rm c,E} = \dot{K}_{\rm csw}$ \\
$\dot{M}_{\rm h,E} = \dot{M}_{\rm hsw}$ \\
$\dot{E}_{\rm h,E} = \dot{E}_{\rm hsw}$ \\
$\dot{M}^Z_{\rm c,E} = \dot{M}^Z_{\rm csw}$ \\
$\dot{M}^Z_{\rm h,E} = \dot{M}^Z_{\rm hsw}$ \\
\hline
Black hole flows (equations~\ref{eq:bhaccr})\\
\hline
$\dot{M}_{\rm BH} = \min ( \dot{M}_{\rm visc}\, , \, M_{\rm BH}/t_{\rm Ed} )$  \\
$\dot{M}_{\rm resv} = \dot{M}_{\rm lowJ} - \dot{M}_{\rm BH} $ \\
\hline
Cooling radius (equation~\ref{eq:drcool})\\
\hline
$\dot{r}_{\rm cool} = (\dot{M}_{\rm co,H}-\dot{M}_{\rm hw,H})/(4\pi \rho_g(r_{\rm cool})r_{\rm cool}^2)$ \\
\hline
\end{tabular}
\caption{The system of equations that is integrated by the code.}
\end{table}


\begin{table}
\label{table:flows}
\begin{center}
\begin{tabular}{lll}
\hline
Flow & Comment & Reference \\
\hline
& {\bf Mass flows} & \\
$\dot{M}_{\rm co,H}$ & cooling flow              & eq. \ref{eq:coolingflow}\\
$\dot{M}_{\rm in,H}$ & infall from halo          & eq. \ref{eq:infall}\\
$\dot{M}_{\rm csw} $ & cold super-wind           & eq. \ref{eq:coldwind_m}\\
$\dot{M}_{\rm cw,H}$ & cold wind to halo         & eq. \ref{eq:conditions}\\
$\dot{M}_{\rm csat}$ & cold wind from satellites & eq. \ref{eq:satflows}\\
$\dot{M}_{\rm cosm}$ & cosmological infall       & eq. \ref{eq:cosm}\\
$\dot{M}_{\rm hsw} $ & hot super-wind            & eq. \ref{eq:hotwind_m}\\
$\dot{M}_{\rm hw,H}$ & hot wind to halo          & eq. \ref{eq:conditions}\\
$\dot{M}_{\rm hsat}$ & hot wind from satellites  & eq. \ref{eq:satflows}\\
$\dot{M}_{\rm in,B}$ & infall to bulge           & eqs. \ref{eq:infall_b1}, \ref{eq:infall_b2} \\
$\dot{M}_{\rm sf,B}$ & star formation in bulge   & eq. \ref{eq:sfbulge}\\
$\dot{M}_{\rm rs,B}$ & restoration in bulge      & eq. \ref{eq:sfbulge}\\
$\dot{M}_{\rm hw,B}$ & hot wind from bulge       & eqs. \ref{eq:hwbmass}, \ref{eq:kinwind_hm1}, \ref{eq:kinwind_hm2}\\ 
$\dot{M}_{\rm cw,B}$ & cold wind from bulge      & eqs. \ref{eq:kinwind_cm}, \ref{eq:kinwind_hm3}\\
$\dot{M}_{\rm in,D}$ & infall to disc            & eqs. \ref{eq:infall_d1}, \ref{eq:infall_d2}\\
$\dot{M}_{\rm sf,D}$ & star formation in disc    & eq. \ref{eq:disc_flows}\\
$\dot{M}_{\rm rs,D}$ & restoration in disc       & eq. \ref{eq:disc_flows}\\
$\dot{M}_{\rm hw,D}$ & hot wind from disc        & eq. \ref{eq:disc_flows}\\ 
$\dot{M}_{\rm cw,D}$ & cold wind from disc       & eq. \ref{eq:disc_flows}\\
\\
& {\bf Kinetic energy of cold halo gas} & \\
$\dot{K}_{\rm co,H}$ & energy of cooling flow         & eq. \ref{eq:dkco}\\
$\dot{K}_{\rm in,H}$ & energy lost by infall          & eq. \ref{eq:kinfall}\\
$\dot{K}_{\rm csw} $ & energy lost by cold super-wind & eq. \ref{eq:coldwind_k}\\
$\dot{K}_{\rm w,H}$  & energy acquired by winds       & eq. \ref{eq:kinfbdisc}, \ref{eq:kinfbbulge}, \ref{eq:kinwind_ck}\\
$\dot{K}_{\rm csat}$ & energy acquired by satellites  & eq. \ref{eq:satflows}\\
$\dot{K}_{\rm ad}  $ & adiabatic expansion            & eq. \ref{eq:kad}\\
\\
& {\bf Thermal energy of hot halo gas} & \\
$\dot{E}_{\rm cosm}$ & shock-heating of infalling IGM & eq. \ref{eq:shock}\\
$\dot{E}_{\rm co,H}$ & cooling of hot gas             & eq. \ref{eq:coolingenergy}\\
$\dot{E}_{\rm hsw} $ & energy lost by hot super-wind  & eq. \ref{eq:hotwind_e}\\
$\dot{E}_{\rm hw,H}$ & hot wind to halo               & eqs. \ref{eq:hwdisc}, \ref{eq:hwbulge}, \ref{eq:jetfb}\\
$\dot{E}_{\rm hsat}$ & hot wind from satellites       & eq. \ref{eq:satflows}\\
$\dot{E}_{\rm ad}  $ & adiabatic expansion            & eq. \ref{eq:ead}\\
\\
& {\bf Newly generated metals} & \\
$\dot{M}^Z_{\rm yi,B}$ & new metals in bulge cold gas & eq. \ref{eq:yield_bd} \\
$\dot{M}^Z_{\rm yi,D}$ & new metals in disc cold gas & eq. \ref{eq:yield_bd} \\
$\dot{M}^Z_{\rm hw,H}$ & new metals injected to the halo & eq. \ref{eq:yield_h} \\
\\
& {\bf BH flows} & \\
$\dot{M}_{\rm lowJ}$ & loss of angular momentum  & eq. \ref{eq:lowJ}\\
$\dot{M}_{\rm visc}$ & viscous accretion rate    & eq. \ref{eq:visc}\\
\hline
\end{tabular}
\end{center}
\caption{Mass and energy flows in the system of equations
(table~1). Metal flows are not reported.}
\end{table}


\subsection{Physical processes and flows}
\label{section:physproc}

The present paper is dedicated to a detailed description of all the
physical processes included in the code.  Here we give only a global
view of these processes with references to the section of the paper
where they are described.  We also report the system of equations that
is integrated by the code, and a list of the mass, energy and metal
flows.

Baryonic matter flows between the components and the phases as
illustrated in figure~\ref{fig:flows}.  Within each component the
three phases may exchange mass through the following flow terms:
\begin{itemize}
\item evaporation of gas from the cold to the hot phase ($\dot{M}_{\rm
ev}$);
\item cooling of gas from the hot to the cold phase ($\dot{M}_{\rm
co}$);
\item star formation from the cold to the star phase ($\dot{M}_{\rm
sf}$);
\item restoration from the star to the hot phase ($\dot{M}_{\rm rs}$).
\end{itemize}
As matter of fact, these flows are formally present in the code but
they are never active at the same time.  Indeed, in the halo component
of central galaxies star formation (and then evaporation and
restoration) are not active.  In the bulge and disc components the hot
phase is kept void by equating the restoration and cooling flow from
the one hand, the evaporation and hot wind flow on the other hand;
this is indicated in figure~\ref{fig:flows} by connecting, in the
bulge and disc component, the arrows corresponding to the equated
flows.  In the following we will restrict ourselves to the flows that
are used in the present formulation of the model, with the {\em
caveat} that the other flows may be easily activated whenever
required.

Baryonic matter flows between the components as follows:

\begin{itemize}
\item primordial gas flows to the halo together with the accreted DM
mass (cosmological infall, $\dot{M}_{\rm cosm}$);
\item cold gas infalls from the halo to the disc or bulge
($\dot{M}_{\rm in}$);
\item cold and hot gas are expelled by the bulge or disc to the halo
in a galactic wind ($\dot{M}_{\rm cw}$ and $\dot{M}_{\rm hw}$);
\item both hot and cold gas are allowed to leave the halo in a
galactic super-wind ($\dot{M}_{\rm hsw}$ and $\dot{M}_{\rm csw}$);
\item this expelled material is allowed to get back to the halo
together with the cosmological infall;
\item winds from satellites are injected in the halo component of
their central galaxy ($\dot{M}_{\rm csat}$ and $\dot{M}_{\rm hsat}$).
\end{itemize}

Mass conservation implies the following relations:

\bea
\dot{M}_{\rm in,H} &=& \dot{M}_{\rm in,B} + \dot{M}_{\rm in,D} \nonumber \\
\dot{M}_{\rm cw,H} &=& \dot{M}_{\rm cw,B} + \dot{M}_{\rm cw,D} \label{eq:conditions} \\
\dot{M}_{\rm hw,H} &=& \dot{M}_{\rm hw,B} + \dot{M}_{\rm hw,D} \nonumber 
\eea

As illustrated above (section~\ref{section:algorithm}), in the cases
of instabilities (disc instabilities and quasar-triggered galaxy
winds), galaxy mergings, tidal stripping, tidal destruction and
removal of halo baryonic component from new satellites, baryonic
masses, energies and metals are moved between components and galaxies
outside the integration routine.  These flows are named in this paper
{\it external flows}.  It is worth noting that, as star formation in
the halo is not allowed, stars flow to the halo component only through
external flows (namely tidal stripping and galaxy mergings).

The system of equations integrated by the galaxy evolution code is
reported in table~1.  For all the equations, the
suffixes c, h and s in the right hand sides denote the cold, hot and
star phases.  In all pedices the letters H, B, D and E following the
comma denote the flows relative to the halo, bulge and disc
components, and the matter ejected out of the DM halo by super-winds.

Table~2
gives a list of all mass flows, with a quick
explanation and a reference to the equation where they are defined;
metal flows follow trivially mass flows (see
section~\ref{section:metals}) with the exception of the newly
generated metals, so for sake of brevity we only report these in the
table.  Finally, table~3 gives the list of
processes modeled in this paper, with reference to the relative
section and the list of the related mass and energy flows.


\begin{table*}
\label{table:processes}
\begin{center}
\begin{tabular}{lcl}
\hline
Physical process implemented in the integration & Section & Flows modeled in that section \\
\hline
Galaxy winds from satellite galaxies           & \ref{section:satellite} & $\dot{M}_{\rm csat}$, $\dot{K}_{\rm csat}$, $\dot{M}_{\rm hsat}$, $\dot{E}_{\rm hsat}$ \\
Hydrostatic equilibrium for the hot halo phase & \ref{section:hot}       & $-$ \\
Shock heating of cosmological infalling gas    & \ref{section:shock}     & $\dot{M}_{\rm cosm}$, $\dot{E}_{\rm cosm}$\\
Radiative cooling of hot halo gas              & \ref{section:cooling}   & $\dot{M}_{\rm co,H}$, $\dot{E}_{\rm co,H}$, $\dot{K}_{\rm co,H}$\\
Infall of cold gas on the galaxy               & \ref{section:infall}    & $\dot{M}_{\rm in,H}$, $\dot{K}_{\rm in,H}$, $\dot{M}_{\rm in,B}$, $\dot{M}_{\rm in,D}$ \\
Galaxy super-winds                             & \ref{section:winds}     & $\dot{M}_{\rm hsw}$, $\dot{E}_{\rm hsw}$, $\dot{E}_{\rm ad}$, $\dot{M}_{\rm csw}$, $\dot{K}_{\rm csw}$, $\dot{K}_{\rm ad}$ \\
Disc structure                                 & \ref{section:discs}     & $-$ \\
Bulge structure                                & \ref{section:bulges}    & $-$ \\
Star formation and feedback in discs           & \ref{section:thin}      & $\dot{M}_{\rm sf,D}$, $\dot{M}_{\rm rs,D}$, $\dot{M}_{\rm hw,D}$, $\dot{M}_{\rm cw,D}$, $\dot{E}_{\rm hw,H}$, $\dot{K}_{\rm w,H}$\\
Star formation and feedback in bulges          & \ref{section:thick}     & $\dot{M}_{\rm sf,B}$, $\dot{M}_{\rm rs,B}$, $\dot{M}_{\rm hw,B}$, $\dot{M}_{\rm cw,B}$, $\dot{E}_{\rm hw,H}$, $\dot{K}_{\rm w,H}$\\
Metal enrichment                               & \ref{section:metals}    & $\dot{M}^Z_{\rm hw,H}$, $\dot{M}^Z_{\rm yi,B}$, $\dot{M}^Z_{\rm yi,D}$ \\
Accretion onto BHs                             & \ref{section:accretion} & $\dot{M}_{\rm visc}$, $\dot{M}_{\rm lowJ}$ \\
AGN feedback                                   & \ref{section:jets}      & $\dot{E}_{\rm hw,H}$ \\
\hline
Processes implemented as external flows & Section \\
\hline
Disc instabilities                              & \ref{section:discs} \\
Quasar-triggered galaxy winds                   & \ref{section:quasar} \\
Decay of satellite orbits by dynamical friction & \ref{section:dynfriction} \\
Tidal stripping at the periastron               & \ref{section:dynfriction} \\
Tidal destruction of satellites                 & \ref{section:dynfriction} \\
Galaxy mergers                                  & \ref{section:galmergers}\\
Stripping of halo component of satellites       & \ref{section:satellite}\\
\hline
\end{tabular}
\end{center}
\caption{List of physical processes implemented in the code, with
reference to the relevant section, and relative mass, energy and metal
flows.}
\end{table*}


\subsection{Super-winds and satellite galaxies}
\label{section:satellite}

The matter ejected out of the DM halo by super-winds is collected into
the variables denoted in table~1 by the ``,E'' pedix.  After each
integration interval these variables are stored in a vector, so as to
be re-accreted at a later time together with the cosmological infall
term; this is explained in section~\ref{section:winds}.

In the present version of the model, satellite galaxies do not retain
their baryonic halo component.  However, satellite galaxies
continually produce galaxy winds as long as star formation is active.
The related flows are given to the halo component of the main DM halo
as follows. First, the (cold and hot) winds and superwinds flows are
equated:

\bea
\dot{M}_{\rm csw} &=& \dot{M}_{\rm cw,H} \nonumber \\
\dot{K}_{\rm csw} &=& \dot{K}_{\rm cw,H} \nonumber \\
\dot{M}_{\rm hsw} &=& \dot{M}_{\rm hw,H} \label{eq:satwinds} \\
\dot{E}_{\rm hsw} &=& \dot{E}_{\rm hw,H} \nonumber 
\eea

\noindent
Then, at the end of the integration over the time interval the ejected
matter is added to a vector aimed to contain the contribution of all
the satellites to the halo component of the central galaxy.  We call
these vectors $M_{\rm csat}$, $M_{\rm hsat}$ and so on.  The central
galaxy is always evolved as the last one; when this happens the
content of the satellite vectors is injected to the halo phase at a
rate equal to the total mass or energy divided by the $\Delta t$ time
interval:

\bea
\dot{M}_{\rm csat} & = & M_{\rm csat}/\Delta t \nonumber\\
\dot{K}_{\rm csat} & = & \max(K_{\rm csat},M_{\rm csat}V_{\rm disp}^2/2)/\Delta t \nonumber\\
\dot{M}_{\rm hsat} & = & M_{\rm hsat}/\Delta t \label{eq:satflows} \\
\dot{E}_{\rm hsat} & = & E_{\rm hsat}/\Delta t \nonumber
\eea

\noindent
It is worth noticing that the kinetic energy of the cold wind is
recomputed using the velocity dispersion of the DM halo as long as
this velocity is higher than the kinetic velocity of the cold ejected
gas.


\subsection{Initial conditions}
\label{section:ic}

At the appearance time all DM halos are assumed to be as large as 10
particles\footnote{
Some DM halos grow larger than 10 particles by a merger, but this
detail is neglected.
}.  
All the bayons present in these primordial DM halos
are assumed to be in the hot halo phase, whose thermal energy,
acquired by gravitational shocks, is computed with the model described
in section~\ref{section:shock}.  An issue with this setting is the
quick start of cooling in these initial halos.  This is in part a
numerical artifact due to the lack of sampling of the tree; if this
were resolved with a smaller particle mass, the DM halo would then
possibly contain some heating source at that time.  To limit this
resolution effect, it is assumed that the halos have just suffered a
major merger at their appearance time, so that the onset of cooling is
delayed by a few sound-crossing times (see section~\ref{section:shock}
for details).  On the other hand, cooling in small halos is hampered
by the ionizing background, and this is implemented (following Benson
et al. 2002) by quenching cooling in all halos with circular
velocities smaller than 50 {\kms} (section~\ref{section:cooling}).

Overcooling is connected to the very general problem of the stability
of model predictions with respect to mass resolution.
Appendix~\ref{section:stab} presents a very simple convergence test of
the model; the main conclusion is that model predictions {\em do not}
converge and that the 50 {\kms} cutoff motivated by the reionization
guarantees that convergence is forced at low redshift and high
masses. This opens a very delicate topic, which has rarely been
addressed in the literature (see Hatton et al. 2003); we leave a
deeper analysis and discussion to further work, and limit ourselves to
pointing out which of the results presented here are more sensitive to
mass resolution.  In any case, as the lack of convergence regards star
formation at very high redshift and faint galaxies, we can safely
conclude that the building of bright galaxies does not depend strongly
on mass resolution.


\section{DM halos}
\label{section:halos}


\subsection{DM merger trees}
\label{section:trees}

\begin{figure}
\centerline{
\includegraphics[width=8cm]{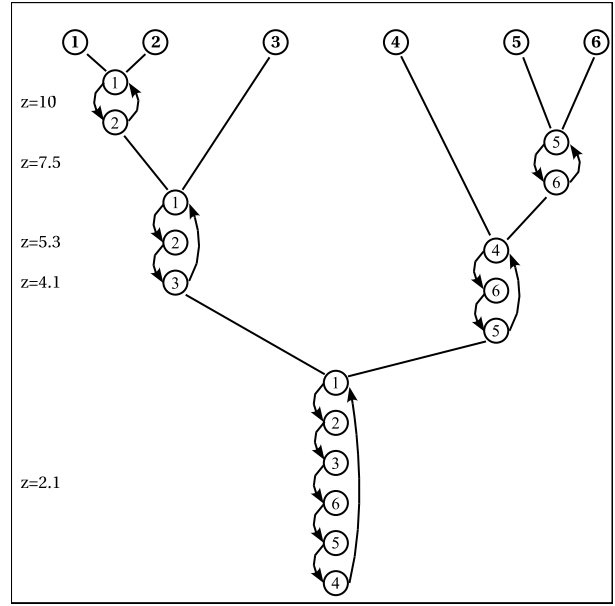}
}
\caption{Example of ordering of the {\pin} merger trees.  Circles with
numbers denote DM progenitors, DM halos are denoted by their linking
lists.  In all the mergings shown in this picture, the DM halo coming
from the left is more massive than that coming from the right.}
\label{fig:ordering}
\end{figure}

As mentioned above, we use the {\pin} code to generate the merger
trees.  The use of {\pin} is motivated by its ability to give,
even with modest computer resources, an adequate description of the
hierarchical formation of DM halos, in excellent agreement with the
results of N-body simulations (see also Zhao et al. 2003; Li, Mo \&
van der Bosch 2005); for instance, the mass function of DM halos is
recovered to within a 5-10 per cent accuracy (Monaco et al. 2002a),
while the mass function of progenitors of DM halos in a specified mass
interval is recovered within a 10-20 per cent error (Taffoni et
al. 2002).   This code uses a scheme based on Lagrangian
Perturbation Theory (Moutarde et al. 1991; Buchert \& Ehlers 1993):
starting from a Gaussian density contrast field sampled on a grid
(very much similarly to the initial conditions of an N-body
simulation) and smoothed over a set of smoothing radii, the earliest
{\em collapse time} (the time at which the first orbit crossing takes
place) is computed for each particle.  Collapsed particles are then
gathered into DM halos, where each halo is seeded by a peak of the
{\em inverse collapse time} field.  This procedure allows a detailed
reconstruction of the DM halos, with known positions, velocities and
angular momenta, and of their merger trees.  The {\pin} merger trees
are equivalent to those given by N-body simulations, with a further
advantage (shared with the less accurate Extended Press \& Schechter
approach; see Bond et al. 1992; Lacey \& Cole 1993) of a very fine
time sampling that allows to track the merging times without being
restricted to a fixed grid in time (or scale factor).  Other notable
differences with respect to N-body trees is the impossibility of
{\pin} DM halos to decrease in mass, a condition which is not strictly
valid for the N-body simulations.  At variance with extended Press \&
Schechter merger trees, {\pin} allows for multiple mergers of DM
halos.

The format of the {\pin} outputs (in the updated 2.1 version,
available at {\tt http://adlibitum.oats.inaf.it/monaco/pinocchio/})
are such that only the output at the final redshift is needed to
reconstruct the merger trees of a realization, and from them compute
the galaxy properties at all times.  In {\pin}, each DM halo retains
its identity even when it disappears by merging with another larger
halo.  The output files contain, for each DM halo that has ever
existed with at least 10 particles, the following information: (i) ID
number, (ii) ID of the halo it belongs to at the final redshift, (iii)
linking list of the halos, (iv) ID of the halo it has merged with, (v)
mass of the halo at the merging time, (vi) mass of the halo it has
merged with (before merging) (vii) merging redshift, (viii) appearance
redshift.  Halos that exist at the final redshift have field (i) and
(ii) equal, and merging redshift equal to -1; field (iv) is also set
to -1, while field (v) contains the mass of the halo at the final
redshift and field (vi) is 0.

Similarly to {\sc galform} (Cole et al. 2000), the linking list
provided by the 2.1 version of {\pin} is organized in
such a way that merged halos are accessed so as to preserves
chronological order.  This is illustrated in the example of
figure~\ref{fig:ordering}.  Each time a halo appears from a peak, its
linking list points to itself.  Suppose now that halo 1 merges with the
smaller halo 2 at \zm=10, and that both halos have no substructure.
Then the linking list is updated so that halo 1 points to halo 2 and
vice-versa.  At \zm=5.3 halo 1 (which contains halo 2 as a
substructure) merges with halo 3, which has no substructure.  Then the
last halo of the chain, halo 2, is linked to halo 3, and this back to
halo 1.  Halos 5 and 6 merge at \zm=7.5, and their fate is similar to
halos 1 and 2.  Halo 5 then merges with the larger halo 4 at \zm=4.1;
in this case halo 4 does not link directly to halo 5, which is put at
the end of the chain, but to halo 6.  At \zm=2.1 halos 1 and 4 merge.
In this case, the last element of the chain of halo 1 (halo 3) is
linked to the second of the chain of halo 4 (halo 6).  The final
sequence is then 1-2-3-6-5-4.  In more general terms, the two groups
are linked in the following order: first the chain of the surviving DM
halo, then the chain of the disappearing one, with the first element
(the main halo before merging) put as last.  It is clear that,
starting from the {\em second} element of the final chain, two
subsequent events are always accessed in chronological order.  This
does not imply a strict chronological order of all the mergings: in
our example halo 3 (\zm=5.3) is accessed before halo 6 (\zm=7.5), but
the two event are on independent branches of the tree.


\subsection{DM halo properties}
\label{section:haloprop}

The physical properties of the DM halos (which are not predicted by
{\pin}) are computed at each integration time-step (see
figure~\ref{fig:flowchart}) as follows.  The density  profile of
the DM halo is assumed to follow Navarro, Frenk \& White (1996;
hereafter NFW), according to which a halo with a virial radius $r_{\rm
H}$ is characterized by a scale radius $r_s$ and a concentration
$c_{\rm nfw}=r_{\rm H}/r_s$.  Defining the quantity $\delta_c\equiv
200c_{\rm nfw}^3(1+c_{\rm nfw})/(3(1+c_{\rm nfw}) \ln(1+c_{\rm
nfw})-c_{\rm nfw})$, the NFW profile of a halo at redshift $z$ is:

\be
\rho_{\rm DM}(r) = \rho_c(z)
\frac{\delta_c}{c_{\rm nfw}x (1+c_{\rm nfw}x)^2}\, ,
\label{eq:nfw}\ee

\noindent
where $\rho_c(z)$ is the critical density at the redshift $z$ and
$x\equiv r/r_s$.  The virial radius $r_{\rm H}$ of a halo of circular
velocity $V_{\rm H}$ is computed assuming that its average density its 200
times the critical density (see, e.g., Mo, Mao \& White 1998):

\be
\begin{array}{lll}
V_{\rm H} &=& (10 M_{\rm H} G H(z))^{1/3} \\
r_{\rm H} &=& V_{\rm H}/10 H(z)
\end{array}
\label{eq:dmhalo}\ee

\noindent
where $M_{\rm H}$ is the total halo mass\footnote{
In the following we denote by $M_{\rm H}$ the {\em total} halo mass,
including DM and baryons, and implicitly assume that the sub-dominant
baryons do not influence the mass profile.  This assumption will be
relaxed in the computation of galaxy discs,
section~\ref{section:discs}.
} and $H(z)$ is the Hubble constant
at $z$.

The concentration $c_{\rm nfw}$ is computed, as a function of $M_{\rm H}$
and $z$, following Eke, Navarro \& Steinmetz (2001).  Given the mass
$M_{\rm H}$, redshift $z$ and concentration $c_{\rm nfw}$ we compute the
gravitational binding energy of the halo ($U_{\rm H}=\int\rho\Phi d^3r$) as:

\be
U_{\rm H} = \frac{G M_{\rm H}^2}{r_{\rm H}} \delta_c^2 \left(\ln^2 (1+c_{\rm nfw})
-\frac{c_{\rm nfw}^2} {1+c_{\rm nfw}}\right)\; .
\label{eq:uh}\ee

\noindent
From it and the virial theorem the velocity dispersion of the halo
(such that the total kinetic energy of halo particles is $M_{\rm H} V_{\rm
disp}^2/2 = -U_{\rm H}/2$) can be computed as:

\be
V_{\rm disp} = \sqrt{-U_{\rm H}/M_{\rm H}}
\label{eq:vdisp}\ee

If $\lambda$ is the spin parameter of the DM halo, its specific
angular momentum is computed as:

\be
\frac{J_{\rm H}}{M_{\rm H}} = G M_{\rm H} \lambda \sqrt{-0.5 U_{\rm H}/M_{\rm H}}\, .
\label{eq:Jh}\ee

\noindent
The DM halo parameters ($M_{\rm H}$, $c_{\rm nfw}$, $r_h$, $V_{\rm H}$, $U_{\rm H}$,
$J_{\rm H}$) are re-computed at each time-step along the integration.

The spin parameter $\lambda$ of DM halos is in principle provided by
{\pin}, which is able to predicted the angular momenta of halos.  The
predicted momenta show some (modest) degree of correlation with those
of simulated halos (the spin directions tend to be loosely aligned
within 60$^\circ$), and their statistics is reproduced at the cost of
adding free parameters (Monaco et al. 2002b).  While even a modest
correlation with the N-body solution is an advantage with respect to
drawing random numbers, the complications involved in reconstructing
the spin history of halos weight more than any possible practical
advantage. As a consequence, we prefer to randomly assign a $\lambda$
value to each DM halo, drawing it from the log-normal distribution:

\be
P_{\log\lambda}(\log\lambda)d\log\lambda = \frac{1}{\sqrt{2\pi}\theta} 
\exp\left(-\frac{(\log\lambda-\log\lambda_0)^2}{2\theta^2}\right)
\label{eq:spindist}\ee

\noindent
where we use values $\lambda_0=0.05$ and $\theta=0.3$ (Cole et
al. 2000 use the slightly lower value of 0.23 for $\theta$; we follow
Monaco, Salucci and Danese 2000 using 0.3, which is a better fit to
many spin distributions available in the literature and cited in that
paper).  As explained in section~\ref{section:discs}, a variation of
$\lambda$ with time (for instance at major mergers) creates problems
with the model of disc structure.  To avoid these problems, the
$\lambda$ parameters are held constant during the evolution of each
halo.  This does not imply a constant specific angular momentum for
the halos, as both $M_{\rm H}$ and $U_{\rm H}$ in equation~\ref{eq:Jh}
change with time.

As for DM satellites, it is assumed that their properties (mass, scale
radius, concentration) remain constant after their merging.
Satellites are however subject to tidal stripping, as explained in
section~\ref{section:dynfriction}.  Tidal stripping is only taken into
account in two cases: to strip baryonic mass from discs and bulges, in
case of extreme stripping, and to compute the merging time for
substructure after a merger.  In all the other cases the unstripped
mass of the satellite is used.


\section{Galaxy mergers, destruction and stripping}
\label{section:mergers}


\subsection{Dynamical evolution of satellites}
\label{section:dynfriction}

The computation of merging times for satellites follows the model of
Taffoni et al. (2003).  In the simplest case, two DM halos without
substructure merge, so that the smaller halo becomes a satellite of
the larger one.  The properties of the two halos at the merging time
are computed as explained in section~\ref{section:haloprop}, and the
subsequent evolution of the main halo is neglected (up to the next
major merger).  The orbital parameters are extracted at random\footnote{
We have verified that, as for the angular momentum of DM halos,
{\pin} is able to estimate the orbital parameters of merging halos,
with roughly correct statistics.  Again, we prefer to extract these
parameters from suitable distributions.
} from
suitable distributions, in particular the eccentricity of the orbit
(defined as $\epsilon = J/J_{\rm cir}$, where $J$ is the initial
angular momentum of the orbit and $J_{\rm cir}$ the angular momentum
of a circular orbit with the same energy)
is extracted from a Gaussian distribution with mean 0.7 and variance
0.2, while the energy of the orbit, parameterized with $x_c$ (defined
as $x_c=r_c/r_{\rm H}$, where $r_c$ is the radius of a circular orbit with
the same energy and $r_{\rm H}$ the halo radius) is taken to be 0.5 for all
orbits.  These numbers are suggested by an analysis of the orbit of
satellites in a high-resolution N-body simulation (Ghigna, private
communication), and are slightly different from the results of Tormen
(1997), who found a lower value (0.5) for $\epsilon$ and did not
publish the distribution of $x_c$ (he gave the distribution of the
radius of the first periastron, but this quantity is affected by
dynamical friction in a subtle way); however, we have verified that
choosing Tormen's value does not induce significant differences in the
results.  The choice of $x_c=0.5$ is equivalent to the implicit choice
of Somerville \& Primack (1999), who use a flat distribution for
$\epsilon$, while Cole et al. (2000) extract a combination of the two
parameters, $\epsilon^{0.78}x_c^2$, from a log-normal distribution and
Kauffmann et al. (1999) extract $\epsilon$ again from a flat
distribution and use $x_c=1$.

Galaxy mergings are due to the decay of the orbit of their host DM
satellite by dynamical friction; in this scheme galaxy satellites can
only merge with their central galaxy.  Tidal shocks can lead to a
complete disruption of satellites; in these rather unlikely events all
the (dark and baryonic) matter of the satellite is dispersed into the
halo of the central galaxy.  Taffoni et al. (2003) give fitting
formulae, accurate to the 15\% level, for the merging and destruction
times for substructures that take into account dynamical friction,
mass loss by tidal stripping and tidal shocks.  We use a slightly
different version of those fitting formule, with updated parameters
that slightly improve the fits to simulations; these are given in
appendix~\ref{section:taffoni}.

Though the merger and disruption times of Taffoni et al. (2003)
include a rather sophisticated treatment of tidal stripping, we
implement this process at a very simple level. Tidal stripping is
applied at the first periastron of the satellite.  The tidal radius is
computed as the radius at which the density of the unperturbed
satellite is equal to the density of the main DM halo at the
periastron. All the mass external to the tidal radius is then
considered as unbound from the satellite.  However, while the
stripping radius is recorded, the halo mass is not updated but kept
fixed in the following evolution.  Indeed, disc structure (see
section~\ref{section:discs}) is computed using the DM halo
profile at of the satellite at its merger time (i.e. ignoring
stripping), while the recomputation of merging times after a major
merger is done using the stripped mass.

Tidal stripping  at periastron affects also the fraction of
stellar and gas mass of disc and bulge that lies beyond the tidal
radius. Notice that, for simplicity, neither the DM halo nor the disc
and bulge are assumed to be perturbed by this process beyond the
decrease of mass.  Within the model structure, stripping is applied
outside the numerical integration, so it is an external flow
(section~\ref{section:algorithm}).

In the more general case of two substructured halos that merge, we
distinguish between major and minor mergers as follows:

\be
 {\rm Major\ merger\ of\ DM\ halos:}\ \ \ \ M_{\rm sat}/M_{\rm main} > f_{\rm hmm}
\label{eq:mmcond} \ee

\noindent
(we recall that the main DM halo includes the satellite).  N-body
simulations show that when this condition, with $f_{\rm
hmm}\simeq0.2$, is satisfied the perturbation induced by the satellite
leads to a reshuffling of all the orbits. At a major merger we then
re-extract the orbital parameters and re-compute the merger and
destruction times for all the satellites of the main DM halo, as if
they just entered the halo. As a consequence, some galaxies near to
their merging time can be moved to a different orbit that does not
lead to merging, while some other galaxies can suffer tidal stripping
more than once.  Clearly the re-computation of the merging and
destruction times for a substructure may not be very accurate,
especially for satellites that have suffered strong mass loss.  In
this case we keep the scale radius and concentration of the satellite
fixed, but use (as mentioned above) the stripped mass to compute the
merging and destruction times.  As these times are rather long for
small satellites, the final results is that the galaxies just don't
merge and the accuracy of the prediction is not an issue.

Minor mergers do not influence the evolution of the satellites of the
main DM halo, but do affect the satellites of the smaller DM halo
(going itself to become a satellite), for which there is no difference
between minor and major merger.


\subsection{Galaxy merger trees}
\label{section:galtree}

The galaxy merger trees are constructed, analogously to the DM halo
merger trees, by specifying for each galaxy (i) the galaxy where its
stars lie at the the final time, (ii) the merging redshift, (iii) the
galaxy it has merged with, (iv) a linking list for the merged
galaxies, (v) and (vi) the masses of each pair of merged galaxies.
Destroyed galaxies are recorded by assigning a negative value to field
(i).  The construction of galaxy trees is performed as follows: at
each halo merger the merging and destruction times for the galaxies
are computed (section~\ref{section:dynfriction}), then the galaxies
are merged or destroyed at that time if the DM halo they belong to has
not been involved in a major merger nor has become a satellite in the
meantime.  While multiple mergers are allowed by {\pin}, galaxy
mergers are all binary.


\subsection{Galaxy mergers}
\label{section:galmergers}

When two galaxies merge, their fate depends again on the ratio of
their masses. Major mergers of galaxies are defined as:

\be
{\rm Major\ merger\ of\ galaxies:}\ \ \ \ M_{\rm sat}/M_{\rm cen} > f_{\rm gmm}
\label{eq:mmgal} \ee

\noindent
where the parameter $f_{\rm gmm}$ is suggested by simulations to take
a value of 0.3 (Kauffmann et al. 1999; Cole et al. 2000).  In this
specific case baryonic masses are used (i.e. the mass in hot, cold and
star phases of the bulge and halo components), and the central galaxy
does not include the satellite, so this condition is similar to that
of equation~\ref{eq:mmcond} with a value of 0.25.  While the condition
on DM halos can be computed directly from the {\pin} trees and without
running the galaxy evolution code, the condition of
equation~\ref{eq:mmgal} must be computed at the merger time, after the
evolution code has determined the baryonic galaxy masses.

At minor galaxy mergers the whole satellite is added to the bulge,
while the disc remains unaffected.  This is at variance with Cole et
al. (2000), that give the stars to the bulge and the gas to the
disc. Their choice is however questionable, as the dissipative matter
is more likely to go to the bulge; this is why we prefer to give
everything to the bulge. A more accurate solution of this problem will
clearly depend on the orbit of the merger, but an implementation
of these second-order effects would require a large set of N-body
simulations to quantify and parameterize these dependences.  In any
case,  our tests have revealed no strong difference between the two
cases (see section~\ref{section:bulges} for more comments).

At a major galaxy merger all the gas and stars of the two merging
galaxies are given to the bulge of the central one.  Due to the
shorter time-scale of star formation in bulges (see
section~\ref{section:feedback}), a starburst is stimulated by the
presence of gas in the bulge component.

Stripping and galaxy destruction (which is an extreme stripping event)
bring stars to the halo of the central galaxy. As star formation
in the halo is inactive, this is the main way to have stars to the
halo.  We anticipate that this mechanism is not very effective in
galaxy clusters, where only a few per cent of the stellar mass is
stripped to the halo, at variance with the 10-40 per cent found in
observed clusters (Feldmeier et al. 2003; Arnaboldi et al. 2003;
Gal-Yam et al. 2003; Zibetti et al. 2005).  Murante et al. (2004;
2006) have performed hydro simulations to address this problem,
coming to the conclusion that the high fraction of cluster stars can
be reproduced, but the main mechanism is not tidal stripping but
violent relaxation in major mergers.  Very recently, Monaco et
al. (2006), based on {\gal} and on the N-body results of Murante et
al. (2006), proposed that the construction of the stellar diffuse
component of galaxy clusters, i.e. the halo star phase, is related to
the apparent lack of evolution of the most massive galaxies since
$z=1$.  They showed that the evolution driven by galaxy mergers alone
is larger than what is observed in large galaxy samples, and that this
discrepancy can be solved if a fair fraction of stars is scattered to
the halo star component at each merger.  However, to reproduce the
observed increase of the stellar diffuse component with halo mass, the
fraction of scattered stars must depend strongly on the properties of
the DM halo and of the merging galaxies.  This dependence should be
provided by simulations, but this is a work in progress.  In this
paper we simply assume that a fraction $f_{\rm scatter}$ of the star
mass of the merging galaxies is scattered to the halo at each major
merging. With this simple rule we are able to produce a low fraction
of halo stars in galactic halos and a higher one in galaxy cluster.

Interactions between satellites, like binary mergers, flybies (that
stimulate star formation) or galaxy harassment (Moore et al. 1996)
are not included at the moment. We know that these events can have an
impact on the evolution of galaxies (Somerville \& Primack 1999;
Balland, Devriendt \& Silk 2003; Menci et al. 2004) and AGNs
(Cavaliere \& Vittorini 2000). We plan to introduce such events in the
future.


\section{Halo gas}
\label{section:halogas}


\subsection{Equilibrium model for the hot phase}
\label{section:hot}

The hot halo phase is assumed to be (i) spherical, (ii) in hydrostatic
equilibrium in an NFW halo (Suto, Sasaki \& Makino 1998), (iii)
filling the volume from a cooling radius {\rcool} to the virial radius
{\rh}, (iv) pressure-balanced both at {\rcool} and {\rh}, (v)
described by a polytropic equation of state with index \gammap, a
parameter that is observed to take a value of about 1.2 (see, e.g., De
Grandi \& Molendi 2002); we use the best-fit value of 1.15.  The
equilibrium configuration of the hot halo gas is computed at each time
step as described below, under the assumption that in the absence of
major mergers the gas re-adjusts quasi-statically to the new
equilibrium configuration.  We {\em do not} assume that the thermal
energy $E_{\rm H}$ of the hot halo phase is a fixed fraction of the
virial energy; on the contrary, we follow its evolution through the
equation given in table~1.  This is done to incorporate the response
of the hot halo phase to the thermal feedback from galaxies.  Under
the assumptions given above, the constraints on the profile are then
the mass and thermal energy of the gas.

The equation of hydrostatic equilibrium:

\be
\frac{dP_g}{dr} = -G\frac{\rho_g M_{\rm H}(<r)}{r^2}
\label{eq:hydro} \ee

\noindent
(where $P$ is the gas pressure, $r$ is the radius, $M_{\rm H}(<r)$ is the
halo mass\footnote{
Gravity is supposed to be dominated by DM, so we use for $M_{\rm H}(<r)$ the
NFW mass profile, as if the baryons were distributed as the DM.  The
error induced in this assumption is not likely to be relevant; a more
sophisticated treatment would not allow to obtain an analytic solution
for the gas profile.
} within $r$ and $\rho_g$ the gas density) is easily solved
in the case of an NFW density profile (equation~\ref{eq:nfw}).  The
solution is:

\bea
\rho_g(r)  &=& \rho_{g0}\left(1-a\left(1-\frac{\ln(1+c_{\rm nfw}x)}{c_{\rm nfw}}
\right)\right)^{1/(\gamma_p-1)} \nonumber\\
P_g(r)     &=& P_{g0}\left(1-a\left(1-\frac{\ln(1+c_{\rm nfw}x)}{c_{\rm nfw}}
\right)\right)^{\gamma_p/(\gamma_p-1)} \label{eq:hydro_sol} \\
T_g(r)     &=& T_{g0}\left(1-a\left(1-\frac{\ln(1+c_{\rm nfw}x)}{c_{\rm nfw}}
\right)\right) \nonumber 
\eea

Defining the virial temperature of the halo as $T_{\rm vir}=\mu_{\rm
hot} m_p V_c^2/3k$ (where $\mu_{\rm hot} = 4/(8-5Y-6Z)$ is the
mean molecular weight of the hot gas; $Z=M^Z_{\rm h,H}/M_{\rm h,H}$ is
the metallicity of the hot halo gas and $Y$ assumes the solar value of
0.25), $\eta=T_g(r)/T_{\rm vir}$ and $\eta_0$ its extrapolation at
$r=0$, the constant $a$ is defined as:

\be
a = \frac{\gamma_p-1}{\gamma_p}\frac{3}{\eta_0}
\frac{c_{\rm nfw}(1+c_{\rm nfw})}{(1+c_{\rm nfw})\ln(1+c_{\rm nfw})-c_{\rm nfw}}
\label{eq:adef} \ee

\noindent
The constants $T_{g0}$, $\rho_{g0}$ and $P_{g0}$ are defined as the
extrapolation of the density and temperature profiles to $r=0$, even
though the gas is assumed to be present only beyond {\rcool}.  As
mentioned above, they are fixed by requiring that the total mass and
energy of the gas correspond to $M_{\rm h,H}$ and $E_{\rm h,H}$.   The
first condition can be solved explicitly if the energy is specified.
Calling ${\cal I}$ the integral:

\be
{\cal I}(\alpha) = \int_{r_{\rm cool}/r_s}^{c_{\rm nfw}} \left( 1-a\left(1-\frac{\ln(1+t)}{t}
\right)\right)^{\alpha}t^2 dt
\label{eq:integral} \ee

\noindent
(where for simplicity we declare only the dependence on the $\alpha$
exponent) we have:

\be
\rho_{g0} =\frac{M_{\rm h,H}}{4\pi r_s^3} 
\times \frac{1}{{\cal I} (1/(\gamma_p-1))}
\label{eq:condition1}
\ee

\noindent
The second condition is:

\be
E_{\rm h,H} = \frac{6\pi k T_{g0}\rho_{g0}r_s^3}{\mu_{\rm hot}m_p} 
\times {\cal I} (\gamma_p/(\gamma_p-1))
\label{eq:condition2}
\ee

\noindent
This equation cannot be solved explicitly, as the coefficient $a$
depends on the energy itself through $\eta_0$. To find a solution to
these two equations it is necessary to use an iterative algorithm. As
a consequence, the computation of the two integrals contained in
equations~\ref{eq:condition1} and \ref{eq:condition2} is the most
time-consuming computation of the whole code.  A dramatic speed-up (at
the cost of a negligible error) is obtained by computing the integrals
in a grid of values of $a$, $r$ and $\gamma_p$; the solution is then
found by linearly interpolating the table.

The function $1-a(1-\ln(1+c_{\rm nfw}x)/c_{\rm nfw})$ in
equations~\ref{eq:hydro_sol} becomes negative at large radii.  In this
case density, pressure and temperature are not defined.  Usually this
happens beyond the virial radius {\rh}, unless the central temperature
is lower than the following limit:

\be 
\eta_0 < 3 \frac{\gamma_p-1}{\gamma_p} \frac{c_{\rm nfw}-\ln(1+c_{\rm nfw})}
{\ln(1+c_{\rm nfw})-c_{\rm nfw}/(1+c_{\rm nfw})}
\label{eq:lowt_cond} \ee

\noindent
This condition can be met at high redshift, when \cnfw-values are
high.  In this case no gas is assumed to be present beyond the point
of zero density, so that the gas is bound to the inner part of the
halo and its pressure at the virial radius is null.


This model for the hot halo gas is very similar to that proposed by
Ostriker, Bode \& Babul (2005) to model the hot component of galaxy
clusters, with two remarkable differences: first, they do not follow
cooling, so the hot gas is present since $r=0$, instead of
$r=r_{\rm cool}$.  Clearly this does not make any difference in cases
like galaxy clusters without cool cores, where $r_{\rm cool}\simeq 0$.
Second, they assume an external pressure, computed on the basis of a
fiducial infall velocity of cold gas, and extrapolate the gas profile
until its thermal pressure equals the external one.  This implies that
the hot gas can extend beyond the virial radius.  Our choice is to
remove this outlier gas, and this allows to describe galaxy
super-winds (section~\ref{section:winds}).  Clearly the two criteria
should be equivalent in the case of a small cooling radius and a
thermal energy roughly similar to the virial one; we plan to deepen
this point and to compare the predicted properties of the hot halo gas
with simulations in the future.


\subsection{Shock heating}
\label{section:shock}

The equilibrium model does not specify the amount of thermal energy of
the hot gas.  This is acquired by the infalling gas through shocks.
The cosmological infall mass flow is computed by linearly
interpolating the DM halo mass between the branch ends (whose distance
in time is $\Delta t$) and assuming that a fraction
$\Omega_b/\Omega_0$ of that mass is in IGM:

\be
\dot{M}_{\rm cosm} = \frac{\Omega_b}{\Omega_0}
\frac{ \Delta M_{\rm H} }{ \Delta t }
\label{eq:cosm} \ee

\noindent
We then assume that this gas acquires an energy equal to $f_{\rm
shock}$ times that suggested by the virial theorem:

\be
\dot{E}_{\rm cosm}  = f_{\rm shock} (-0.5\, U_{\rm H})  \dot{M}_{\rm cosm} / M_{\rm H}
\label{eq:shock} \ee

\noindent
where the binding energy of the halo $U_{\rm H}$ is given by
equation~\ref{eq:uh}.  The parameter $f_{\rm shock}$ is suggested by
hydro simulations to be slightly higher than 1 (Wu, Fabian \& Nulsen
2000).  We adopt a value of 1.2.

A similar heating is applied in the following external flows:
\begin{itemize} 
\item the hot gas contained in the DM halos at the appearance time
(see section~\ref{section:ic});
\item the hot halo gas of satellite DM halos at their merging time'
\item the hot halo gas at major mergers.\end{itemize} In all these
cases:

\be
E_{\rm H}  = f_{\rm shock} (-0.5\, U_{\rm H})  M_{\rm h,H} / M_{\rm H}
\label{eq:shock2} \ee

Hydro simulations suggest that any cold gas present in the halo is
re-heated by shocks during a major merger. Accordingly, we allow
shock-heating to affect also the cold halo phase at major mergers.
This option is active in all the results presented here, but can
be switched off on request.

Cases (ii) and (iii) refer to gravitational heating due to merging
events.  This heating is of course not instantaneous; the energy is
redistributed to the whole halo gas in a few crossing times. This
behaviour is implemented by quenching cooling for a number $n_{\rm
quench}$ of crossing times $r_{\rm H}/V_{\rm H}$; after the quenching
(which ends at some time $t_q$), the cooling flow is allowed to start
gradually as $\exp\{-[r_{\rm H}/V_{\rm H}(t-t_q)]^2\}$.  The parameter
$n_{\rm quench}$ is very important to control the cooling of gas,
especially at high redshift.  The same quenching is applied to DM
halos at their appearance time (section~\ref{section:ic}); it
amounts to assuming that the appearing halos have just formed by a
major merger.

Using hydro simulations, some authors (Kravtsov 2003; Keres et
al. 2005) have recently pointed out that cold gas can flow directly to
the core of small-mass halos at high redshift by radiating very
quickly the energy acquired by shocks.  Following this idea, Dekel \&
Birnboim (2006; see also Cattaneo et al. 2006) have proposed that this
lack of shock heating is a likely responsible for the observed
transition in the behaviour from dwarf to bright galaxies.  Besides,
Croton et al. (2006) implement this idea by equating the infall mass
flow with the cosmological infall one in the infall-dominated halos.
A similar view is taken by Bower et al. (2006), who suppose that a
short cooling time prevents the formation of a hydrostatic hot halo
phase.  These considerations question the validity of the shock
heating and hydrostatic equilibrium hypotheses in infall-dominated
halos; however, these are also the cases where the infall of gas on
the galaxy does not depend much on the cooling time, so we choose to
retain these hypotheses as they are accurate in the cases where they
are most relevant.


\subsection{Cooling}
\label{section:cooling}

The cooling radius is defined as the radius within which the hot halo
gas has cooled down.  In most semi-analytic models the hot gas profile
is computed at a major merger; the time-dependent cooling radius is
then computed as the radius at which the cooling time of a gas shell
is equal to the time since the merger.  In the present model the
cooling radius is instead treated as a dynamical variable.  This
allows to re-compute the gas profile at each time-step, and to
incorporate the heating effect of the hot wind coming from the central
galaxy.

The cooling rate of a shell of gas of width $\Delta r$ at a radius $r$
is computed as:

\be \Delta \dot{M}_{\rm co,H}(r) = \frac{\Delta M_{\rm h,H}(r)}{t_{\rm cool}(r)}\, ,
\label{eq:coolshell} \ee

\noindent
where $\Delta M_{\rm h,H}=4\pi r^2 \rho_g(r)\Delta r$ is the shell mass and
$t_{\rm cool}(r) = 
3kT_g(r) \mu_{\rm hot} m_p / 
2\rho_g(r) \Lambda_{\rm cool}(T_g(r))$
is the cooling time at radius $r$.  For the cooling function
$\Lambda_{\rm cool}$, we use the metallicity-dependent function
tabulated by Sutherland \& Dopita (1993). The cooling time depends on
density and temperature, but the density dependence is by far
stronger, both intrinsically and because the temperature profile is
much shallower than the gradient profile.  So the integration in $r$
can be performed by assuming $T_g(r)\simeq T_g(r_{\rm cool})$.
The resulting cooling flow is:

\be
\dot{M}_{\rm co,H} = \frac{4 \pi r_s^3\rho_{g0}}{t_{\rm cool,0}} 
\times {\cal I}(2/(\gamma_p-1))
\label{eq:coolingflow} \ee

\noindent
where ${\cal I}$ is defined in equation~\ref{eq:integral}.  In this
equation the cooling time $t_{\rm cool,0}$ is computed using
$\rho_{g0}$ for the density (the dependence of density on radius is
taken into account by the integrand), and $T_g(r_{\rm cool})$ for the
temperature, as explained above.  Analogously to the integrals of
equations~\ref{eq:condition1} and \ref{eq:condition2}, the integral in
equation~\ref{eq:coolingflow} is computed on a grid of parameter
values and then estimated by linear interpolation on the table.  The
rate of energy loss by cooling is computed analogously:

\be
\dot{E}_{\rm co,H} = \frac{3kT_g(r_{\rm cool})}{2\mu_{\rm hot}m_p}
\frac{4 \pi r_s^3\rho_{g0}}{t_{\rm cool,0}} 
\times {\cal I} (2/(\gamma_p-1))
\label{eq:coolingenergy} \ee

\noindent
The cooled gas carries with it a kinetic energy:

\be
\dot{K}_{\rm co,H} = \frac{1}{2}\dot{M}_{\rm co,H} V_{\rm disp}^2
\label{eq:dkco}\ee

\noindent
where $V_{\rm disp}$ is the velocity dispersion of the DM halo defined
in equation~\ref{eq:vdisp}.

When a heating source is present, these two terms behave differently.
While the energy radiated away by the hot gas at a given density and
temperature does not change, the amount of cooled mass depends on how
much of this energy is replaced by the heating source.  We then
compute the cooling time as:

\be
t_{\rm cool,0} = \frac{3kT_g(r_{\rm cool}) \mu_{\rm hot} m_p}{2\rho_{g0}
(\Lambda_{\rm cool}-\Gamma_{\rm heat})}
\label{eq:cool_heat} \ee

\noindent
This cooling time is used in equation~\ref{eq:coolingflow} to compute
the mass cooling flow.  A negative value implies a net heating of the
source, in which case the mass cooling flow $\dot{M}_{\rm co,H}$ (but
not $\dot{E}_{\rm co,H}$) is set to zero.

The source of heating is the central galaxy, which hampers cooling
through the hot wind energy flow, $\dot{E}_{\rm hw,H}$.  This flow
carries the energy produced by SNe, both in the bulge and in the disc,
and by the AGN.  Satellites instead are assumed to orbit on average in
the external regions, so that the energy contributed by their winds is
injected beyond the cooling radius and does not interact directly with
the cooling flow.  To compute the heating term it is necessary to
specify how this heating is distributed.  For simplicity we assume
that heating and cooling affect the same gas mass $4\pi
r_s^3\rho_{g0}{\cal I}(2/(\gamma_p-1))$, i.e. the inner shell at
$r_{\rm cool}$ that is effectively cooling.  The heating term is then
computed as:

\be 
\Gamma_{\rm heat} = \frac{\dot{E}_{\rm hw,H}}{4\pi r_s^3 
{\cal I}(2/(\gamma_p-1))} 
\left(\frac{\mu_{\rm hot}m_p}{\rho_{g0}}\right)^2
\label{eq:heating}\ee

\noindent
Once the cooling and heating sources are fixed, the evolution of the
cooling radius is computed by inverting the usual relation,
$\dot{M}_{\rm co,H} = -4\pi \rho_gr_{\rm cool}^2 dr_{\rm cool}/dt$,
taking into account that the hot wind mass flow $\dot{M}_{\rm hw,H}$
is adding to the hot halo phase at the cooling radius:

\be
\dot{r}_{\rm cool} = \frac{\dot{M}_{\rm co,H}-\dot{M}_{\rm hw,H}}
{4\pi \rho_g(r_{\rm cool})r_{\rm cool}^2}\, .
\label{eq:drcool} \ee

\noindent
In this way, the cooling radius decreases if the hot wind term
overtakes the cooling term.

Equation~\ref{eq:drcool} shows that the cooling radius should not
vanish.  Moreover, in the case of very strong cooling flows the
Runge-Kutta integrator may try some calls with $r_{\rm cool}>r_{\rm
H}$, giving rise to numerical problems.  The cooling radius is then
forced to lie between a small value (taken to be 0.01 times the scale
radius \rs) and 90 per cent of the virial radius \rh.  This is done by
gradually setting $\dot{r}_{\rm cool}$ to zero when the limits are
approached.  The presence of a small lower limit for $r_{\rm cool}$
does not influence the results, because of the flat density profile in
the central regions.  The upper limit can instead influence
significantly the behaviour of cooling, but this happens when most of
the hot gas has already cooled, a situation in which a precise
modeling is not very important and the validity of the model itself is
more doubtful (see the discussion in section~\ref{section:hot}).

The existence of a ``cooling hole'' at the centre of the halo
requires that, as we assumed in section~\ref{section:hot}, pressure is
balanced at {\rcool}.  However, this assumption is clearly unphysical,
as the cooled gas cannot give sufficient pressure support, so this
assumption will work only as long as $\dot{r}_{\rm cool}>c_s$, where
$c_s$ is the sound speed of the hot gas.  In general, the gas at
{\rcool} will be pushed toward the centre by its pressure.  This can
be modeled very simply as follows:

\be
\dot{r}_{\rm cool}' = \dot{r}_{\rm cool} - c_s
\label{eq:close_hole} \ee

\noindent
We have noticed that this feature induces an excessive and unphysical
degree of cooling in the later stages of evolution, when, due to the
decreasing temperature gradient, the residual thermal energy of the
gas gets lower than the virial energy.  To avoid it we consider the
sound speed term only when the gas thermal energy $E_{\rm h,H}$ is
higher than the virial value.  This term is included on request, and
is used in the results presented here.

Finally, after reionization the ionizing background is likely to
prevent the cooling of any halo whose circular velocity is smaller
than $\sim$50 \kms.  As suggested by Benson et al. (2002), to mimic
the effect of the UV background we simply suppress any cooling in all
halos with $V_c<50$ \kms.


\subsection{Infall}
\label{section:infall}

The dynamical time of the halo at a radius $r$ is defined as the time
required by a mass particle to free-fall to the centre:

\bea \lefteqn{t_{\rm dyn}(r) =
\frac{1}{\sqrt{2}V_c}\left(\frac{3\delta_c} {200c_{\rm nfw}^2}
\right)^{-1/2} r_s } \label{eq:tdyn} \\ &&\times \int_0^{r/r_s} \left[
\frac{\ln(1+y)}{y} -\frac{\ln(1+r/r_s)}{r/r_s} \right]^{-1/2}\, dy
\nonumber \eea

%

\noindent
The cold phase is unstable to collapse and flows to the central
galaxy.  Starting with White \& Frenk (1991), many semi-analytic
codes, at variance with {\gal}, unify the processes of cooling and
infall by computing an infall radius for the gas as the radius at
which the infall time of a gas shell is equal to the time since last
major merger, then using the smallest between the cooling and infall
radii to compute the cooling flow.  This choice implies an assumption
of no difference between the hot halo gas and the cooled gas that is
infalling toward the central galaxy.  The hot wind ejected by a galaxy
acts preferentially on the most pervasive hot phase, affecting in a
much weaker way the cold infalling gas, which naturally fragments into
clouds with a low covering factor.  So, we deem that treating the
infalling gas as belonging to a different phase is a step forward in
the physical description of galaxy formation, especially in those
infall-dominated halos where a high fraction of halo gas is in the
cold phase.  As a matter of fact, the recipe by Croton et al. (2006)
discussed above (section~\ref{section:shock}) goes in the same
direction, because with their assumption of $\dot{M}_{\rm
cosmo}=\dot{M}_{\rm in,H}$ (valid for infall-dominated halos) the
cooled gas is not affected by feedback.

The cold gas is let infall to the central galaxy on a number $n_{\rm
dyn}$ of dynamical times computed at the cooling radius \rcool:

\be
\dot{M}_{\rm in,H} = \frac{M_{\rm c,H}}{n_{\rm dyn}t_{\rm dyn}(r_{\rm cool})}
\label{eq:infall} \ee

\noindent
The corresponding loss of kinetic energy is:

\be
\dot{K}_{\rm in,H} = \frac{K_{\rm H}}{n_{\rm dyn}t_{\rm dyn}(r_{\rm cool})}
\label{eq:kinfall} \ee


The infalling cold gas is distributed between the disc and the bulge
as follows.  As a first option, all the infalling gas is given to the
disc, under the assumption that it has the same specific angular
momentum as the DM halo:

\bea
\dot{M}_{\rm in,B} &=& 0\label{eq:infall_b1} \\
\dot{M}_{\rm in,D} &=& \dot{M}_{\rm in,H}  \label{eq:infall_d1} 
\eea

\noindent
In presence of a significant or dominant bulge the formation of such a
disc by infall implies that the bulge has no influence on it, even
when a large fraction of it is embedded in the bulge.  This is a
rather strong assumption, as the hot pressurized phase pervading the
bulge (section~\ref{section:thick}) can lead to significant loss of
angular momentum of the gas by ram pressure.  Then, as a second option
we let gas infall on the bulge by a fraction equal to the mass of the
disc contained within the half-mass radius of the bulge:

\bea
\dot{M}_{\rm in,B} &=& \dot{M}_{\rm in,H} \times \left[1-\exp\left(-\frac{R_{\rm B}}{2R_{\rm D}}\right)
\left(1+\frac{R_{\rm B}}{2R_{\rm D}}\right)\right] \label{eq:infall_b2} \\
\dot{M}_{\rm in,D} &=& \dot{M}_{\rm in,H} \times \exp\left(-\frac{R_{\rm B}}{2R_{\rm D}}\right)
\left(1+\frac{R_{\rm B}}{2R_{\rm D}}\right) \label{eq:infall_d2} 
\eea

\noindent
Here and in the following, $R_{\rm B}$ denotes the half-mass radius of the
bulge, while $R_{\rm D}$ is the scale radius of the disc.

This option has a significant effect on the ability of quenching
cooling at low redshift by AGN jets (section~\ref{section:jets}): if
the infalling gas has to wait for an external trigger (like a merger)
to get into the bulge component, and from there to accrete onto the
central BH, the feedback from the AGN would be activated with a
significant delay with respect to the start of the cooling flow, while
the activation is much quicker if the infalling gas is allowed to flow
directly into the bulge.

An interesting multi-phase model for cooling and infall has been
proposed by Maller \& Bullock (2004).  While part of the gas that
resides within the cooling radius cools and fragments into clouds, a
fair fraction of it remains at the same temperature and, due to the
drop in density, with a long enough cooling time to prevent its
cooling.  The infall of the clouds to the galaxy is then followed in
detail, taking into account the physical processes (mainly cloud
collisions and ram-pressure drag) that make the gas loose enough
kinetic energy to fall into the galaxy.  This process leads to a
significant slowing down of the infall.  Unfortunately, the Maller \&
Bullock model does not take into account the effect of the residual
pressure on the evolution of the cooling radius.  Clearly our $n_{\rm
dyn}$ parameter gives a poor representation of this complexity, and a
further sophistication of the modeling of this process in {\gal} may
be worth performing in the future.


\subsection{Galaxy super-winds and cosmological fall-back}
\label{section:winds}

Whenever the gas phases of the halo component are too energetic to be
bound to the DM halo, they are allowed to escape as a galaxy
super-wind.

The hot gas is let flow away whenever its energy overtakes the virial
one by a factor $f_{\rm wind}$:

\be 
{\rm hot\ wind\ condition:}\ \ \ E_{\rm hot,H} > f_{\rm wind} E_{\rm vir}\, ,
\label{eq:hotwindcond} \ee

\noindent
where $E_{\rm vir}=(-0.5)U_{\rm H} M_{\rm h,H}/M_{\rm H}$ (see
section~\ref{section:shock}).  The parameter $f_{\rm wind}$ is
inserted to avoid the excessive escape of gas that we have noticed
when it is set to unity; the results are rather stable when $f_{\rm
wind}\sim 2$, so we adopt the best-fit value 1.7 for the rest of the
paper. Clearly this parameter should be fixed by a careful comparison
to hydro simulations.  Calling $t_{\rm sound}=r_h/c_s$ the
sound-crossing time of the halo, the hot wind mass and energy flows
are then computed as:

\bea
\dot{M}_{\rm hsw} &=& \left(1-\frac{f_{\rm wind} E_{\rm vir}}{E_{\rm hot,H}}\right) 
\frac{M_{\rm hot,H}}{t_{\rm sound}}  \label{eq:hotwind_m} \\
\dot{E}_{\rm hsw} &=& \frac{3kT_g}{\mu_{\rm hot}m_p} \dot{M}_{\rm hsw} \label{eq:hotwind_e}
\eea

\noindent
A quick computation can show that the loss of thermal energy by
adiabatic expansion of the hot gas due to the mass loss should be
equal to 2/3 of the energy loss: if $P=\rho kT/\mu_{\rm hot}m_p$ and
$dV=\rho dM$ then $PdV = kT dM/\mu_{\rm hot}m_p$. So we set:

\be
\dot{E}_{\rm ad} = \frac{2}{3} \dot{E}_{\rm hsw}
\label{eq:ead} \ee

A similar model is applied to the cold wind (with $\sigma_{\rm H}^2
=2K_{\rm H} /M_{\rm c,H}$ the velocity dispersion of cold halo clouds
and $t_{\rm kin}= r_h/\sigma_{\rm H}$).  The cold gas is ejected
out of the halo if:

\be
{\rm cold\ wind\ condition:}\ \ \ 
K_{\rm H} > f_{\rm wind} \frac{1}{2} M_{\rm c,H}V_{\rm disp}^2 \, .
\label{eq:coldwindcond} \ee

\noindent
The resulting mass and energy flows are:

\bea
\dot{M}_{\rm csw} &=& \left(1-\frac{f_{\rm wind}V_{\rm disp}^2}{\sigma_{\rm H}^2}\right) 
\frac{M_{\rm c,H}}{t_{\rm kin}} \label{eq:coldwind_m}\\
\dot{K}_{\rm csw} &=& \left(1-\frac{f_{\rm wind}V_{\rm disp}^2}{\sigma_{\rm H}^2}\right) 
\frac{K_{\rm H}}{t_{\rm kin}} \label{eq:coldwind_k}\\
\dot{K}_{\rm ad} & = &\frac{2}{3} \dot{K}_{\rm csw}
\label{eq:kad} \eea

The mass ejected by the DM halo is then re-acquired back by it at a
later time.  We estimate the fall-back time as follows.  The cold and
hot gas phases escape because their typical velocity (kinetic or
thermal) is larger than the escape velocity of the halo they belong
to.  At the end of the integration over a time interval, we then
scroll the merger tree forward in time and compute the time at which
the parent DM halo has a larger escape velocity than the typical
velocity of the ejected gas.  We then let this gas fall back to the DM
halo by adding it to the cosmological infall flow
(equation~\ref{eq:cosm}).  However, while the large-scale structure
outside a DM halo is clustered, galactic super-winds are emitted in a
much more isotropic way.  As a consequence, much mass could be ejected
into voids and never fall back to the DM halo.  We take this into
account by letting only a fraction $f_{\rm back}$ of the ejected gas
fall back to the DM halo.  Our results are remarkably insensitive of
the value of this parameter; we use 0.5 in the following.


\section{Bulge and Disc structure}
\label{section:structure}

For each exponential disc we record its scale radius $R_{\rm D}$ and its
velocity $V_{\rm D}$ at the optical radius, defined as $3.2R_{\rm D}$ (Persic,
Salucci \& Stel 1996).  The half-mass radius of the disc is then equal
to $1.6783 R_{\rm D}$.  For each bulge we record its half-mass radius $R_{\rm B}$
(equal to 1.35 effective radii) and its circular velocity defined as
$V_{\rm B}^2=GM_{\rm B}/R_{\rm B}$.  These quantities are sampled in the time grid
defined in section~\ref{section:algorithm}.


\subsection{Discs}
\label{section:discs}

The size of galaxy discs is computed following an extension of the
model by Mo, Mao \& White (1998) that takes into account the presence
of a bulge.  It is assumed that the hot gas has the same specific
angular momentum as the DM halo, and that this angular momentum is
conserved during the infall.  Moreover, it is assumed that the disc is
exponential.  The angular momentum of the disc is:

\be
J_{\rm D} = 2\pi \int_0^{\infty} V_{\rm rot}(r)\Sigma_{\rm D}(r)r^2dr \, .
\label{eq:Jd} \ee

\noindent
where $\Sigma_{\rm D}(r)=M_{\rm D}\exp(-r/R_{\rm D})/2\pi R_{\rm D}^2$ is the exponential
profile for surface density.  The rotational velocity given in this
formula contains contributions from the DM halo, bulge and disc:
$V_{\rm rot}^2 = V_{\rm H}^2 + V_{\rm B}^2 + V_{\rm D}^2$.  The halo contribution is
simply $V_{\rm H}^2(r) = GM_{\rm H}(r)/r \times (1-\Omega_b/\Omega_0)$\footnote{
In this case the density profiles of DM and baryons are so different
that they need to be treated differently, so that, at variance with
the computation of the hot halo gas profile
(section~\ref{section:hot}) and in agreement with Mo, Mao \& White
(1998), we exclude baryons from the computation of the DM density
profile.
} and an analogous expression is valid for the bulge, for which we
assume a Young (1976) density profile (whose projection gives the
observed de Vaucouleurs profile).  The disc contribution is as usual:

\be
V_{\rm D}^2 = \frac{GM_{\rm D}}{R_{\rm D}} y^2(I_0(y)K_0(y)-I_1(y)K_1(y))\, ,
\label{eq:vdisc}\ee

\noindent
where $y=r/2R_{\rm D}$ and the functions contained in the equation are the
standard Bessel functions.  The specific angular momentum must be
equal to that of the DM halo, $J_{\rm D}/M_{\rm D} = J_{\rm H}/M_{\rm H}$.  This translates
into an equation for $R_{\rm D}$ that must be solved iteratively, starting
from the approximate solution $R_{\rm D}\sim 0.71 \lambda r_{\rm H}$ (the simplest
case described by Mo, Mao \& White 1998).  This computation is a
bottleneck for the whole code, especially if disc structure is updated
at each time-step as the profile of the hot halo gas is.  To speed up
the code, disc structure is updated each time the disc grows in mass
by some fraction, set to 1 per cent.  We have verified that this
approximation reproduces with fair accuracy a disc which is growing in
mass by continuous infall.  Because of feedback, a disc that receives
no infalling gas decreases in mass by ejecting a wind to the halo.
This mass ejection presumably leads to a decrease of surface density
with no change in the radius.  We then decide to re-compute the radius
only when the disc mass increases with respect to the value at the
last update.

Adiabatic contraction of the DM halo as the baryons settle in the
centre is introduced (again following Mo, Mao \& White 1998) by
assuming that the adiabatic invariant $GM(r)r$ is constant.  This
implies that the DM mass within a radius $r$ comes from a larger
radius $r_i$ such that:

\be
M_{\rm DM}(r_i) + M_{\rm D} (r) + M_{\rm B} (r) = \frac{r_i}{r} M_{\rm DM}(r_i)
\label{eq:adcon} \ee

\noindent
(here $M_{\rm DM}$ is the unperturbed mass density profile of the DM
halo, and only the non-baryonic DM is considered).
Equation~\ref{eq:adcon}, which must be solved together with
equations~\ref{eq:Jd} and \ref{eq:vdisc}, introduces a second
iteration in the computation, and then a further slow-down.  A
solution for this would be to solve the equation in a 4D grid of
values of the parameters $\lambda$, $c_{\rm nfw}$, $M_{\rm D}/M_{\rm H}$ and
$M_{\rm B}/M_{\rm H}$, then interpolating the solution on the table.  This
improvement is in project.  In the meantime, the computation of
adiabatic contraction is switched on only on request.  We find that,
at variance with many other authors, its introduction does not
influence strongly the results of the code, and we ascribe this
difference to the introduction of the bulge in the computation of the
angular momentum of the disc (equation~\ref{eq:Jd}), which alone leads
to more concentrated discs.

The specific angular momentum of DM halos changes during their
evolution, both in modulus and in direction.  The change in the latter
quantity, for instance, drives those precessions of discs that are
commonly found in N-body simulations.  As mentioned above, {\pin} can
provide information on the angular momentum of DM halos, so it could
be used to track its evolution.  A simpler choice would be that of
re-drawing $\lambda$ from its distribution
(equation~\ref{eq:spindist}) at each major merger of DM halos.
However, major changes in $J_{\rm H}$ are not handled easily within
the Mo, Mao \& White model if a disc is already formed.  In fact, as
the DM halo grows in mass the main contribution to its angular
momentum is given by the most recently accreted mass shells, while the
model does not take into account the internal distribution of angular
momentum.  As a consequence, the change induced by last accreted and
not yet cooled shell would force a recomputation of the whole disc
structure.  This can lead to unphysical events like sudden bursts of
star formation due to a decrease of $\lambda$ after a DM halo major
merger.  Such events can be avoided either by simply keeping fixed the
value of $\lambda$ for each DM halo, which is our choice, or by using
a more sophisticated algorithm for disc structure, able to consider
the contribution to the angular momentum of the disc by accreting
shells of gas.

A connected sophistication lies in taking into account the internal
distribution of angular momentum, which behaves like $J_{\rm
D}(r)\propto M(r)^\alpha$ (Warren et al. 1992; Bullock et al. 2001),
with the exponent $\alpha$ ranging from 1 to 1.3.  This has important
implications in the modeling of discs, as it implies that the first
cooled gas has a low angular momentum, and then settles into a more
compact, higher density disc.  However, a straighforward
implementation of this criterion is not easy, as it leads to a
coupling of $J_{\rm D}$ to the amount of cooled gas and then to the
disc mass. This is determined by the combined action of cooling and
feedback, which in turn depends on the surface density of cold gas and
then on $R_{\rm D}$ and $J_{\rm D}$.  This leads to nasty oscillations
or numerical instabilities in the integration.

This illustrates an extreme case of another issue which is not
addressed by this code, namely that the re-heated gas of galaxy winds
is assumed to have the same specific angular momentum as the halo,
something that may be unrealistic in many cases.  Clearly the
distribution of angular momentum of gas is a topic that needs much
attention.  Hydro simulations are the right tool to address this
issue, and yet the angular momentum of stellar discs is only recently
showing some hints of numerical convergence in the biggest simulations
(see, e.g., Governato et al. 2006, but see also D'Onghia et al. 2006).
From this point of view, it is remarkable that the simplest
assumptions on the distribution of angular momentum give discs whose
properties are not so different from reality.


\subsection{Bulges}
\label{section:bulges}

If discs contain the baryons that have retained their angular
momentum, bulges contain the baryons that have lost most of it.  In
{\gal} there are four ways to let gas lose angular momentum and flow
to the bulge.  The first is direct infall from the halo.  Indeed, as
suggested by Granato et al. (2004) and mentioned above, the first
fraction of gas that collapses is likely to have a low angular
momentum, and is then doomed to become a spheroid.  We do not
implement this idea directly, to avoid the numerical problems
mentioned in the previous subsection.  However, as explained in
section~\ref{section:infall}, we allow gas to infall in an existing
bulge by a fraction equal to the fraction of disc mass embedded in a
bulge.

The second mechanism is bar instability.  Whenever a disc embedded in
a DM halo has a sufficiently high surface density, it becomes unstable
to bar formation.  The bar brings a fair fraction of the disc mass
into the bulge; we assume this fraction to be $f_{\rm bar}=0.5$.  For
the condition for disc stability we use the following (Efstathiou,
Lake \& Negroponte 1982; Christodoulou, Shlosman \& Tohline 1995; Mo,
Mao \& White 1998):

\be
\epsilon \simeq V_{\rm D} \sqrt{\frac{R_{\rm D}}{GM_{\rm D}}} > \epsilon_{\rm limit} \, .
\label{eq:discstab}\ee

\noindent
The value for $\epsilon_{\rm limit}$ ranges from 1.1 for stellar discs
to 0.9 for gaseous disc; as this effect is important at high redshift,
when discs are mostly gaseous, we use the lower value.  This condition
is checked at each integration time-step, and whenever it is not
satisfied the integration is interrupted.  At this point a fraction
$f_{\rm bar}$ of the disc mass (that we assume to be 0.5) is given to
the bulge, and the integration is started again; in other words, bar
instability is implemented as an external flow.  No starburst is
explicitly connected with this event, but the presence of gas in the
bulge gives rise to a stronger burst of star formation (see
section~\ref{section:thick}).  The results are not strongly
sensitive to the parameter $f_{\rm bar}$, as the unstable discs
typically acquire so much mass that they undergo a series of
consecutive bar instabilities; changing the value of $f_{\rm bar}$
influences a bit the pattern of repeated starbursts but does not
influence much the final mass in the bulge.

The third mechanism is of course the merging of galaxies.  The
implementation of mergers has already been described in
section~\ref{section:galmergers}.  It is worth repeating here that
the
uncertainty in the fate of gas in minor mergers (we put the whole
satellite in the bulge, other authors make different choices) does not
influence much the results.

In both cases of disc instability and mergers the radius $R_{\rm B}$ of the
bulge formed by the merger of objects 1 and 2 is computed as suggested
by Cole et al. (2000):

\be
\frac{(M_1+M_2)^2}{R_{\rm B}} =\frac{M_1^2}{R_1} + \frac{M_2^2}{R_2}
+\frac{f}{c} \frac{M_1M_2}{R_1+R_2}\, .
\label{eq:merger}\ee

\noindent
Here $R_1$ and $R_2$ are the half-mass radii of the merging galaxies,
computed assuming (as done above) an exponential profile for the
disc and a Young (1976) profile for the bulge.  For the parameter
$f/c$ we use the values of 2 for mergers and 4 for disc instabilities,
as suggested by Cole et al. (2000).

The fourth mechanism to form a bulge is through feedback.  A necessary
condition for a gaseous disc not to transfer angular momentum outwards
is that the dissipative cold gas has a low enough velocity dispersion,
a condition which is satisfied in observed nearby discs.  However, the
kinetic energy of the cold phase is determined by feedback from SNe.
We anticipate (see section~\ref{section:kinfeed}) that for discs with
a high enough surface density of cold gas the velocity dispersion of
cold clouds can be much higher.   This process is observed to be
at play in high redshift galaxies (Genzel et al. 2006)
and implies a loss of angular momentum and a corresponding thickening
of the disc into a bulgy object.  We model this event by simply
forcing a bar instability (i.e. transferring a fraction $f_{\rm bar}$
of mass to the bulge) whenever the surface density of cold gas
$\Sigma_{\rm cold,D}$ overtakes a limiting value:

\be
\Sigma_{\rm cold,D} > \Sigma_{\rm limit}
\label{eq:thickdisc} \ee

\noindent
This mechanism is used in Fontanot et al. (2006a) to enhance the
number of bright quasars, but is not used in the results presented in
this paper.


\section{Star Formation and Feedback}
\label{section:feedback}

The treatment of feedback in the present version of {\gal}
follows the multi-phase model of M04.  As anticipated in
section~\ref{section:baryons}, instead of implementing a full
multi-phase model of the ISM (which would be straightforward as far as
the coding is concerned, but would lead to a number of numerical
problems) we prefer to insert the main results of the M04 model as
simple recipes, so as to simplify the numerical integration and to
gain a more immediate grasp on the physical processes inserted.
Anyway, we stress that the insertion of a physically motivated model
for feedback in place of a set of phenomenological recipes (as done in
most other galaxy formation models) is one of the most important
features of {\gal}.




M04 studied the evolution of the ISM under the following assumptions:
\begin{itemize}
\item the ISM is composed by two phases, a hot and a cold one, in
thermal pressure equilibrium; \item collapsing and star-forming clouds
arise from the cold clouds by kinetic aggregations; \item type II SNe
exploding within a star-forming cloud give rise to a single
super-bubble; \item the super-bubble propagates in the most pervasive
hot phase; \item the super-bubble expands until either it is stopped by
the external pressure or it blows out of the structure.  \end{itemize} 
Four possible
self-regulated feedback regimes follow naturally from this setting,
depending on whether the super-bubbles stop by pressure confinement or
blow-out, before or after the internal hot gas has started cooling (so
as to form a pressure-driven snowploughs, hereafter PDS).

\begin{figure}
\centerline{
\includegraphics[width=8cm]{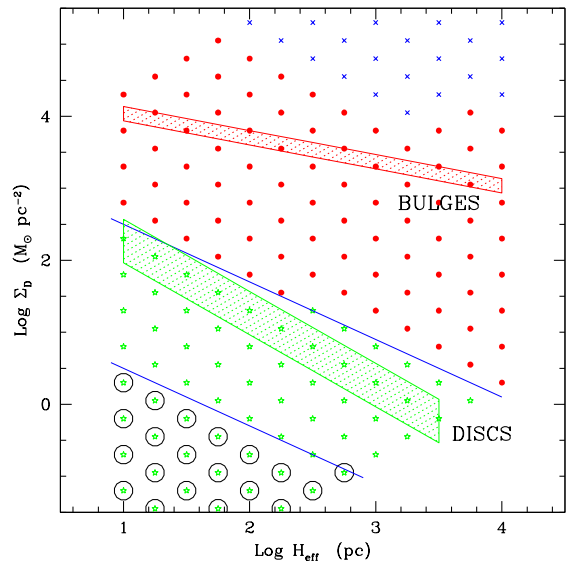}
}
\caption{Feedback regimes as a function of surface mass density and
vertical scale-length.  Green stars denote systems in the adiabatic
blow-out regimes (the encircled points highlight the unstable systems;
see M04), red dots denote systems in the adiabatic confinement regime,
blue crosses denote systems in the PDS confinement regime.  The blue
continuous lines separate the regions with different feedback regimes.
The green and red shaded areas give the typical regions occupied by
galaxy discs and bulges. This is an adapted version of figure~3a of
M04.  }
\label{fig:regimes}
\end{figure}

In M04 it was shown that the dynamics of feedback depends
mainly on the total surface density of the disc $\Sigma_{\rm D}$ and
vertical scale-height {\heff} of the system.  In realistic systems
feedback can take place in the regimes where super-bubbles blow-out or
are confined in the adiabatic stage.  
Figure~\ref{fig:regimes} shows that the adiabatic blow-out and
confinement regimes take place in two different regions of the
$\Sigma_{\rm D}$-{\heff} space, separated by the relation:

\be
\Sigma_{\rm D} = 8 \left(\frac{H_{\rm eff}}{1000\ {\rm pc}}\right)^{-0.8}
\ \ {\rm M}_\odot\ {\rm pc}^{-2} 
\label{eq:fbregime}
\ee

\noindent
The numerical constants in this relation depend on the (uncertain)
values of the parameters used in the model, so they are to be
considered as indicative.

Galaxy discs and bulges tend to stay respectively in the adiabatic
blow-out and confinement regimes, so at a basic level there is no much
doubt on how to apply the feedback regimes within a galaxy formation
code: discs are in the adiabatic blow-out regime, bulges are in the
adiabatic confinement regime.  However, the central region of a disc
can be in the adiabatic confinement regime for at least two reasons:
(i) it is embedded in a bulge, so that its ISM is pressurized by the
bulge hot phase; (ii) its surface density is high enough to cross the
limit of equation~\ref{eq:fbregime} for adiabatic confinement.  This
happens at a typical surface density of $\Sigma_{\rm D} \sim100 - 300$
\surf.  Point (i) has been used to justify the direct infall of gas
from the halo to the bulge (section~\ref{section:infall}).  Point (ii)
has been mentioned in section~\ref{section:bulges} to argue for
feedback-induced loss of angular momentum, and the effects of such
feedback have been modeled by forcing bar formation when the surface
density of cold gas is high (equation~\ref{eq:thickdisc}).  We
show more details and results on this mechanisms in Fontanot et
al. (2006a).

To fully implement the M04 feedback recipes in {\gal} it is necessary
to assess the role of the velocity dispersion of cold clouds, a
quantity which is left as a free parameter in that paper. This is done
in the following subsection. After that we explain how we implement
feedback in the case of thin or thick systems.  


\subsection{Kinetic feedback and the velocity dispersion of cold clouds}
\label{section:kinfeed}

We do not include here a full self-consistent treatment of the kinetic
energy of cold clouds, but try to take into account its effect as
follows.  The velocity dispersion $\sigma_{\rm cold}=\sqrt{2K/M_{\rm
cold}}$ of cold clouds in the ISM of a star-forming galaxy is
determined as a first approximation (i.e. neglecting all mass flows)
by the equilibrium between the injection of kinetic energy by SNe and
the dissipation by turbulence:

\be \dot{K} = \dot{K}_{\rm SN} - \dot{K}_{\rm ds}
\label{eq:kinetic} \ee

The injection of kinetic energy is:

\be
\dot{K}_{\rm SN} = \epsilon_k \frac{\dot{M}_{\rm sf}}{M_{\rm \star,SN}} 
E_{\rm SN} 
\label{eq:kininj} \ee

\noindent
where $E_{\rm SN}$ is the energy of a SN, $M_{\rm \star,SN}$ is the
mass of newly formed stars per SN and $\epsilon_k$ is its fraction
given to the cold phase as kinetic energy.  The rate of energy
dissipation is observed in hydro simulation to be of the order (Mac
Low 2003):

\be
\dot{K}_{\rm ds} = \frac{1}{2} \frac{M_{\rm cold}\sigma_{\rm cold}^3}{L_{\rm drive}}
\label{eq:turb} \ee

\noindent
where $L_{\rm drive}$ is the scale at which turbulence is driven,
suggested to be twice the diameter of the typical size of the
super-bubble.  To determine $\sigma_{\rm cold}$ we assume that an
equilibrium configuration is quickly reached, so that $\dot{K}=0$.  We
then obtain:

\be
\sigma_{\rm cold} = \sigma_0\,  \left(\frac{t_\star}{1\ {\rm Gyr}}\right)^{-1/3}
\label{eq:kinfb} \ee

\noindent
where $\sigma_0 = [2L_{\rm drive}\epsilon_k E_{\rm SN}/ M_{\rm
\star,SN}(1\ {\rm Gyr})]^{1/3}$ and the time-scale for star formation
is $t_\star=M_{\rm cold}/\dot{M}_{\rm sf}$.  In a thin system like the
Milky Way the fraction of (thermal and kinetic) energy injected into
the ISM is predicted by M04 to be about 5 per cent, so $\epsilon_k$
will be of order of 0.01, while the driving scale should be of order
of the vertical scale-length of the system, $\sim100$ pc.  The
numerical value of $\sigma_0$ is then:

\bea
\lefteqn{\sigma_0 = 9.3 \left(\frac{L_{\rm drive}}{100\ {\rm pc}}\right)^{1/3}}\label{eq:sigma0}\\
&& \times
\left(\frac{\epsilon_k}{0.01}\right)^{1/3} E_{51}^{1/3}
\left(\frac{M_{\rm \star,SN}}{120\ {\rm M}_\odot }\right)^{-1/3}\ {\rm km\ s}^{-1}
 \nonumber\eea

The predicted value of $\sigma_{\rm cold}$ for a Milky Way-like galaxy
(with $t_\star\sim 2$ Gyr) is then $\sim7$ \kms, in remarkable
agreement with the observed value which ranges between 6 and 9 \kms.
Much higher values of the velocity dispersion are expected in thick
systems, where $\epsilon_k$ is much higher (all the SN energy is
retained by the system) and $t_\star$ much lower.  This is in
agreement with the sparse data available (see, e.g., Dib, Bell \&
Burkert 2006 for local galaxies, and Genzel et al. 2006
for galaxies at $z\sim2$).  As a consequence, we neglect kinetic
feedback in discs but take it into account in bulges.  Given the great
uncertainty in the driving scale of turbulence in mergers, the
normalization constant $\sigma_0$ will be left as a free parameter.

We stress that equation~\ref{eq:sigma0} is valid under the assumption
that turbulence is driven by SNe and not by gravitationally induced
motions like differential rotation or tidal disturbances in mergers.
In the case of our Galaxy, Mac Low (2003) argues from energetic
arguments that SNe are the most likely drivers of turbulence, and the
good agreement of the predicted value of $\sigma_{\rm cold}$ given
above with the observed one suggests that they are at least a
significant contributor.  In thick systems, like gas-rich spheroids or
mergers, gravitationally-induced turbulence will be very important.
However, the most relevant consequence of a high value of $\sigma_{\rm
cold}$ for our purposes is the massive removal of cold gas during
episodes of strong star formation. It is clear that if a condition
$\sigma_{\rm cold}>V_{\rm B}$ is reached in a gas-rich bulge, then the
driver of turbulence must be star formation and not gravity, while in
the case of a low value of $\sigma_{\rm cold}$ no mass is removed.
So, in this case the assumption will work in the range where its
results are most influential.


\subsection{Discs as thin systems}
\label{section:thin}

In thin systems, super-bubbles blow out of the disc soon after they
form and while they are still in the adiabatic stage; their porosity
remains low.  The fraction of the SN energy that is injected into the
ISM is limited to a few per cent of the total budget, while most
energy is directly injected into the halo through a tenuous,
metal-rich hot wind.  The ISM self-regulates to a configuration that
is very similar to that of the Milky Way (M04).  The star formation
regulates to a level similar to that found in nearby galaxies by
Schmidt (1959) and quantified more recently by Kennicutt (1998).

In M04 the resulting star-formation timescales were incorrectly scaled
with the infall time and the total surface density $\Sigma_{\rm D}$ of the
system.  A more careful analysis reveals that, for a fixed set of
parameters and at fixed \heff, $t_\star$ depends mostly on the cold
gas surface density $\Sigma_{\rm cold,D}$.  This is expected, as the
star-formation timescale is determined by the intrinsic properties of
the ISM, not by the rate at which mass is acquired by it or by the
amount of stars present.  A better fit of the results of M04 in the
adiabatic blow-out regime (for the standard set of parameters defined
in that paper) gives:

\be
t_\star = 6\, 
\left(\frac{\Sigma_{\rm cold,D}}{1\ {\rm M}_\odot\ {\rm pc}^{-2}}\right)^{-0.4} 
\left(\frac{H_{\rm eff}}{100\ {\rm pc}}\right)^{0.65}\ {\rm Gyr}
\label{eq:fbthin1}\ee

In the simple case of a constant value of {\heff} for all galaxies,
the Schmidt-Kennicutt law is recovered with a very similar
normalization (however, observations refer to integrated quantities,
this prediction refers to a section of the disc).  But the assumption
of a constant {\heff} is rather artificial, as the vertical
scale-length of the cold gas is determined by the vertical
gravitational equilibrium as $H_{\rm eff}=\sigma_{\rm cold,D}^2/\pi
G\Sigma_{\rm D}$, so a constant value would imply a tuned variation of
$\Sigma_{\rm D}$ and $\sigma_{\rm cold,D}$.  The other simple
possibility is to assume a constant value for $\sigma_{\rm cold,D}$.
This leads to the prediction $t_\star \propto \Sigma_{\rm
cold,D}^{-1.05} f_{\rm cold}^{0.65}$, where $f_{\rm cold,D}$ is the
fraction of cold gas in the disc.  This relation is apparently steeper
than the observed Schmidt law; however, $\Sigma_{\rm cold,D}$ and
$f_{\rm cold,D}$ are usually correlated in galaxies, in that lower
$\Sigma_{\rm cold,D}$ discs, having lower star-formation rates, retain
a greater fraction of their gas.  For instance, with our reference
choice of parameters (section~\ref{section:results}) we obtain $f_{\rm
cold,D}\propto \Sigma_{\rm cold,D}^{0.8}$, which implies the much
shallower relation $t_\star \propto \Sigma_{\rm cold,D}^{-0.53}$.

The dependence of the velocity dispersion on the star-formation
timescale, equation~\ref{eq:kinfb}, is then inserted, through the
definition of $H_{\rm eff}$, into equation~\ref{eq:fbthin1}, and the
resulting relation is normalized so as to lie on the average
Schmidt-Kennicutt relation at $\Sigma_{\rm cold,D}=13$ \surf (the gas
density of the Milky Way, obtained by assuming $R_{\rm D}=3.5$ kpc and
$M_{\rm cold,D}=5\times 10^9$ \msun, see Cox 2000).  The resulting
relation for the star-formation timescale in discs is:

\be
t_{\rm\star,D} = 9.1\; \left(\frac{\Sigma_{\rm cold,D}}{1\ {\rm M}_\odot\ {\rm pc}^{-2}}\right)^{-0.73} 
\left(\frac{f_{\rm cold,D}}{0.1}\right)^{0.45}\ {\rm Gyr}
\label{eq:fbthin2}\ee

\noindent
This relation is again apparently steeper than the Schmidt-Kennicutt
one, but when the correlation between $\Sigma_{\rm cold,D}$ and
$f_{\rm cold}$ is taken into account we obtain $t_{\rm \star,D}
\propto \Sigma_{\rm cold,D}^{-0.37}$ or $\Sigma_{\rm sfr,D} \propto
\Sigma_{\rm cold,D}^{1.37}$, in excellent agreement with the observed
Schmidt law.

Other complications can be introduced in this picture: if the driving
length $L_{\rm drive}$ is equal to twice the diameter of the
blowing-out super-bubbles, it will scale with {\heff}.  Moreover, the
efficiency of feedback is known in M04 to increase slightly with
$\Sigma_{\rm D}$ (and then with $\Sigma_{\rm cold,D}$). Introducing these
dependences we obtain a scaling of $t_\star$ with $\Sigma_{\rm cold,D}$
and $f_{\rm cold,D}$ similar to equation~\ref{eq:fbthin2}, with slightly
different exponents.  We have verified that these somewhat different
descriptions of feedback do not give significantly different results,
so we keep equation~\ref{eq:fbthin2} as a reference.

According to M04, in the adiabatic blow-out regime the mass
ejection rate due to the action of super-bubbles is modest, while
much gas belonging to the hot phase leaks out of the halo
simply because it is too hot to be confined by the disc.  This
unavoidable mass flow, observed as a hot corona surrounding the
star-forming discs (see, e.g., Fraternali et al. 2002), amounts to a
mass ejection rate very similar to the star-formation rate.  We then
set the disc hot wind term $\dot{M}_{\rm hw,D}$ as equal to the star
formation rate $\dot{M}_{\rm sf,D}$.  This is at variance with most
other galaxy formation models, where the rate of gas re-heating,
equivalent to our $\dot{M}_{\rm hw,D}$, is related to the
star-formation rate through a $\beta$ parameter, assumed to scale as a
power of the disc velocity.  Our assumption is equivalent to
$\beta=1$; the higher ejection efficiency in small discs will then be
due to galaxy super-winds triggered by the energy injection from the
galaxy.  Finally, we neglect any cold wind term, due to the results of
M04 and to the predicted low level of turbulence in discs
(section~\ref{section:kinfeed}).

Using the instantaneous recycling approximation (see also
section~\ref{section:metals}), we let a fraction $f_{\rm rest}$ of the
mass of newly formed stars to be restored to the ISM.  Summing up, the
star-formation, restoration and wind mass flows in discs are:

\be
\begin{array}{lll}
\dot{M}_{\rm sf,D} &=& \dot{M}_{\rm cold,D}/t_{\rm \star,D} \\
\dot{M}_{\rm rs,D} &=& f_{\rm rest} \dot{M}_{\rm sf,D} \\
\dot{M}_{\rm hw,D} &=& \dot{M}_{\rm sf,D}  \\
\dot{M}_{\rm cw,D} &=& 0 
\end{array}
\label{eq:disc_flows}\ee

The hot wind mass flow does not carry much thermal energy: assuming a
typical temperature of $10^6$ K (M04) and for $M_{\rm \star,SN}=120$
\msun\ and $10^{51}$ erg per SN, the fraction of SN energy carried
away by this wind is $\simeq0.05$.  However, most SN energy is blown
out to the halo by the blowing-out super-bubbles, and this energy
mixes with the thermal energy of the hot outflowing wind.  We then
assume that a fraction $f_{\rm th,D}$ of a $10^{51}$ erg SN is carried
by the hot wind gas; this fraction is estimated to be $\sim0.8$ by
Monaco (2004b).  However, this number is highly uncertain because (i)
the loss of SN energy within the star-forming cloud could be higher,
as many authors working on the physics of ISM have often claimed, (ii)
the SN energy could be well in excess of $10^{51}$ erg, (iii) some
energy could be lost in the interaction with the hot halo gas.
Moreover, there is a degeneracy between this parameter and $M_{\rm
\star,SN}$, which depends on the IMF.  It is then wise to leave
$f_{\rm th,D}$ as a free parameter; it is however remarkable that its
best-fit value turns out to be 0.5, not far from the value suggested
by Monaco (2004b).

To the energy of type II SNe we add a contribution from type Ia SNe as
one per year per $10^{12}$ {\msun} of stars.  This is done in order to
test the energetic effect of these SNe.  In the case of thin systems
with a continuous star formation, and given the uncertainty on $f_{\rm
th,D}$, the contribution from type Ia SNe is almost irrelevant.  As
suggested for instance by Recchi et al. (2002), type Ia SNe may have a
higher efficiency of energy injection, as they don't explode in the
dense molecular clouds. As a consequence, one might want to have a
different efficiency $f_{\rm th,Ia}$ for the two SN types.  However,
given the modest importance of these objects we prefer not to add a
further parameter.

The contribution of the disc to the hot wind energy term is then:

\be
\left. \dot{E}_{\rm hw,H}\right|_{\rm disc} = f_{\rm th,D} E_{\rm SN}
\left(\frac{\dot{M}_{\rm sf,D}}{M_{\rm \star,SN}}
+ 1\ {\rm SN\ yr}^{-1} \frac{M_{\rm star,D}}{10^{12}\ {\rm M}_\odot} \right)
\label{eq:hwdisc}\ee

\noindent
The outflowing hot gas will interact with the halo cold phase in a way
that is difficult to model.  In order to be able to constrain the
effect of this process, we allow for a fraction of the thermal energy
to be given to the halo cold phase:

\be
\left. \dot{K}_{\rm w,H}\right|_{\rm disc} = f_{\rm kin} 
\left. \dot{E}_{\rm hw,H}\right|_{\rm disc}
\label{eq:kinfbdisc}\ee

\noindent
For simplicity we do not subtract this energy from the hot wind
budget.  We stress that this term is not due to a contribution of cold
wind but to the interaction of the hot wind with the cold halo phase.

Star formation in the external parts of galaxy discs is known to be
truncated (see, e.g., Kennicutt 1989), and this is likely due to the
differential rotation, even though also the thermal transition from
warm to cold gas may play a crucial role, as suggested by Shaye
(2004).  This author also suggests that the threshold for star
formation may be simply expressed in terms of a threshold for gas
surface density, which scales with the disc and ISM properties and,
very importantly, with the ionizing UV background.  We implement this
threshold $\Sigma_{\rm thr}$ for star formation simply by increasing
the star-formation timescale $t_{\rm\star,D}$ by the inverse of the
fraction of disc mass for which $\Sigma_{\rm cold,D}>\Sigma_{\rm
thr}$.  Clearly, a better but more complicated choice would be to
divide the disc into three zones, corresponding to the bulge, the
star-forming disc and the gas beyond the star-formation threshold;
this is left to future work.  We have also tried to modulate this
threshold with redshift, following the increase by an order of
magnitude of the ionizing background from $z=0$ to 1 (Bianchi,
Cristiani \& Kim 2001).  However, the dependence of the threshold on
the ionizing background, is so weak that this modulation does not
influence the results appreciably.


\subsection{Bulges as thick systems}
\label{section:thick}

In the model by M04 the gravitational perturbations to the ISM are
neglected.  This assumption is questionable for spiral discs, but the
coincidental similarity between the timescale for kinetic aggregations
and the disc timescale, that drives the sweeping by spiral arms, makes
the kinetic aggregation mechanism a good substitute for spiral arms in
creating massive clouds, at least at an order-of-magnitude level.
Things change in the case of thick systems like bulges, where
most energy is efficiently pumped into the hot phase so that the ISM
is much more pressurized, the star-forming clouds are much smaller and
denser, and kinetic aggregations are disfavored.  This leads to the
prediction of a lower level of star formation compared with the
Schmidt-Kennicutt law (figure 7 of M04).  However, tidal disturbances
will be much stronger in a bulgy object, at least when it is formed
through a disc instability or a merger, so that the formation of
collapsing clouds will not be determined by kinetic aggregations.
Moreover, in a major merger the pressurization due to the onset of the
adiabatic confinement regime will cause a quick drop of the Jeans
mass, and this will make most clouds present in the merging discs
collapse and form stars in a short time.  These transient effects have
not been properly modeled by M04, so we prefer to determine the
star-formation timescale directly from the Schmidt law:

\be
t_{\rm \star,B} = 4\;  \left(\frac{\Sigma_{\rm cold,B}}{1\ {\rm M}_\odot\ {\rm pc}^{-2}}\right)^{-0.4} \ {\rm Gyr}
\label{eq:schmidt} \ee

\noindent
The resulting star-formation and restoration rates in bulges are:

\be
\begin{array}{lll}
\dot{M}_{\rm sf,B} & = & M_{\rm c,B}/t_{\rm \star,B} \\
\dot{M}_{\rm rs,B} & = & f_{\rm rest} \dot{M}_{\rm sf,B} 
\end{array}\label{eq:sfbulge} \ee

As for the thin system case, the rate at which hot gas flows to the
halo is similar to the star formation rate, so that we retain this
prediction.  While in thin systems most energy (though not most mass)
is injected to the halo mainly by blowing-out super-bubbles, in thick
systems the energy is ejected mainly through this hot wind.  In the M04
model the typical temperature of the hot phase is found to be of
$\sim10^7$ K, higher by an order of magnitude than the thin system
case; the fraction of SN energy carried away by the hot wind is then
$f_{\rm th,B}\simeq0.5$, slightly lower than the thin system case.
Moreover, while a disc is unable to confine a gas phase with a
temperature as high as $\sim10^6$ K, a massive bulge can
gravitationally confine its hot phase, whose temperature corresponds
roughly to the virial temperature of a bulge with $V_{\rm B}\sim V_{\rm hot}
= 300$ \kms.  We then limit the hot wind as follows:

\be
\beta = \frac{\sqrt{V_{\rm hot}^2-V_{\rm B}^2}}{V_{\rm hot}}
\label{eq:betab} \ee

\noindent
where of course $\beta=0$ if $V_{\rm B}>V_{\rm hot}$.  The resulting hot wind
rate is:

\be
\left.\dot{M}_{\rm hw,B}\right|_{\rm therm} =  \beta \dot{M}_{\rm sf,B}
\label{eq:hwbmass} \ee

\noindent
while the hot wind energy is:

\bea
\lefteqn{\left. \dot{E}_{\rm hw,H}\right|_{\rm bulge} }\\ &=& f_{\rm th,B} E_{\rm SN}
\beta \left(\frac{\dot{M}_{\rm sf,B}}{M_{\rm \star,SN}}
+ 1\ {\rm SN\ yr}^{-1} \frac{M_{\rm star,B}}{10^{12}\ {\rm M}_\odot} \right) \nonumber
\label{eq:hwbulge}\eea

\noindent
As for the thin systems, we allow some energy to accelerate the cold
halo phase:

\be
\left. \dot{K}_{\rm w,H}\right|_{\rm bulge,hw} = f_{\rm kin} 
\left. \dot{E}_{\rm hw,H}\right|_{\rm bulge}
\label{eq:kinfbbulge}\ee

Because in thick systems all the energy of SNe is injected to the ISM
and star-formation timescales are much shorter (in virtue of the
increased surface density), the injection of kinetic energy leads to
velocity dispersions of the cold phase much larger than the $\sim$7
\kms\ found in discs.  In this case some cold clouds may get enough
kinetic energy to be able to leave the potential well of the bulge.
The probability that a cold cloud is unbound, i.e. it has a velocity
larger than the escape velocity of the bulge, $\sqrt{2}V_{\rm B}$, is
computed under the hypothesis of a Maxwellian distribution of
velocities with rms $\sigma_{\rm cold,B}$:

\be
P_{\rm unb}={\rm erfc} \left(\frac{V_{\rm B}}{\sigma_{\rm cold,B}}\right)+\frac{2}{\sqrt{\pi}}
\frac{V_{\rm B}}{\sigma_{\rm cold,B}}\exp\left(-\frac{V_{\rm B}^2}{\sigma_{\rm cold,B}^2}\right)
\label{eq:punbound}\ee

\noindent
The average velocity of the unbound clouds is:

\be
v_{\rm unb} = \sigma_{\rm cold,B} \frac{4}{\sqrt{2\pi}}
\frac{\left(1+\left(\frac{V_{\rm B}}{\sigma_{\rm cold,B}}\right)^2\right)
\exp\left(-\frac{V_{\rm B}^2}{\sigma_{\rm cold,B}^2}\right)
}{P_{\rm unb}}
\label{eq:vunb} \ee

\noindent
The velocity of the clouds after being ejected out of the bulge
will be:

\be
v_{\rm out} = \sqrt{v_{\rm unb}^2 - 2V_{\rm B}^2}
\label{eq:vout} \ee

\noindent
This process will generate a wind flux roughly equal to the unbound
mass, $P_{\rm unb} M_{\rm c,B}$, divided by the crossing time
$R_{\rm B}/v_{\rm unb}$.  This outflow is naturally identified with a cold
wind, adding mass to the cold halo phase.  The cold wind mass flow is:

\bea
\dot{M}_{\rm cw,B} & = &  M_{\rm c,B} P_{\rm unb}
\frac{v_{\rm unb}}{R_{\rm B}} \label{eq:kinwind_cm}\\
\left. \dot{K}_{\rm w,H} \right|_{\rm bulge,cw} & = & \frac{1}{2}
\dot{M}_{\rm cw,B} v_{\rm out}^2 \label{eq:kinwind_ck}
\eea

\noindent
However, the fate of the cold clouds ejected by a bulge may be
different.  The exit from the pressurized bulge to the halo would lead
to an expansion of those clouds, making them much more sensitive to
the bow shock generated in their interaction with the hot halo gas and
to thermal evaporation.  As a result, the cold clouds ejected by the
bulge could be heated and mix with the hot phase.  This is analogous
(though not in the details) to what happens when kinetic feedback is
implemented in SPH simulations: a cold particle neighbouring a
star-forming region is given some kinetic energy, but this energy is
quickly thermalized by the interaction with the other particles.  If
this is the case, the cold and hot wind flows should be given as
follows:

\bea
\left.\dot{M}_{\rm hw,B}\right|_{\rm kin} & = &  M_{\rm c,B} P_{\rm unb} 
\frac{v_{\rm unb}}{R_{\rm B}} \label{eq:kinwind_hm1}\\
\dot{M}_{\rm hw,B} & = & \left.\dot{M}_{\rm hw,B}\right|_{\rm therm} +
\left.\dot{M}_{\rm hw,B}\right|_{\rm kin}
\label{eq:kinwind_hm2} \\
\dot{M}_{\rm cw,B} & = & 0 \label{eq:kinwind_hm3}
\eea

\noindent
From the energetic point of view, the kinetic energy of this outflow
will be typically not enough to heat the gas to a high temperature, so
the heating will be done with the same energy budget of
equation~\ref{eq:hwbulge}; in other words, no energy is added to the
$\dot{E}_{\rm hw,H}$ term.  We have implemented both possibilities,
and the choice between the two has been left free.  The results
presented here are obtained giving the outflowing gas to the hot
phase.


\section{Metals}
\label{section:metals}

The evolution of metals is given by the equations reported in
table~1.  In this set of equations most metal flows
are obtained from their related mass flow as follows:

\be
\dot{M}^Z_{\rm flow} = \frac{M^Z_{\rm source}}{M_{\rm source}}
\dot{M}_{\rm flow}
\label{eq:mfgeneric} \ee

\noindent
where $M_{\rm source}$ and $M^Z_{\rm source}$ are the gas and metal
masses of the source phase.  The cosmological infalling gas is assumed
to have a metallicity $Z_{\rm pre}\sim 10^{-6}$ due to pre-enrichment
by sources that are below the mass resolution limit (from popIII stars
to very small primeval galaxies).  These flows take into account the
transfer of metals among phases and components, but not their
production.

We assume that (i) metals are produced by newly formed stars
in the instantaneous recycling approximation, i.e. their production
follows instantaneously star formation; (ii) the new metals are
instantaneously mixed with the ISM.  Newly metals are spread into the
ISM mainly by SNe, so analogously to the energy they are likely to be
selectively ejected to the halo.  This is clearly true in the
blowing-out thin systems, but even in thick systems metals are first
mixed with the hot gas that escapes the halo at a rate equal to the
star-formation rate.  To model this effect without using
explicitly a multi-phase description of the ISM, a fraction $f_{\rm
Zej}$ of the new metals is ejected directly to the halo through hot
winds. In the case of bulges, this fraction is multiplied by the
$\beta$ factor of equation~\ref{eq:betab}, to take into account the
ability of massive bulges to retain the outflowing hot gas. We have:

\be                     
\left. \dot{M}^Z_{\rm hw,H}\right|_{\rm ej} = f_{\rm Zej} Y 
\left(\beta\dot{M}_{\rm sf,B}+\dot{M}_{\rm sf,D}\right)
\label{eq:yield_h}\ee

\noindent
where $Y$ is the fraction of mass in newly produced metals per
generation of stars.  This term is added to the $\dot{M}^Z_{\rm hw,H}$
metal flow; satellite galaxies will inject their metals to the halo
component of the main DM halo they belong to through the satellite
metal flows.  The other metals will be given to the ISM of the
component they belong to:

\bea
\dot{M}^Z_{\rm yi,B} &=& (1-\beta f_{\rm Zej}) Y \dot{M}_{\rm sf,B}\, ,\nonumber\\
\dot{M}^Z_{\rm yi,D} &=& (1-f_{\rm Zej}) Y \dot{M}_{\rm sf,D}\, ,
\label{eq:yield_bd} \eea

\noindent
A value of $f_{\rm Zej}=0.5$ will be used in the following.


\section{AGN activity}
\label{section:agn}

\subsection{Accretion onto black holes}
\label{section:accretion}

The modeling of black hole accretion in {\gal} has already been
described, though in a slightly different way, by Monaco \& Fontanot
(2005), and will be briefly summarized here.

A seed black hole of mass $M_{\rm seed}$ is put at the centre of each
DM halo.  These black holes may be generated by the collapse of the
first stars (e.g., Volonteri, Haardt \& Madau 2003).  Seed masses
should be of order of tens to hundreds \msun; however, they start
growing very soon during the early evolution of baryons in DM halos.
This happens at times and for DM halo masses that are not sampled in
the typical runs used in galaxy formation. We then use a higher value
for the seed mass, $M_{\rm seed}=1000$ \msun; the results are quite
stable for reasonable variations of this parameter.

The accretion of gas onto the black holes is possible only if this gas
has lost nearly all of its angular momentum.  The first step in this
loss is the same that leads to the formation of bulges; we then base
our computation of black hole accretion on the cold bulge gas.  As the
amount of accreted gas is small, we do not remove the accreted mass
from the matter budget.  In other words, the mass in black holes does
not obey a mass conservation constraint as the mass of all the other
components.

The gas is assumed to lose angular momentum at a rate proportional to
the star-formation rate.  This is justified by the radiation drag
mechanism proposed by Umemura (2001) and used by Granato et
al. (2004).  However, a connection between loss of angular momentum
and star formation likely has a more general validity, as many
mechanisms able to cause a loss of angular momentum (turbulence,
kinetic aggregations etc.) are directly or indirectly driven by
massive stars and SNe.  The simplest way of modeling the loss of
angular momentum is then:

\be
\dot{M}_{\rm lowJ} = f_{\rm lowJ} \dot{M}_{\rm sf,B}
\label{eq:lowJ} \ee

\noindent
More generally, the accreted mass could scale as a power law of the
star-formation rate; for instance, angular momentum loss driven by
encounters would likely scale with the square of the driving force.
This is described in detail in Fontanot et al. (2006a); we describe
here only the simplest choice.

We follow Granato et al. (2004) by assuming that this gas does not flow
directly onto the black hole but settles on a reservoir of low-angular
momentum gas, whose mass is $M_{\rm resv}$.  From this reservoir the
gas accretes onto the black hole at a rate regulated by the viscous
time-scale of the accretion disc, modeled by Granato et al. (2004)
as:

\be
\dot{M}_{\rm visc}=k_{\rm accr}\frac{\sigma_{\rm B}^3}{G}
\left( \frac{M_{\rm res}} {M_{\rm BH}} \right)^{3/2}
\left( 1+ \frac{M_{\rm BH}} {M_{\rm resv}} \right)^{1/2}
\label{eq:visc} \ee

\noindent
where the constant $k_{\rm accr}$ is suggested by the authors to take a
value\footnote{Because of a misprint, the value is
indicated to be $10^{-4}$ in their paper, while the correct value is an
order of magnitude larger.} $\sim$0.001, $\sigma_{\rm B}\simeq 0.65 V_{\rm B}$ is the 1D
velocity dispersion of the bulge and $M_{\rm BH}$ is of course the BH
mass.

Accretion is always limited by the Eddington-Salpeter rate, whose
time-scale (for a radiation efficiency of 0.1) is $t_{\rm Ed}=0.04$
Gyr.  The evolution of the black hole-reservoir system is then:

\be
\begin{array}{lll}
\dot{M}_{\rm BH} &=& \min \left( \dot{M}_{\rm visc}\, , \, \frac{M_{\rm BH}}{t_{\rm Ed}}
\right)  \\
\dot{M}_{\rm resv} &=& \dot{M}_{\rm lowJ} - \dot{M}_{\rm BH} 
\end{array}\label{eq:bhaccr} \ee

\noindent
In Monaco \& Fontanot (2005), the accretion rate onto the black hole
was modeled simply as $M_{\rm resv}/t_{\rm Ed}$.  We prefer
equation~\ref{eq:visc} in virtue of its sounder physical motivation.


\subsection{Feedback from jets}
\label{section:jets}

Jets coming from radio-loud AGNs are though to be one of the most
promising mechanisms to stop the cooling flows in galaxy clusters
(McNamara et al. 2005; Voit \& Donahue 2005).  Besides, a successful
reproduction of the high-luminosity cutoff of the luminosity function
of galaxies requires a proper modeling of this kind of feedback
(Benson et al. 2003; Bower et al. 2006; Croton et al. 2006).  The
efficiency of radiation of AGNs is known to decrease when the
accretion rate in units of the Eddington rate $\dot{m}=\dot{M}_{\rm
BH}t_{\rm Ed}/M_{\rm BH}$ is lower than $\sim 10^{-2}$; these slowly
accreting black holes however radiate very efficiently in the radio
(see the discussion in Merloni, Heinz \& Di Matteo 2003).  It is then
reasonable to assume that the efficiency of energy emission in jets is
0.1 if $\dot{m}$ is small, and lower for higher accretion rates.  As
$\sim$10 per cent of bright QSOs are radio-loud, we estimate the
efficiency of jet emission as 0.01 in the case of $\dot{m}>0.01$.

As suggested by Croton et al. (2006), the efficiency with which this
energy heats the hot halo gas component should scale with the hot gas
temperature to the power 3/2.  In order to keep the expression simple
and avoid possible unwanted numerical instabilities in the
integration, we scale the efficiency with the circular velocity of the
halo:

\be
f_{\rm jet} = f_{\rm jet,0} \left(\frac{V_c}{1000\, {\rm km\ s}^{-1}}\right)^3
\label{eq:effjet} \ee

\noindent
The energy injected into the halo hot gas is then:

\be
\left. \dot{E}_{\rm hw,H}\right|_{\rm jets} = \left\{
\begin{array}{lll} 
f_{\rm jet} 0.1\,  \dot{M}_{\rm BH} c^2 & {\rm if} & \dot{m}<0.01\\
f_{\rm jet} 0.01\, \dot{M}_{\rm BH} c^2 & {\rm if} & \dot{m}>0.01 \end{array}
\right. \label{eq:jetfb} \ee

\noindent
This mechanism benefits much by the direct infall of gas to the bulge,
described in section~\ref{section:bulges}, but, as demonstrated in
appendix~\ref{section:stab}, works only for particle masses not higher
than $10^9$ \msun.


\subsection{Quasar-triggered galaxy winds}
\label{section:quasar}

A bright quasar shining into a star-forming bulge can inject a great
amount of energy into the ISM, leading to a massive removal of cold
gas.  This mechanism has been described in detail by Monaco \&
Fontanot (2005).  These winds have a modest effect on the formation of
a galaxy, but influence the formation and evolution of AGNs. This is
described in full detail in Fontanot et al. (2006a), where models
with quasar-triggered winds are presented.  We do not use these winds
in the results presented in this paper, as the introduction of winds
influences deeply the number of bright quasars but only modestly their
host galaxies.


\section{Parameters}
\label{section:parameters}

\begin{table*}
\begin{tabular}{clllll}
& Name & reference & Comment & Constraint & Equation/\\
&      & model     &         &            & Section\\
\hline
&&& {\bf  mergers}\\
&$f_{\rm hmm}$ &  0.2 & major merger condition for DM halos & N-body & 
                          eq.~\ref{eq:mmcond} \\
&$f_{\rm gmm}$ &  0.3 & major merger condition for galaxies & N-body &
                          eq.~\ref{eq:mmgal} \\
&$f/c$         &  2.0/4.0 & bulge formation in mergers/disc instabilities & N-body & 
                          eq.~\ref{eq:merger} \\
!&$f_{\rm scatter}$ & 0.1 & fraction of stars scattered at a galaxy major merger    & N-body & 
                          sect.\ref{section:galmergers} \\
&&& {\bf halo component}\\
!&$\gamma_p$    & 1.15  & polytropic index of the hot gas & observ. & 
                          eq.~\ref{eq:hydro_sol}\\
&$f_{\rm shock}$& 1.2 & shock heating factor               & N-body & 
                          eqs.~\ref{eq:shock}, \ref{eq:shock2} \\
!&{\tt heat cold gas}& YES & switch for heating cold halo gas at major mergers & N-body &
                          sect.~\ref{section:shock} \\
!&$n_{\rm quench}$ & 1.0 & no. of crossing times for quenching cooling & free & 
                          sect.~\ref{section:shock} \\ 
!&{\tt close hole} & YES & switch for closing the cooling hole & free & 
                          eq.~\ref{eq:close_hole}\\
!&{\tt infall on bulge} & YES & switch for allowing infall on the bulge & free &
                          eqs.~\ref{eq:infall_b1}, \ref{eq:infall_b2}\\
!&$n_{\rm dyn}$ & 1.0   & no. of dynamical times for infall  & free & 
                          eqs.~\ref{eq:infall}, \ref{eq:kinfall}\\ 
!&$f_{\rm wind}$ & 1.7   & energy factor to trigger a super-wind & free &
                          eqs.~\ref{eq:hotwindcond}, \ref{eq:coldwindcond}\\
&$f_{\rm back}$ & 0.5 & fraction of super-wind mass that falls back & free & 
                          sect.~\ref{section:winds}\\ 
&&& {\bf disc structure}\\
&$\epsilon_{\rm limit}$ & 0.9 & limit for bar instability  & N-body & 
                          eq.~\ref{eq:discstab}\\
&$f_{\rm bar}$ & 0.5 & fraction of disc that goes to bulge & free & 
                          sect.~\ref{section:bulges} \\
&{\tt adiabatic contr.} & NO & switch for adiabatic contraction & free &
                          eq.~\ref{eq:adcon} \\
&&& {\bf stars and metals}\\
&$M_{\rm \star,SN}$ & 120 \msun &stellar mass per SN            & IMF  & 
                          eqs.~\ref{eq:hwdisc}, \ref{eq:hwbulge} \\
&$f_{\rm rest}$  & 0.4    & fraction of restored mass      & IMF  & 
                          eqs.~\ref{eq:disc_flows}, \ref{eq:sfbulge}\\
!&$Y$             & 0.03   & metal yield per generation           & observ. & 
                          eqs.~\ref{eq:yield_h}, \ref{eq:yield_bd} \\
&$Z_{\rm pre}$  & $10^{-6}$ & metallicity due to pre-enrichment   & free & 
                          sect. \ref{section:metals} \\
!&$f_{\rm Zej}$  & 0.5  & fraction of metals ejected to halo & free & 
                          eqs.~\ref{eq:yield_h}, \ref{eq:yield_bd} \\
&&& {\bf star formation and feedback}\\
!&$f_{\rm th,D}$  & 0.5 & thermal efficiency of feedback in thin systems    & free & 
                          eq.~\ref{eq:hwdisc}\\
!&$f_{\rm th,B}$  & 0.5 & thermal efficiency of feedback in thick systems    & free & 
                          eq.~\ref{eq:hwbulge} \\
!&$f_{\rm kin}$   & 0   & kinetic energy from hot winds                      & free & 
                          eqs.~\ref{eq:kinfbdisc}, \ref{eq:kinfbbulge} \\
!&$\sigma_0$      & 60 \kms  & turbulent velocity of clouds      & free & 
                          eqs.~\ref{eq:kinfb}, \ref{eq:sigma0} \\
!&$\Sigma_{\rm thr}$   & 0 \surf   & gas surface density threshold for star formation & observ. &
                          section~\ref{section:thin} \\
!&$\Sigma_{\rm limit}$ & $\infty$ \surf & critical gas surface density for discs & free&
                          eq.~\ref{eq:thickdisc} \\
!&{\tt hot kin. fb} & YES & switch for heating cold gas by kinetic feedback &free&
                          eqs.~\ref{eq:kinwind_cm}-\ref{eq:kinwind_hm3} \\
&&& {\bf AGN}\\
&$M_{\rm seed}$ & 1000 \msun & seed black hole mass         & theory & 
                          sect.~\ref{section:accretion} \\
!&$f_{\rm lowJ}$ & 0.003 & rate of loss of angular momentum    & free & 
                          eq.~\ref{eq:lowJ} \\
!&$f_{\rm jet,0}$& 1     & efficiency of jet feedback for a 1000 \kms\ halo & N-body & 
                          eq.~\ref{eq:effjet} \\ 
&&& {\bf numerical parameters}\\
!&$M_{\rm part}$ & $10^9$\msun & particle mass                      & --- & 
                          --- \\
&$\Delta_t$    & 0.1 Gyr & numerical interval for the integration & free & 
                          sect.~\ref{section:algorithm}\\
\hline

\end{tabular}
\caption{Model parameters, with their value adopted in the reference
model, brief description, available constraints (independently of the
model) and reference in the text.  Parameters highlighted by a mark
are of primary importance. Cosmological parameters are not included.}
\label{table:parameters}
\end{table*}

Table~\ref{table:parameters} gives a complete list of the parameters
of the model, with the value used to compute the results given in this
paper.  To these the cosmological parameters ($\Omega_0$,
$\Omega_\Lambda$, $\Omega_b$, $H_0$ and $\sigma_8$; see the
Introduction) should be added; these are now fixed with a good
accuracy, with the exception of $\sigma_8$ whose value influences
strongly the number of galactic DM halos at high redshift.  The number
of parameters is high, and this reflects both the number of physical
processes that are included in the galaxy formation code and the level
of uncertainty in many of these processes.  Anyway, galaxy formation
is a problem of complexity, and there is no way to reduce the number
of parameters other than hiding them by fixing them to some value.  On
the other hand, the number of observables that can be used to
constrain these parameters is very high, so their values can be fixed,
with the exception of a few degeneracies.  The most obvious one in
this context is the degeneracy between the energy of a SN, the star
mass per SN $M_{\rm star,SN}$ and the thermal feedback efficiencies
$f_{\rm th,B}$ and $f_{\rm th,D}$ (equations~\ref{eq:hwdisc} and
\ref{eq:hwbulge}).  For this reason we only vary the efficiencies,
leaving $M_{\rm \star,SN}$ fixed and not even including the SN energy
as a parameter.

In practice, many of these parameters are fixed independently by the
results of N-body simulations (like the parameters of merging) or by
the knowledge of the IMF, and their variation within the known
uncertainties does not influence the results strongly.  The remaining
parameters, labeled with a mark in the table, are then of primary
importance.  Their value is fixed by comparing with a set of basic
data (section~\ref{section:results}), we give here a very brief list
of the observable which is most useful to fix each parameter.

$f_{\rm scatter}$: it regulates the amount of halo stars, and is fixed
by requiring a fraction of at least $\sim20$ per cent of halo stars in
groups and clusters ($M_{\rm H}\ga10^{14}$ \msun).

$n_{\rm quench}$ and $n_{\rm dyn}$: they regulate the star formation
density, especially at high redshift; their value is fixed by
reproducing the very uncertain star formation density at $z>4$.  They
are nearly degenerate and their value is especially sensitive to mass
resolution.

$\gamma_p$: it regulates star formation, especially at low redshift,
but its influence is modest when it is varied within the range
suggested by observations.

$f_{\rm wind}$: by regulating the energy level at which a galactic
super-wind is triggered, it influences the relation between stellar
and DM mass.

{\tt heat cold gas}: its switching on gives an effect similar to an
increase of $n_{\rm quench}$ or $n_{\rm dyn}$ by 1.

{\tt close hole}: when it is on, it increases cooling by a significant
amount which depends on $\gamma_p$.  All the parameters of the hot
halo gas and stellar feedback should be re-tuned if this option is
active.

{\tt infall on bulge}: it influences mostly the jet feedback, so it
should be switched on if an efficient quenching of the cooling flow is
wanted.

$Y$: it regulates the amount of metals produced by stars, and is fixed
by requiring a good match of galaxy metallicities.

$f_{\rm Zej}$: it regulates the fraction of metals injected in the
halo component, and is fixed by requiring a good match of the metal
enrichment of the hot halo gas.

$f_{\rm th,D}$: it regulates the stellar mass of small galaxies, and
is fixed by requiring a good match of the power-law part of the stellar
mass function.

$f_{\rm th,B}$, $f_{\rm kin}$ and {\tt hot kin. fb}: they have only a
modest influence on the results, so they are left to their reference
values.

$\sigma_0$: it influences star formation in small bulges, and then the
amount of downsizing of the AGN population; it is fixed as explained
in Fontanot et al. (2006a).

$\Sigma_{\rm thr}$: it influences the mass of cold gas in local discs
and its value is suggested by observations.

$\Sigma_{\rm limit}$: it influences the history of star formation and
BH accretion in a rather subtle way; it is used in Fontanot et
al. (2006a).

$f_{\rm lowJ}$: it regulates the BH--bulge relation at $z=0$, and its
value depends on whether quasar-triggered winds are switched on; see
Fontanot et al. (2006a) for details.

$f_{\rm jet,0}$: its value is suggested to be unity by N-body
simulations, and it allows to quench cooling flows and limit the mass
of elliptical galaxies.


\section{Post Processing}
\label{section:post}

For each tree that is analyzed, all the variables introduced in
section~\ref{section:baryons}, together with the disc and bulge radii
and velocities, star formation rates and accretion rate onto black
holes (at the sampling time, not averaged over the time bin), are
sampled in intervals of time $\Delta_t$ and at the final time.  All
these quantities are output at the end, together with the merger
histories of galaxies, constructed as specified in
section~\ref{section:trees}.

For each time bin this information is issued for all the existing
galaxies of a tree.  However, this is not sufficient to reconstruct
the star-formation history of a galaxy component, because disc
instabilities, mergers, tidal stripping and destruction events move
stars among galaxies and components; as a consequence, the sampled
star formation history of stars that are formed in a galaxy component
does not coincide with the star formation history of the stars
eventually found in that galaxy component.  To reconstruct this last
quantity, all the events that transfer stellar mass from a component or a
galaxy to another are recorded and output at the end.  Star formation
histories are then reconstructed at the post-processing stage by
scrolling the sampled history of galaxies and moving stellar mass among
the components.  This is a very quick process, that can be done at the
reading time.

The output of {\gal} (star formation histories and metallicities
of the cold gas) is then given to the spectro-photometric code {\sc
grasil} (Silva et al. 1998), which is able to predict the SEDs of the
predicted galaxies from the UV to the radio.  This way it is possible
to construct predictions for the luminosity functions at a fixed
redshift or in redshift intervals, galaxy number counts and galaxy
backgrounds.  The {\sc grasil} code has been tested against many local
and distant galaxies and is widely used by the community; however, it
introduces further uncertainties and parameters, so we prefer in this
paper to discuss only the prediction of {\gal} before its interfacing
with {\sc grasil}.  All the details of this interfacing will be given
in Fontanot et al. (2006b).

The star formation rates are then reconstructed in time bins of width
$\Delta_t$, which is set to 0.1 Gyr.  This sampling is fine but for
the last bin considered, as the last formed stars, that dominate the
UV, B and FIR spectra, live for less than the sampling bin.  To
overcome this difficulty, we sample also the value of bulge and disc
star-formation rates evaluated at the end of each time bin.  Then, for
each galaxy component, the time bin corresponding to the time at which
the SED is computed is split into two parts, of widths 0.09 and
0.01 Gyr.  To the second part we assign a star-formation rate equal to
the punctual value at the end of the bin, while to the first part we
assign a star-formation rate such as the integral over the bin gives
the total amount of stars.  As {\sc grasil} re-samples the
star-formation histories on a much finer time grid, this procedure is
accurate as long as the star formation rate does not vary strongly on
timescales smaller than 10 Myr, which is the case for most galaxies.
The accuracy of this procedure with respect to the use of the complete
star-formation rate produced by {\gal} during the integration will be
shown in Fontanot et al. (2006b).

To compute the AGN activity of galaxies, a sampling in a time grid is
not sufficient, due to the low duty cycle of AGNs.  To optimize the
statistical sampling of AGNs without inflating the output files, we
save all significant accretion events at each integration time-step;
this is done whenever the punctual value of the accretion rate is
larger than a minimal value of $1.76\times 10^{-3}$ \msunyr,
corresponding to a bolometric luminosity of $10^{43}$ erg s$^{-1}$.
This very detailed output adds only $\sim$10 per cent to the total
disc space needed by a run.  For the computation of the luminosity
function of AGNs, each of these events is treated as an independent
event of duration equal to the integration interval of time.  This
procedure is explained in detail in Fontanot et al. (2006a).


\section{Results}
\label{section:results}

As already mentioned in the Introduction, this paper is devoted to a
detailed description of the {\gal} code, so in this section we present
only some of the main results, obtained with the set of parameters
given in table~\ref{table:parameters} unless otherwise stated.  For
sake of simplicity we restrict ourselves to predictions that do not
require the use of a spectrophotometric code.  In this way all the
uncertainties involved in the generation of SEDs for the model
galaxies are bypassed.

All the results are based on a single $512^3$ {\pin} run of a 150 Mpc
box ($h=0.7$) with the standard cosmology given in the Introduction.
The particle mass is $1.0\times 10^9$ \msun; then the smallest
considered halo contains 50 particles, for a mass of
$5.1\times10^{10}$ \msun, while the mass of the smallest progenitor is
$1.0\times10^{10}$ \msun.  As a comparison, the much bigger Millennium
Simulation (Springel et al. 2005) has a slightly higher particle mass,
$1.2\times 10^9$ {\msun} for our Hubble constant, and its merger
histories are reconstructed starting from 20 particles, corresponding
to a smallest progenitor of $2.5\times 10^{10}$.  To test the
stability of the results, two more boxes have been produced with the
same number of particles and sizes of 200 and 100 Mpc, for particle
masses a factor of 2.4 worse (higher) and 3.4 better (lower).  The
results of the stability tests are briefly shown in
appendix~\ref{section:stab}; we will highlight in the following which
results are more sensitive to mass resolution.

The {\gal} runs have been performed on simple PCs.  A 150 Mpc run with
nearly 2000 trees needs about 20 hours on a 3GHz PC, the generated
output is about 3 Gb.  The bottleneck of the computation is {\sc
grasil}, which requires about five minutes per galaxy, so the
computation of 10000 galaxies in a mock pencil beam survey requires
about one month of CPU.

\begin{figure*}
\centerline{
\includegraphics[width=17cm]{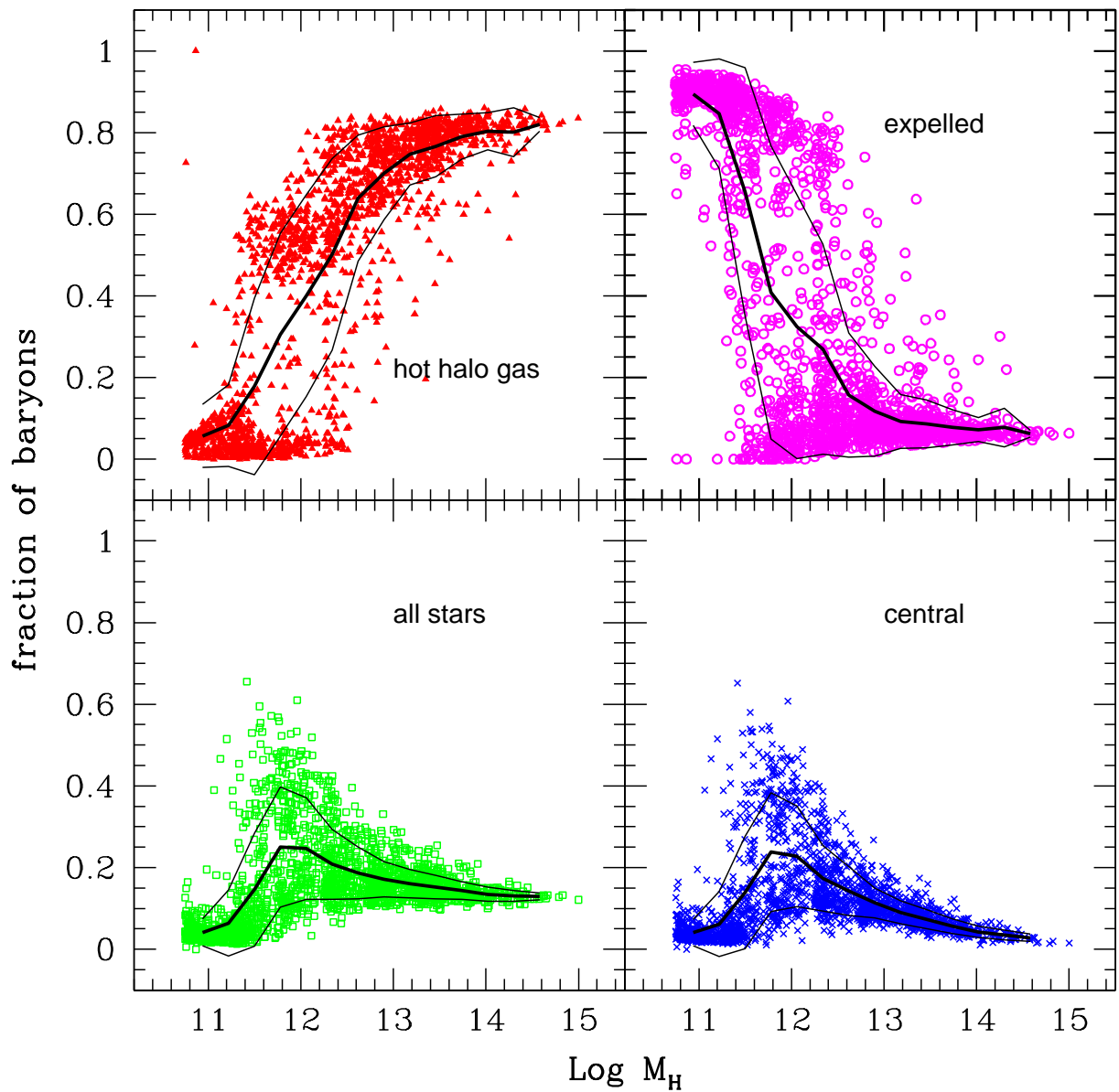}
}
\caption{DM halos at $z=0$: fraction of baryons in hot halo gas (red
triangles, upper left panel), expelled by winds (magenta circles, upper
right panel), in stars in all galaxies (green open squares, lower left
panel), in stars in the central galaxy (blue crosses, lower right
panel).  Averages and rms of the points are shown in all panels.}
\label{fig:feedback}
\end{figure*}

We find that the stellar mass of the typical central galaxy contained
in the smallest DM halos at $z=0$ is $3\times 10^{8}$ {\msun}, so the
stellar mass function is severely incomplete below this limit.  

To limit the size of the output and the computing time, the mass
function of DM halos is sampled by picking no more than 300 halos per
mass bin of 0.5 dex (or all halos in the bin if they are less than
300).  The statistical distributions of galaxies are then computed by
weighting each galaxy by $w_{\rm tree}$, the inverse of the fraction
of the host DM halos picked up in the mass bin.  For each DM halo all
galaxies and satellites are computed, so that for a given stellar mass
the satellites of large DM halos are oversampled with respect to the
central galaxies, which are numerically dominant at all masses.  This
feature shows up when galaxy properties are reported for samples of
model galaxies, like in the Tully-Fisher relation: at fixed stellar
mass, satellite galaxies with a low weight $w_{\rm tree}$ are more
numerous in the plot than central galaxies with a high $w_{\rm tree}$.
We then further sparse-sample the satellites as follows: we compute
the average stellar mass of central galaxies as a function of the DM
halo mass, then assign to each satellite a weight $w_{\rm gal}$ equal
to the ratio between the $w_{\rm tree}$ of the average DM halo hosting a
central galaxy with the same stellar mass and the $w_{\rm tree}$ of
the host DM halo. We then select the satellite with a probability
equal to $1/w_{\rm gal}$.  In this way we obtain a roughly constant
number of galaxies per logarithmic interval of stellar mass.

Figure~\ref{fig:feedback} gives, for each selected DM halo evolved to
$z=0$, the fraction of baryons that are found in hot halo gas, in
stars of all galaxies (but not in the halo), in stars of the central
galaxy, and the fraction of baryons ejected by winds and not fallen
back.  These clearly do not exhaust the list of baryonic components
(halo stars and cold gas components are excluded here), so these
fractions do not add up to one.  However, this figure helps in showing
a number of important features.  There is a clear transition at a DM
halo mass range of $10^{12}-10^{13}$ \msun: smaller halos lose most of
their baryons by ejecting them to the IGM and their stellar content is
dominated by the central galaxy, while larger halos retain most of
their baryons as hot gas and have a small fraction of stars ($\sim20$
per cent) in the central galaxy.  In both cases the fraction of stars
(both total and in the central galaxy) declines quickly with
increasing or decreasing mass, while a maximum is reached at $10^{12}$
\msun.  All these features reproduce nicely the trends suggested in
the seminal papers of galaxy formation (see, for instance, White \&
Rees 1978 or Dekel \& Silk 1986): small halos are evacuated by winds
driven by SNe, while in large halos the large cooling time (aided by
AGN feedback) prevents most gas from cooling.  As a further element,
it is worth noting the presence of many Milky Way-sized halos with
almost no hot gas, in nice agreement with the lack of X-ray emitting
around our Galaxy.

\begin{figure}
\centerline{
\includegraphics[width=8cm]{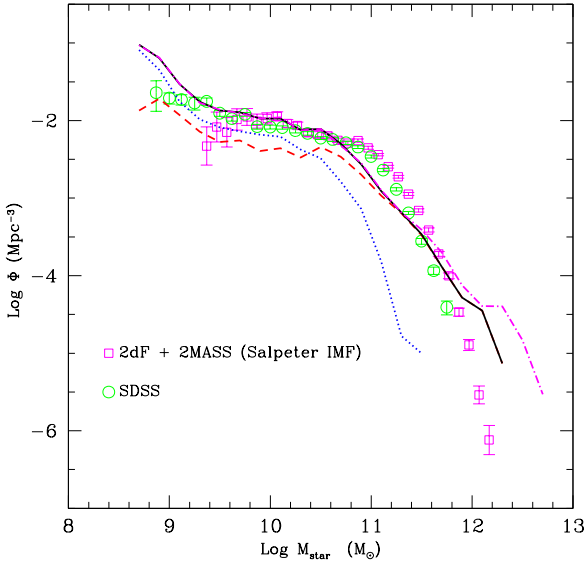}
}
\caption{Stellar mass function of galaxies at $z=0$ (continuous black
line), compared with the results of 2dF$+$2MASS (Cole et al. 2001) and
SDSS (Bell et al. 2003).  The dashed red and dotted blue lines give
respectively bulge-dominated and disc-dominated galaxies, while the
dot-dashed magenta line refers to a model without quenching of the
cooling flows by AGN jets ($f_{\rm jet,0}=0$).}
\label{fig:mfstar}
\end{figure}

The trends of decreasing efficiency of star formation in smaller and
larger halos than $10^{12}$ \msun\ are needed to produce the observed
low-luminosity slope and high-luminosity cutoff of the galaxy LFs.
Figure~\ref{fig:mfstar} shows the predicted stellar mass function of
galaxies, compared with the data by the 2dF$+$2MASS (Cole et al. 2001)
and SDSS (Bell et al. 2003) surveys.  The low-mass slope is nicely
reproduced (though the result is sensitive to resolution, see
appendix~\ref{section:stab}) , even in the slight steepening below
$10^{10}$ \msun.  The high-mass cutoff is not strong enough, and the
biggest ellipticals are too massive by at most a factor of two; 
however, this excess may be connected, as proposed by Monaco et
al. (2006), to the construction of the diffuse stellar component of
galaxy clusters, and would be fixed by a proper modeling of the
scattering of stars during mergers (section~\ref{section:galmergers});
this work is in progress.  We also report in the figure the resulting
mass function without AGN feedback, to show that the quenching of the
cooling flow by AGN jets decreases significantly the discrepancy with
observations.

Figure~\ref{fig:mfstar} shows also the stellar mass function of bulge-
and disc-dominated galaxies (these are very similar to the mass
functions of bulges and discs).  In line with what is observed, bulges
dominate at large masses, while discs are more abundant at small
masses.  However, as noticed also by Croton et al. (2006), the mass
function of bulges does not have a broad peak at $\sim10^{10}-10^{11}$
\msun, as shown by the luminosity function of ellipticals; there is a
definite excess of small bulges.  This is likely connected to the
excess in the predicted number density of $10^{10}$ {\msun} galaxies
at $z\sim1$ reported by Fontana et al. (2006), a problem shared by
many galaxy formation models. This point will be deepened in Fontanot
et al. (2006b).

\begin{figure}
\centerline{
\includegraphics[width=8cm]{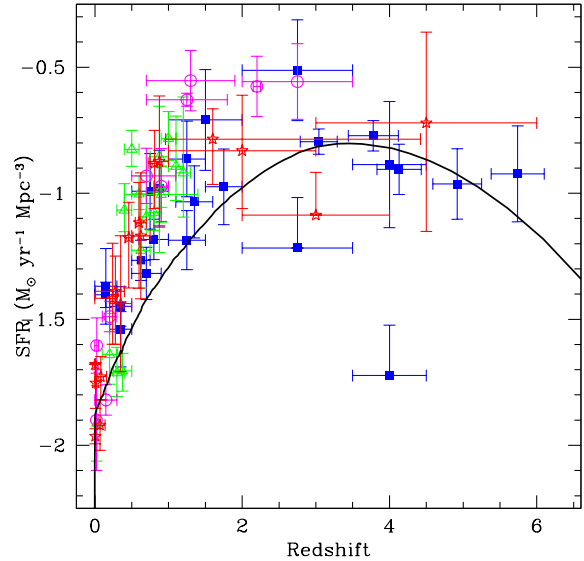}
}
\caption{Star formation rate density as a function of redshift,
compared with the data compiled and homogenized by Hopkins (2004).
Points denote star formation estimates based on rest-frame UV (blue
squares), $[{\rm O}II]$ (green triangles), H$\alpha$ and H$\beta$
(magenta circles), X-ray, FIR and sub-mm (red stars).}
\label{fig:csfr}
\end{figure}

Figure~\ref{fig:csfr} shows the prediction of the star formation rate
density as a function of redshift.  The data have been collected and
homogenized by Hopkins (2004).  The prediction is consistent with the
data, with some possible underestimate at $z\sim1$.  In particular, a
very broad peak of star formation is predicted to be present at
$z\sim3$, in agreement with the estimates based on sub-mm counts.  The
level of star formation is still high at $z\sim6$, only a factor of 2
lower than the peak value; however, this prediction depends
sensitively on the mass resolution.  Most of this star formation takes
place in discs, triggered by the strong cooling flows at high
redshift, while mergers dominate the strongest starbursts; this will
be deepened in Fontanot et al. (2006b).

\begin{figure}
\centerline{
\includegraphics[width=8cm]{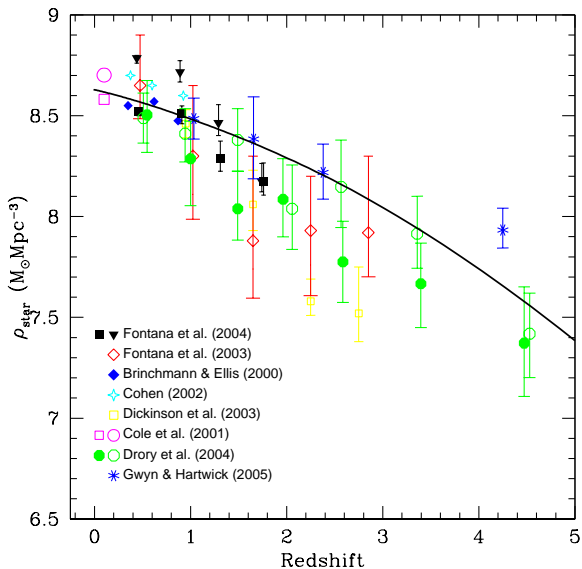}
}
\caption{Stellar mass density as a function of redshift, compared with
the data by Brinchmann \& Ellis (2000), Cole et al. (2001), Cohen
(2002), Dickinson et al. (2003), Fontana et al. (2003; 2004), Drory et
al. (2004), Gwinn \& Hartwick (2005).  The data are {\em not}
homogenized for the different sample completeness.}
\label{fig:sdens}
\end{figure}

Figure~\ref{fig:sdens} shows the cosmic stellar mass density as a
function of redshift.  Though this quantity is related to the cosmic
star formation rate (it is simply its integral in time), it is
compared with a completely different set of data (Brinchmann \& Ellis
2000; Cole et al. 2001; Cohen 2002; Dickinson et al. 2003; Fontana et
al. 2003, 2004; Drory et al. 2004; Gwinn \& Hartwick 2005).  The data
in this figure are not homogenized for the different mass limits, so
each point should be considered as a lower limit.  The agreement with
data is again good from $z\sim4$ to 0.  A more refined analysis 
of the build-up of the stellar mass is reported in Fontana et
al. (2006) and Fontanot et al. (2006b).

\begin{figure*}
\centerline{
\includegraphics[width=8cm]{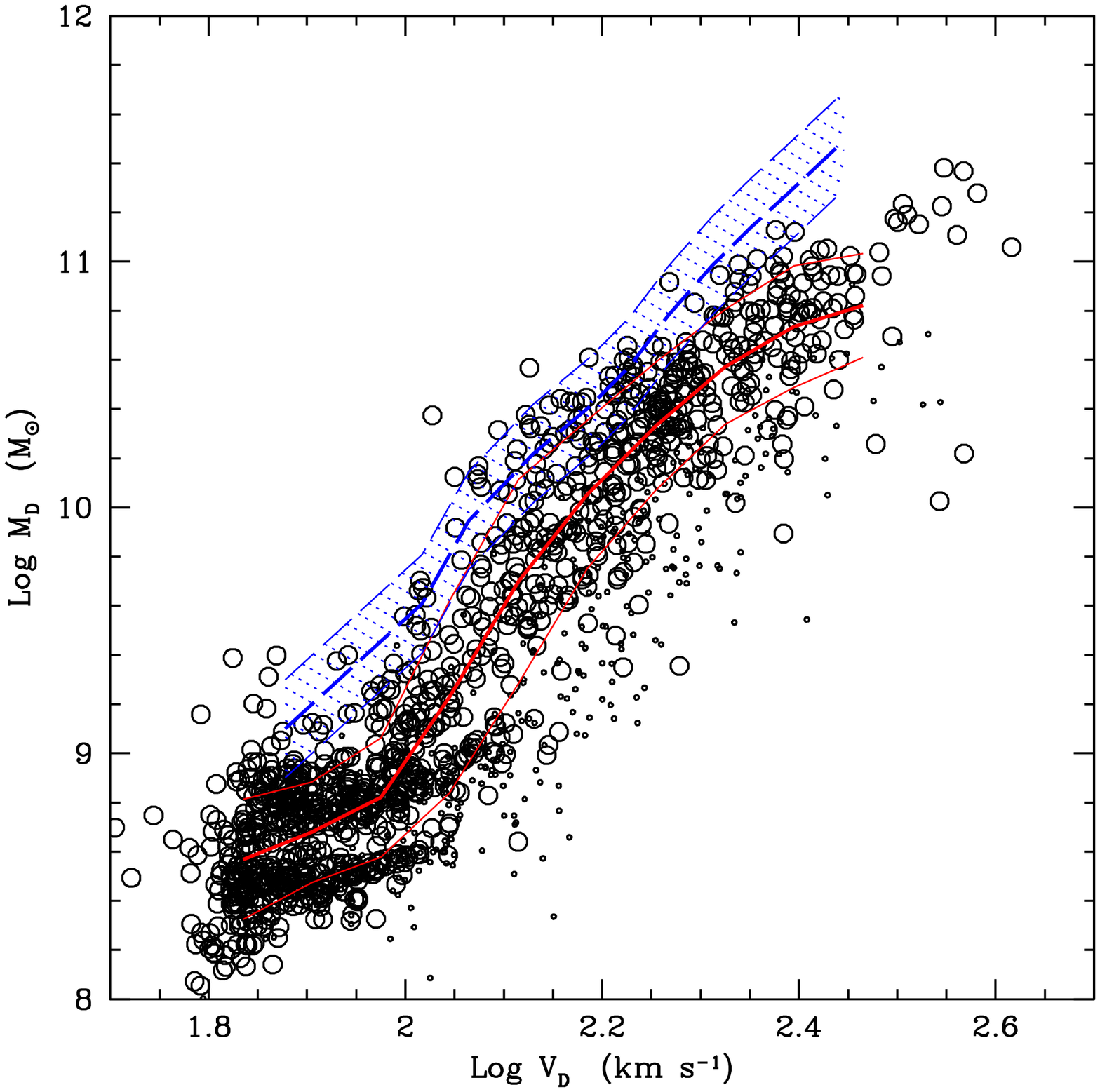}
\includegraphics[width=8cm]{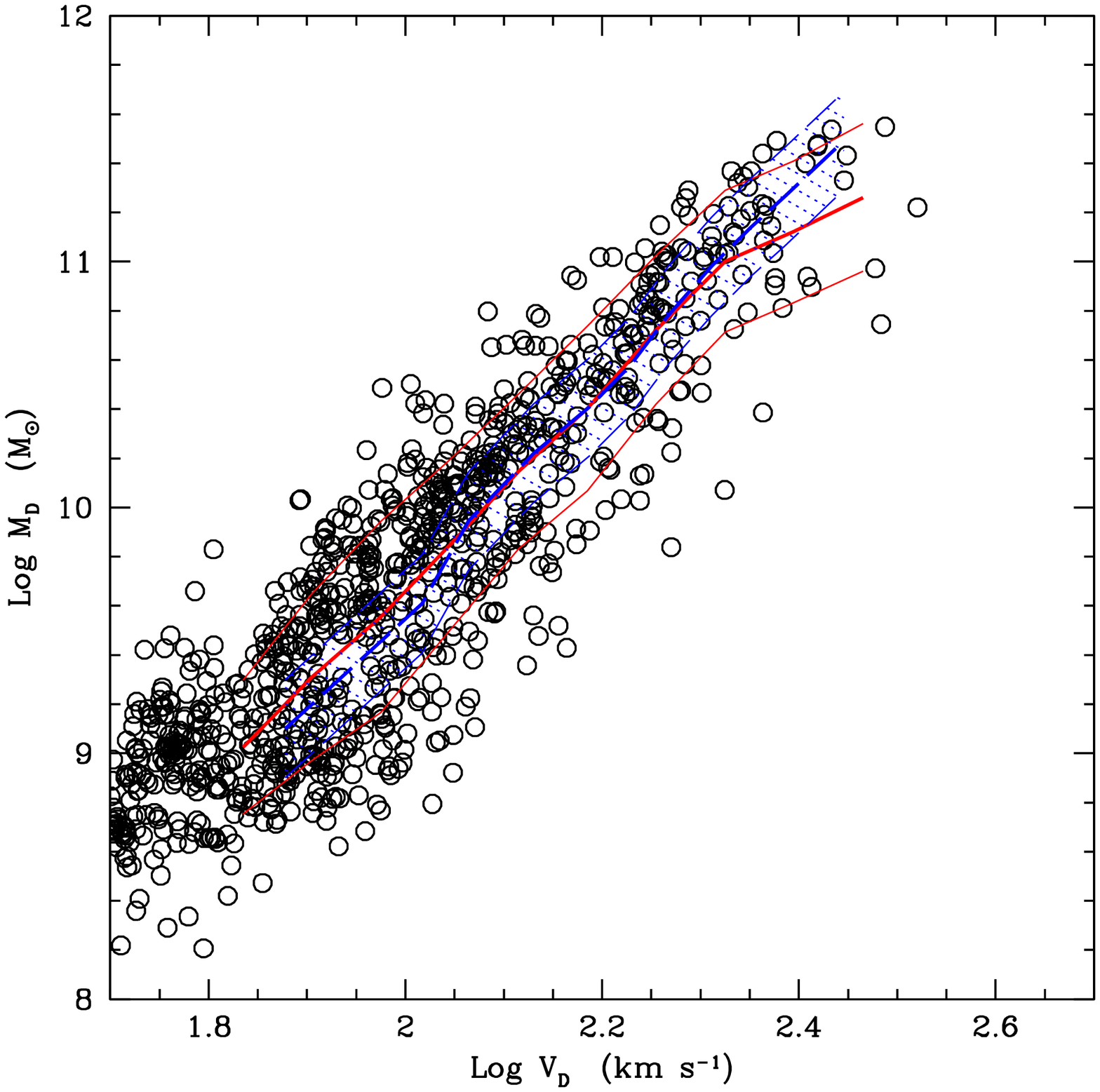}
}
\caption{Baryonic Tully-Fisher relation for disc-dominated galaxies at
$z=0$, compared with the relation obtained from the Universal Rotation
Curve (Persic, Salucci \& Stel 1996; Yegorova \& Salucci 2006), shown
as a shaded area.  Dots and circles correspond respectively to discs
with a gas fraction smaller and higher than 1 per cent.  Continuous
lines give the average (thick line) and average$\,\pm\,$rms (thin
lines) of the circles.  Left panel: standard model; right panel: model
with concentration scaled to $c_{\rm nfw}=4$ for $10^{12}$ {\msun} DM
halos at $z=0$.}
\label{fig:bartf}
\end{figure*}

Figure~\ref{fig:bartf} (left panel) shows the baryonic Tully-Fisher
relation for the disc-dominated galaxies, where disc velocities
$V_{\rm D}$ are computed at the optical radius (3.2$R_{\rm D}$) and
$M_{\rm D}$ is the total disc mass.  This relation is compared with
that obtained from the universal rotation curve (Persic, Salucci \&
Stel 1996, updated by Yegorova \& Salucci 2006); the intrinsic scatter
around this relation is reported by the authors to be $\sim$0.2 dex.
As a first point, there is a tail of rapidly spinning discs that are
not seen in the data.  These discs are formed at high redshift (when
DM halos were denser and discs more compact), then become satellites
so that no further infall on the disc takes place.  Consequently, they
are compact and gas-poor, and thus can hardly be classified as
spirals.  This is demonstrated in the figure, where all objects with a
gas fraction lower than 1 per cent are denoted as small dots; most if
not all outliers are then removed by applying this selection.  As a
second point, the gas-rich discs follow a Tully-Fisher relation
parallel to the observed one but with a higher velocity by 0.1 dex, or
25 per cent, in $V_{\rm D}$.  (Notably, galaxies follow the same
Tully-Fisher relation independently of their surface brightness.)  We
have verified that no realistic combination of feedback parameters
leads to a better zero-point of the relation.  This disagreement is
more likely related to the shape of the halo; for instance, Mo \& Mao
(2000) suggest that a concentration as low as $c_{\rm nfw}=4$ is
required to fit the Tully-Fisher relation.  To test this hypothesis we
then scaled all concentrations by a constant factor so as to take a
value of 4 for a $10^{12}$ {\msun} halo at $z=0$. The resulting
relation (figure~\ref{fig:bartf}, right panel) shows a much better
agreement with the observed one (this change does not affect much the
other galaxy properties, but ellipticals result larger and less dense
as well).  We then conclude that the disagreement in the zero-point of
the baryonic Tully-Fisher relation is likely due to the inner profile
of the DM halo more than to feedback.  As a third and final point, the
scatter around the Tully-Fisher relation is found to be $\sim$0.25
dex, possibly higher than what is suggested by data.

It must be stressed that this model was run without the computation of
adiabatic contraction of the halo (section~\ref{section:discs}).  We
have verified that this further process, besides slowing down the
computation significantly, makes the discs slightly more compact,
without influencing drastically the other results; in fact, taking
into account the presence of a bulge leads to a significant
contraction in many discs, so that the adiabatic contraction of the
halo does add much to this effect.

\begin{figure}
\centerline{
\includegraphics[width=8cm]{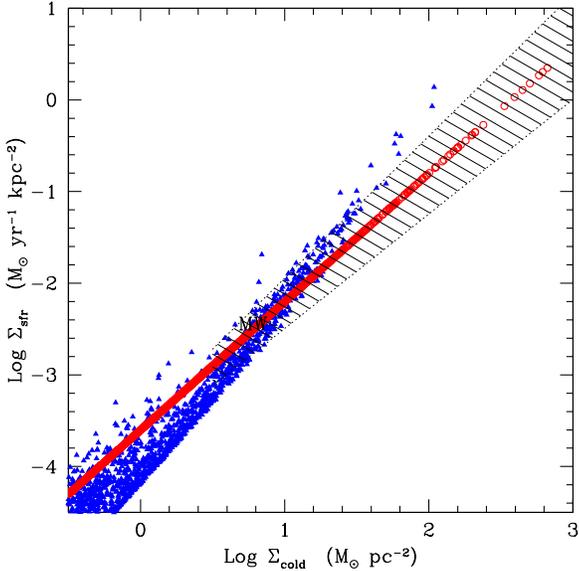}
}
\caption{Schmidt-Kennicutt law for the disc (blue triangles) and bulge
(red circles) components.  The shaded area gives the observed relation
reported by Kennicutt (1998).}
\label{fig:sch}
\end{figure}

Figure~\ref{fig:sch} shows the surface densities of cold gas versus
star formation rate for discs and bulges, compared with the
Schmidt-Kennicutt law (Kennicutt 1998).  By construction, bulges stay
exactly on the average relation, while discs follow the star-formation
timescale of equation~\ref{eq:fbthin2}, which, as anticipated, gives a
relation compatible with the observed one in virtue of the correlation
between cold gas fraction and surface density.  The agreement is very
good both in terms of zero-point (apart from a marginal underestimate
that could be easily fixed by a better tuning) and in terms of
scatter.

\begin{figure}
\centerline{
\includegraphics[width=8cm]{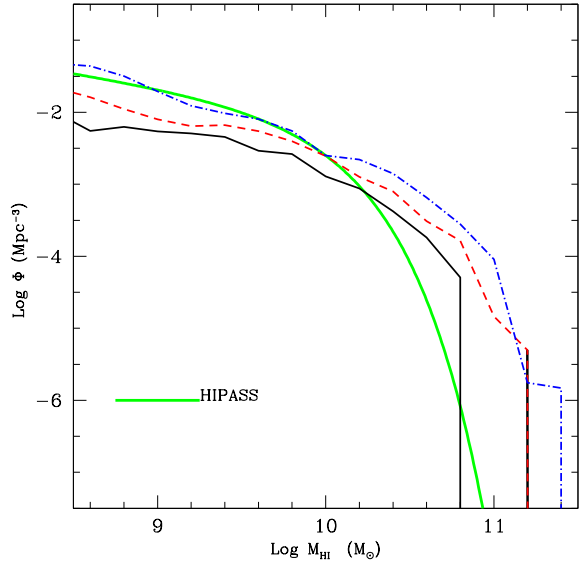}
}
\caption{Mass function of HI gas (black continuous line), compared
with the analytic fit of the results of the HIPASS sample (Zwaan et
al. 2003; green thick line), assuming that the correction for helium
and molecular gas amounts to a factor of 2. The statistical error of
the observed mass function is very small.  The dashed red and
dot-dashed blue lines give the results of models with a threshold for
star formation ($\Sigma_{\rm thr}=10$ \surf); in the blue dot-dashed
model DM halo concentrations $c_{\rm nfw}$ are scaled so as to assume
a value of 4 for a $10^{12}$ {\msun} halo at $z=0$.}
\label{fig:MFgas}
\end{figure}
 
Figure~\ref{fig:MFgas} shows the prediction of the mass function of HI
at $z=0$, compared with that obtained from the HIPASS sample (Zwaan et
al. 2003).  The statistical error of the observational mass function
(well fit by a Schechter function) is so small for this sample that we
do not even report it in the figure; conversely, the transformation
from HI to total cold gas mass is rather uncertain.  Following
Fukugita \& Peebles (2004), the HI and molecular gas densities at
$z=0$ amount to $\Omega_{HI}=(3.5\pm 0.8)\times 10^{-4}$ (Zwaan et
al. 2003) and $\Omega_{H_2}=(1.6\pm0.4)\times 10^{-4}$ (Keres, Yun \&
Young 2003), so that, taking into account the abundance of helium
(another 1.38 factor) the fraction of cold gas results a factor of two
higher than that of HI; this is the conversion factor used in
figure~\ref{fig:MFgas}.  With the standard choice of parameters {\gal}
does not reproduce this mass function either at small or at large
masses.  In appendix~\ref{section:stab} we show that the cutoff of
this function is remarkably sensitive to mass resolution and decreases
with decreasing particle mass, so that at this stage we decide not to
give credit to the discrepancy at large masses.  At smaller masses, we
find that the amount of gas is sensitive both to the threshold for
star formation $\Sigma_{\rm thr}$ introduced in
section~\ref{section:thin} and to the assumed star-formation law.
Given the good reproduction of the Schmidt-Kennicutt law (even with a
marginal underestimate of the star-formation rate, which would imply
an increase of gas masses), the discrepancy in the amount of cold gas
cannot be due to an error in the star-formation time-scale at a given
$\Sigma_{\rm cold,D}$, but can be due to an error in the size of
discs, which influences the value of $\Sigma_{\rm cold,D}$ and then of
$t_{\star\rm ,D}$.  Such an error is also suggested by the
discrepancy, reported above, in the zero-point of the baryonic
Tully-Fisher.  We then show two more models in figure~\ref{fig:MFgas},
both with $\Sigma_{\rm thr}=5$ {\surf}, the second one with
concentrations scaled to take a value $c_{\rm nfw}=4$ for a $10^{12}$
DM halo at $z=0$.  Clearly, adding a threshold for star formation
helps in increasing the amount of cold gas (under the assumption that
the correction for the HI gas given above applies also beyond the
threshold for star formation) but cannot solve the discrepancy, while
the $c_{\rm nfw}=4$ case leads even to a satisfactory prediction of
the low-end of the gas mass function.  We conclude that the mass
function of cold gas gives a very important fine constraint to the
model, and that it gives further support to the idea that the NFW
profile of DM halos leads to too compact discs.  However, the
uncertainty in the relation between HI and cold gas must be better
taken into account before reaching strong conclusions.

\begin{figure}
\centerline{
\includegraphics[width=8cm]{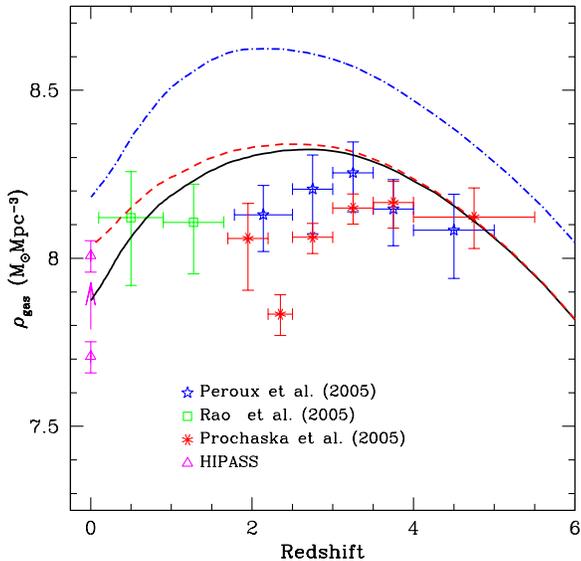}
}
\caption{Evolution of the cold gas density, compared with the HIPASS
point at $z=0$ (with and without the correction for helium and
molecular gas) and to estimates of HI gas density from DLAs, obtained
by P\'eroux et al. (2005), Rao et al. (2005) and Prochaska, et
al. (2005).  Model lines are as in figure~\ref{fig:MFgas}.}
\label{fig:gdens}
\end{figure}
 
Figure~\ref{fig:gdens} shows the evolution of the cold gas density,
compared with the HIPASS point (shown with and without the correction
for molecular gas and helium) and the data from DLA systems (P\'eroux
et al. 2005; Rao et al. 2005; Prochaska, Herbert-Fort \& Wolfe 2005),
to which no correction has been applied.  The same three models as in
figure~\ref{fig:MFgas} are shown.  All the models show the same trend
of a broad peak at $z\sim2-4$ and a decrease by a factor of $\sim5$ at
$z<2$; the $c_{\rm nfw}=4$ model gives a remarkably high
normalization.  The data themselves are characterized by large
errorbars and by a discrepancy between the Prochaska et al (2005) and
P\'eroux et al. (2005) datapoints at $z\sim2.5$.  No strong conclusion
can then be drawn from this figure, apart from a rough qualitative
agreement.

\begin{figure}
\centerline{
\includegraphics[width=8cm]{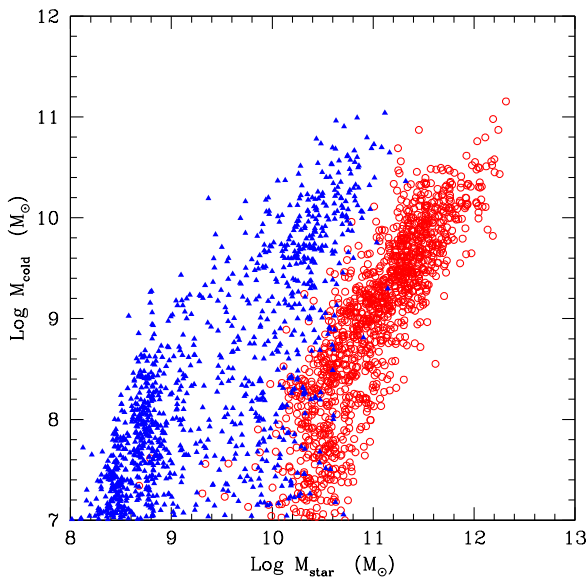}
}
\caption{Gas content of disc-dominated (blue triangles) and
bulge-dominated (red triangles) galaxies as a function of stellar
mass.}
\label{fig:gas_frac}
\end{figure}
 
Figure~\ref{fig:gas_frac} shows, for the standard model, how cold
gas is distributed in disc-dominated or bulge-dominated
galaxies. Regarding the latters, while small ellipticals have a low
gas content, the most massive objects have a significant gas load; at
variance with what appears in the plot, they do not dominate the high
end of gas mass function because they are hosted with the rare but
well sampled massive DM halos.  This excess of cold gas does not
contribute much in terms of mass, but makes these galaxies too blue,
thus highlighting that our quenching of the cooling flow by AGN
feedback is not strong enough.  This point will be addressed in more
detail in a forthcoming paper.

\begin{figure}
\centerline{
\includegraphics[width=8cm]{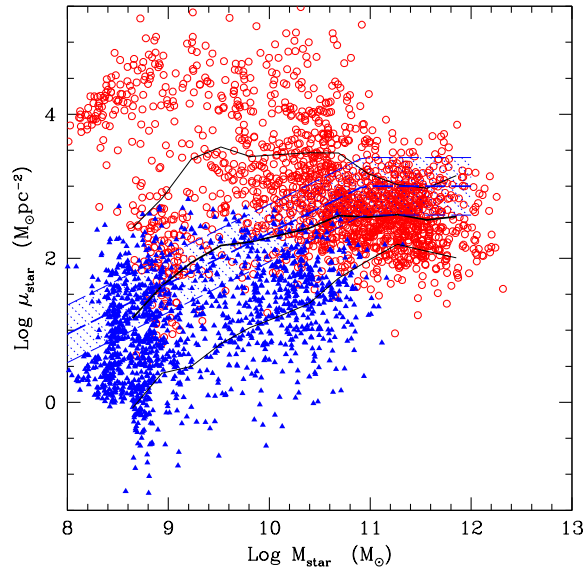}
}
\caption{Stellar mass versus stellar surface density for
disc-dominated (blue triangles) and bulge-dominated (red circles)
galaxies in the model.  The thick and thin continuous lines give the
average and average$\,\pm\,$rms of the galaxies. The shaded area reports the 
observational result of Kauffmann et al. (2003).}
\label{fig:surf_dens}
\end{figure}

Figure~\ref{fig:surf_dens} shows the galaxies in the stellar
mass--surface density plane.  Spiral discs and bulges tend to occupy
different regions of the plane; the scatter is however so strong that,
as observed, no clear bimodality emerges from the distribution of
these galaxies.  The region occupied by SDSS galaxies in this plot
(Kauffmann et al. 2003) is highlighted in the figure.  A decrease of
the surface brightness at masses $<10^{10}$ \msun\ is obtained, in
good agreement with the SDSS data; however, model galaxies are much
more scattered.

\begin{figure}
\centerline{
\includegraphics[width=8cm]{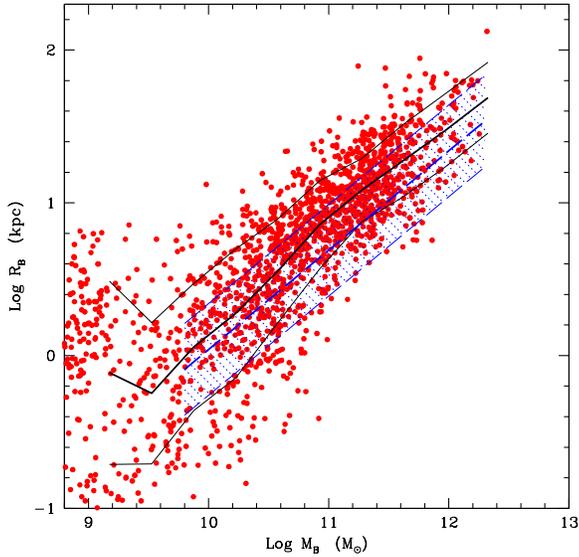}
}
\caption{Half-mass radii of elliptical galaxies as a function of
stellar mass.  Points with errorbars give the average and rms in bins
of the model galaxies.  The lines give the the average and rms values
of the data obtained from Marconi \& Hunt (2003).  }
\label{fig:ell}
\end{figure}

Figure~\ref{fig:ell} shows the bulge-dominated galaxies in the stellar
mass--$R_{\rm B}$ plane, compared with the observed relation obtained
from the data of Marconi \& Hunt (2003), $R_{\rm B} = 4.9 (M_{\rm
B}/10^{11}\ {\rm M}_\odot)^{0.65}$ kpc (see also Monaco \& Fontanot
2005), with a scatter of 0.3 dex.  The data lie nicely within the
observed range, with some flattening at small masses which is
consistent with the findings of Graham et al. (2006).  We do not show
a prediction of the fundamental plane at this stage, because the
virial theorem is implicit in our relation between mass, radius and
velocity dispersion, and the computation of ${\cal M/L}$ ratios need a
spectro-photometric code to be computed.

\begin{figure}
\centerline{
\includegraphics[width=8cm]{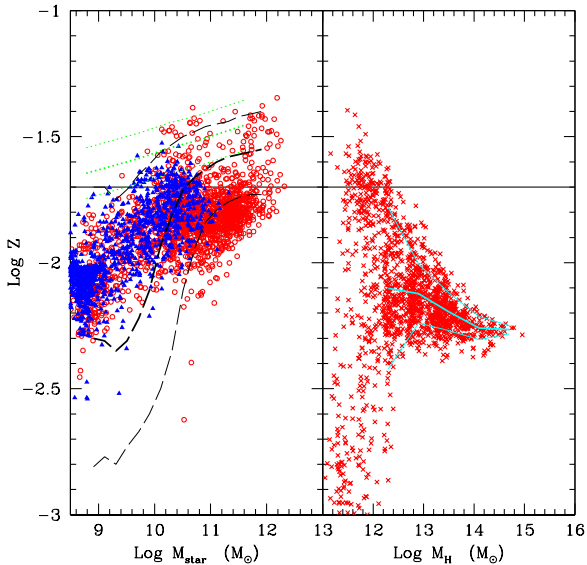}
}
\caption{ Left panel: metallicity of bulge-dominated (red points)
and disc-dominated (blue triangles) galaxies, compared to the average
and rms range of data of SDSS galaxies (Gallazzi et al. 2005, black
dashed lines) and of the NOAO survey cluster ellipticals (Nelan et
al. 2005, green dotted lines).  Right panel: predicted metallicity of
the hot halo component; in this panel, the continuous thick and thin
cyan lines give the average and rms of the points.}
\label{fig:metals}
\end{figure}

Figure~\ref{fig:metals} shows the predicted metallicity of bulges,
discs and hot halo gas.  Elliptical galaxies show a mass-metallicity
relation that steepens at masses smaller than $11^{10}$ \msun, spirals
show a similar though weaker trend.  These metallicities are compared
with the results of Nelan et al. (2005), relative to the cluster
elliptical galaxies of the NOAO fundamental plane survey, and Gallazzi
et al. (2005), relative to SDSS galaxies.  Both datasets suggest an
increase of metallicity with stellar mass, which confirms our
qualitative prediction, but they differ remarkably in the
normalization.  This reflects the intrinsic difficulties in estimating
ages and metallicities in large samples of galaxies.  Taking this into
account, we notice that the region of masses lower than
$\sim3\times10^{10}$ {\msun} and metallicities lower than 0.005 is
populated by SDSS galaxies (presumably by disc-dominated objects) but
not by our model.  Taking this result at face value, these data
suggest that there may be a need of feedback in small-mass discs. In
Figure~\ref{fig:metals} we also show the predicted metallicity of hot
halo gas, which is enriched to a level of roughly $Z_\odot/3$ in
clusters, which raises in galaxy groups and then is dominated by
scatter at smaller DM halo masses (which are also almost devoid of hot
gas).  This is in qualitative agreement with the observed trend
(Baumgartner et al. 2005) of an increase of metallicity from galaxy
clusters to groups of 2-3 keV ($\sim10^{14}$ \msun), possibly followed
by a decrease at smaller masses.


\section{Discussion and conclusions}
\label{section:conclusions}

We have presented and described in detail the code {\gal} for the
formation and evolution of galaxies and AGNs.  The most relevant
features of {\gal} have been mentioned in the Introduction: it
attempts, through its original modeling of cooling, star formation,
feedback, galactic winds and super-winds, AGN activity and AGN
feedback, to move from a phenomenological description of galaxy
formation, based on simple scalings with the properties of the host DM
halo, toward a fully physically motivated one.  The numerical
integration of the mass, energy and metal flows allows a wide set of
physical processes to be straighforwardly implemented, and the
multi-phase description of the inter-stellar and intra-cluster media,
although used only in a limited way, is a step forward towards a more
realistic description of the complex physics involved.

Despite many significant technical differences, the predictions of
{\gal} at $z=0$ are in line with that of the other semi-analytical or
N-body models: most of the figures shown in
section~\ref{section:results}, will be no real surprise for most
experts of galaxy formation.  Besides, a qualitative comparison of our
results with the N-body ones of Tornatore et al. (2003) has shown in
many cases the same trends, as for instance the formation of too
massive ellipticals in clusters or the inability of stripping to
produce a sufficient number of halo stars (see also Monaco et
al. 2006).  Similarly, the hidden dependence of the results on mass
resolution and the lack of a proper convergence is shared by the two
methods.  This implies that the field is reaching an interesting level
of maturity, so that an increase in the level of sophistication in
these models is justified.

In the Introduction we stated that we do not regard this model as a
``theory of everything'' for galaxies, but simply as a powerful tool
to understand the complex nature of galaxies and to bridge in a
realistic way the physical processes in play with the observations
that can constrain them.  It is then important to focus on the
discrepancies with observations found in this paper, and to the
insights on the physical processes that they provide.

The exponential cutoff of the luminosity function is only roughly
reproduced, which implies that the quenching of the cooling flow by
the AGN is not well modeled; a stronger indication is given by the
excess of cold gas in large ellipticals.  This is no surprise given
the poor level of understanding of this process.  Our quenching is
performed by allowing gas to cool, form stars and accrete onto the BH,
at variance with the choice of switching cooling off whenever some
fiducial criterion is satisfied, like e.g. in Hatton et al (2003),
Bower et al. (2006) or Croton et al. (2006), a choice that leads to
better fits to the luminosity functions and correct red colours for
the most massive galaxies.  Besides, if the build-up of massive
galaxies by consecutive mergers is slowed down by the creation of the
diffuse stellar component in galaxy clusters, as proposed by Monaco et
al. (2006), then the high end of the stellar mass function (but not
galaxy colours) may be determined by gravitational processes more than
by AGN feedback.  This suggests that the longly known cutoff of the
stellar mass function may still give surprises in the future.

Another well-known discrepancy arises in reproducing the zero-point of
the baryonic Tully-Fisher relation.  Our results support strongly the
idea that this problem is not due to an unsuitable modeling of
feedback but to the shape of the inner density profile of DM halos;
the cusped NFW halos produce discs that are too compact (with $V_{\rm
D}$ values higher by 25 per cent), while by scaling the halo
concentration so as to assume the low value $c_{\rm nfw}=4$ for a
$10^{12}$ {\msun} DM halo at $z=0$, as suggested by Mo \& Mao (2000),
it is possible to recover the zero-point of the Tully-Fisher relation.
Besides, the formation of galaxy discs is based on the (reasonable but
still not fully demonstrated) assumptions of conservation of angular
momentum and formation of exponential discs, and our model neglects
important problems related to the distribution of angular momentum
within the DM halos.  Given the complexity of this topic, it is
remarkable that the observed Tully-Fisher relation is wrong by only
0.1 dex under the simplest assumptions.

The discrepancy in disc sizes is reflected also in the mass function
of cold gas; model disc galaxies have correct stellar masses and stay
on the correct Schmidt-Kennicutt law, but show a deficit of cold gas
mass.  This lack is partially fixed by inserting a threshold for star
formation, but only the formation of less compact discs (hosted in
flatter DM halos) can solve the discrepancy.  Our analysis also shows
the potentiality of using the mass function of cold gas in galaxies to
test galaxy formation models, although the uncertainty in the relation
between HI and total cold gas hampers strong conclusions.

Another point of disagreement with data is given by the excess of
small bulges at $z=0$.  This is related to the excess of $10^{10}$
{\msun} galaxies at $z\ga1$ (Fontana et al. 2006), and to the
intrinsic difficulty of ``downsizing'' the AGN population (Fontanot et
al. 2006a).  The lack of low-metallicity discs may add to this
evidence.  These connected problems point to some missing feedback
sources that acts in disc-dominated galaxies at high redshift (those
that later merge to form bulges).

All these discrepancies point to the need of a deeper
understanding of the complex process of galaxy formation, that will
necessarily need a focus on the details of the various processes
involved.  In this regard, observations of larger and larger samples
of galaxies will need to be complemented by detailed observations of
objects where feedback is at work. To bridge together the two sets of
evidences it will be necessary to use galaxy formation models that, on
the one hand, are able to generate predictions for large samples of
objects and, on the other hand, contain a sophisticated enough
treatment of gas dynamics to make predictions on the detailed
properties of galaxies. {\gal} is a contribution in this direction.

\section*{Acknowledgments}

We thank Lucia Ballo, Stefano Borgani, Gabriele Cescutti, Gabriella De
Lucia, Carlos Frenk, Gianluigi Granato, Giuseppe Murante, Ezio
Pignatelli, Paolo Salucci, Laura Silva and Tom Theuns for many
discussions, and Stefano Cristiani for his encouragement.  The
anonymous referee has been very helpful in improving the presentation
of the paper.  PM thanks the Institute for Computational Cosmology of
Durham for hospitality.

{}


\appendix
\section[]{Merging and destruction times for galaxies}
\label{section:taffoni}

The merging and destruction times are computed using the physically
motivated analytic fits (accurate to $\sim15$ per cent) of Taffoni et
al. (2003) to the merging and destruction times found using N-body
simulations and analytical models.  We use here a version of the
analytic fits that differs slightly from the one proposed in that
paper in the values of the fitting parameters:

Let the merging halos have mass, circular velocities, virial radii and
concentrations respectively $M_{\rm H}$, $V_{\rm H}$, $r_{\rm H}$,
$c_{\rm H}$ (main halo) and $M_{\rm S}$, $V_{\rm S}$, $r_{\rm S}$,
$c_{\rm S}$ (satellite).  Let $f_{\rm sat}=M_{\rm S,0}/M_{\rm H}$ be
the ratio between the initial mass of the satellite and the halo mass
(we recall that the halo mass includes the satellite), and let
$\epsilon$ and $x_c$ be the orbital parameters of the satellite (see
section~\ref{section:dynfriction}).  In the case of a rigid satellite
that suffers no mass loss, the dynamical friction time (the time
required by an orbit to decay) is:

\be
\tau_{\rm rigid} = 0.46 \frac{r_{\rm H}^2V_{\rm H}}{GM_{\rm S,0}} 
\left(1.7265+0.0416 c_{\rm H} \right) 
\frac{x_c^{1.5}}{\ln\left(1+1/f_{\rm sat}\right)}
\label{eq:rigid}\ee

\noindent
while for a live satellite, i.e. a satellite that suffers significant
mass loss by tidal stripping, the dynamical friction time is:

\bea
\tau_{\rm live} &=& \frac{r_{\rm H}^2V_{\rm H}}{GM_{\rm S,0}}
\times \left[{\cal B} f_{\rm sat}^{0.12} + {\cal C} f_{\rm sat}^2\right] \label{eq:live} \\
&&\times \left[0.25\left(c_{\rm H}/c_{\rm S}\right)^6-0.07(c_{\rm S}/c_{\rm H})+1.123\right]\nonumber\\
&&\times \left[0.4+\left(\epsilon-0.2\right){\cal Q}\right]\nonumber
\eea

\noindent where
\bea
{\cal B} & = & -0.0504+0.3355 x_c +0.3281 x_c^2 \\
{\cal C} & = & 2.151-14.176 x_c +27.383 x_c^2 \\
{\cal Q} & = & 0.9+10^8 \left(f_{\rm sat}-0.0077/(1-1.08 x_c)-0.0362\right)^6\nonumber\\
&&\left(12.84 + 3.04 x_c - 23.4 x_c^2\right) 
\eea

\noindent
The general expression for the dynamical friction time is obtained
by interpolating between the two expressions: for $f_{\rm sat}>0.1$:

\be
\tau_{\rm df} = \tau_{\rm rigid}\, ;
\ee

\noindent for $0.08< f_{\rm sat} \le 0.1$:

\be
\tau_{\rm df} = \tau_{\rm rigid}\left(\frac{f_{\rm sat}-0.08}{0.02}\right)
+ \tau_{\rm live}\left(1-\frac{f_{\rm sat}-0.08}{0.02}\right)\, ; 
\ee

\noindent for $0.007< f_{\rm sat} \le 0.08$:

\be
\tau_{\rm df} = \tau_{\rm live}\, ; 
\ee

\noindent for $f_{\rm sat}\le 0.007$:

\be
\tau_{\rm df} = \max\left(2.1 \tau_{\rm rigid} \epsilon^{0.475 
(1-\tanh(10.3 f_{\rm sat}^{0.33}-7.5 x_c))},\tau_{\rm live}\right)\, .
\ee

For the destruction times we follow closely Appendix B of Taffoni et
al. (2003).

\section[]{Numerical stability}
\label{section:stab}

\begin{figure}
\centerline{\includegraphics[width=7cm]{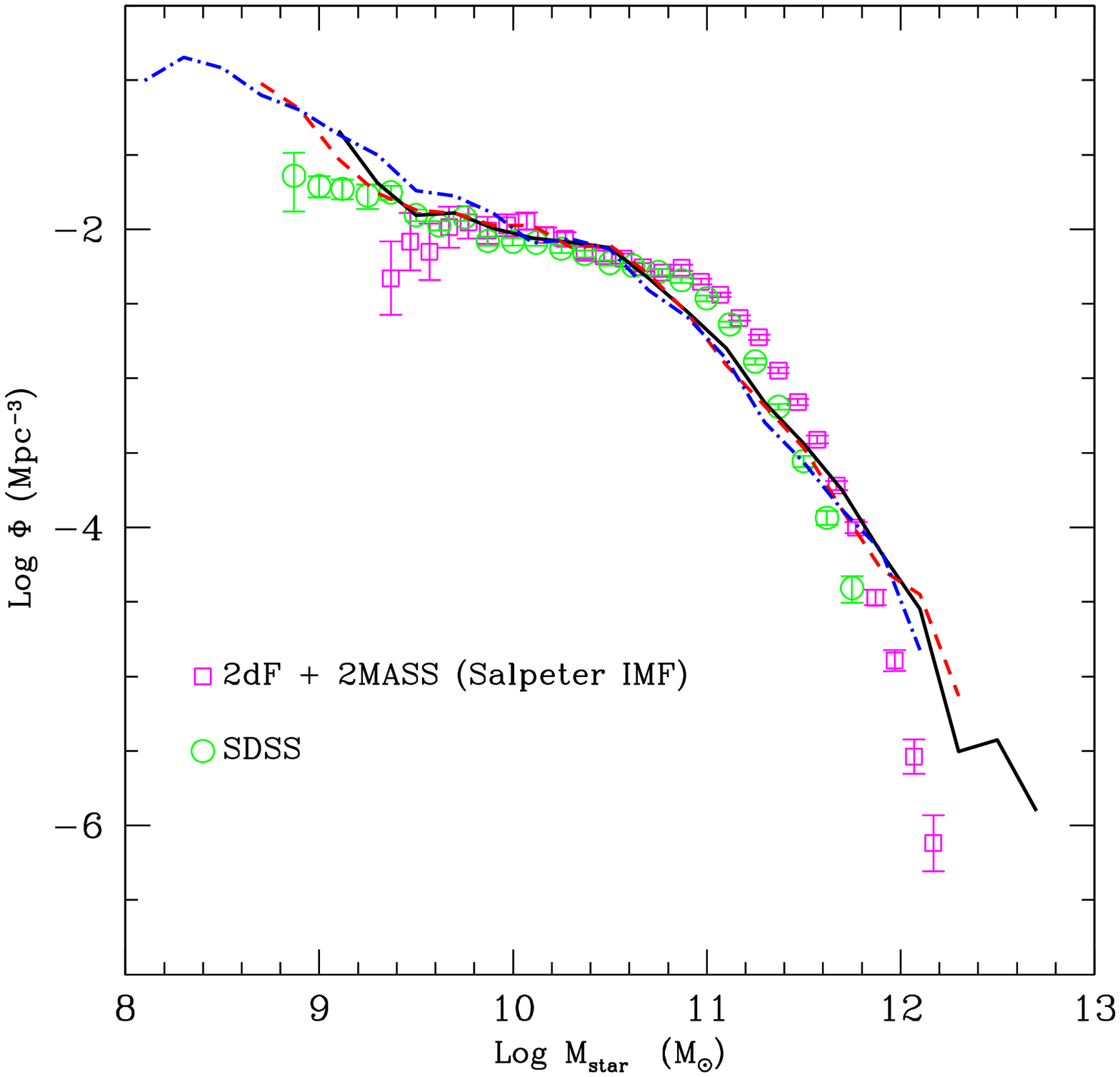}}
\centerline{\includegraphics[width=7cm]{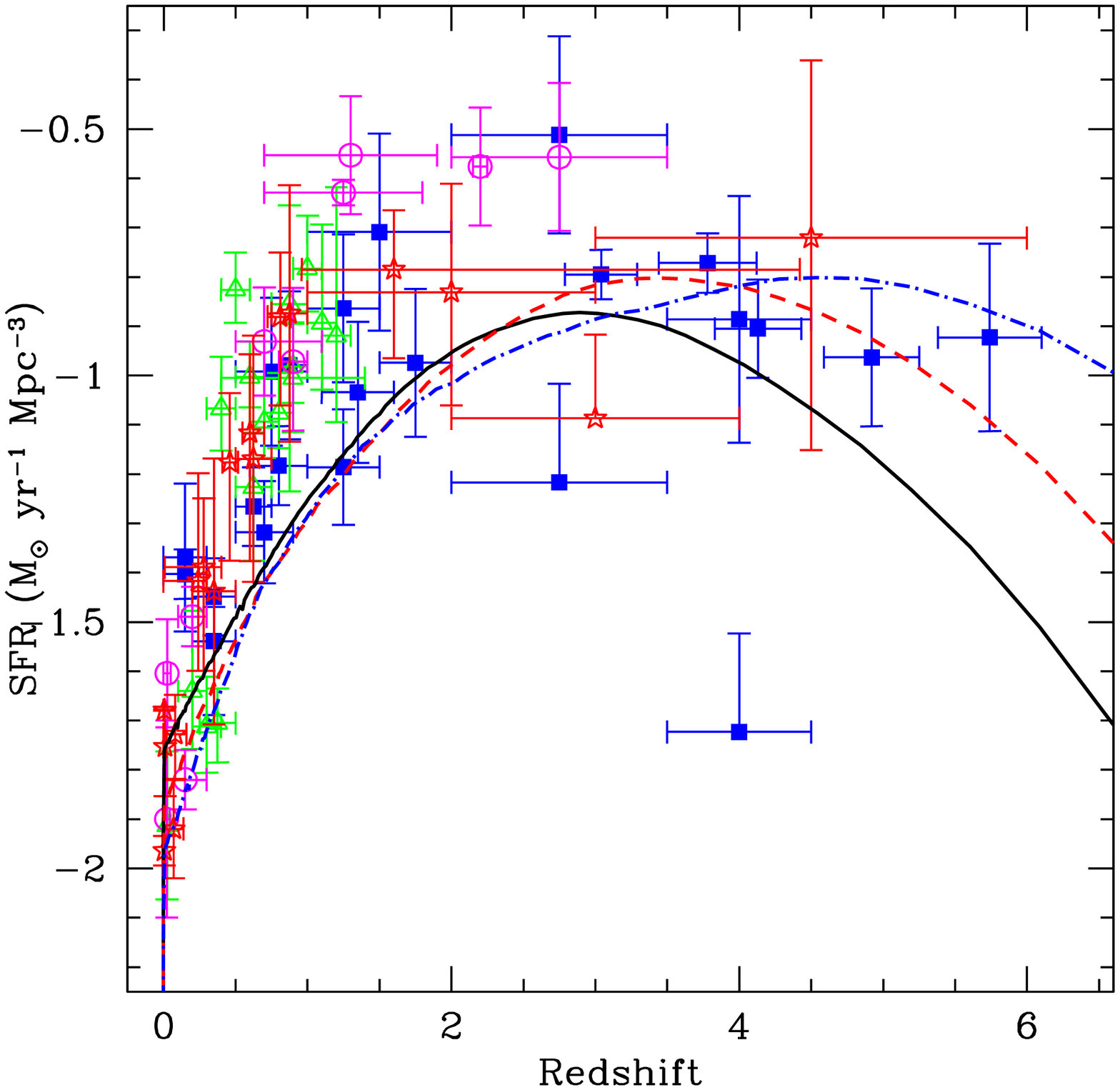}}
\centerline{\includegraphics[width=7cm]{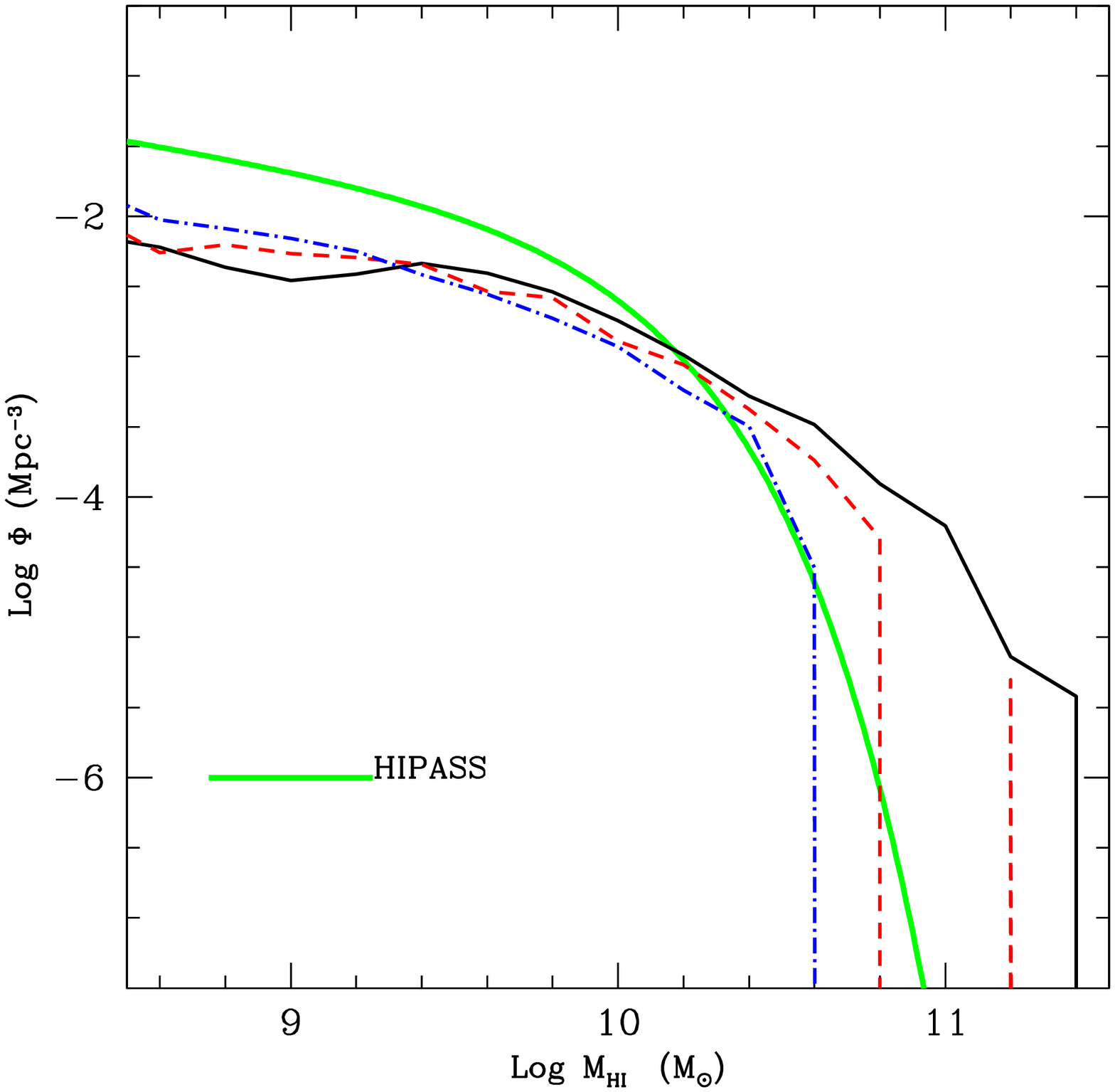}}
\caption{Stellar mass function (upper panel; see
figure~\ref{fig:mfstar}), star formation density (mid panel; see
figure~\ref{fig:csfr}) and cold gas mass function (lower panel; see
figure~\ref{fig:MFgas}) for galaxies in the three runs with particle
mass $3.0\times10^8$ {\msun} (blue dot-dashed lines), $1.0\times10^9$
{\msun} (red dashed line) and $2.4\times10^9$ {\msun} (black
continuous line).}
\label{fig:stability}
\end{figure}

In this Appendix we show how some results are sensitive to the mass
resolution of the merger trees.  This is not to be considered as a
complete convergence study, but only as a first test of the robustness
of the results.  Figure~\ref{fig:stability} gives the stellar mass
function, star formation density and cold gas mass function for the
three boxes introduced in section~\ref{section:results} and for the same set
of parameters given in table~\ref{table:parameters}.

The following conclusions can be drawn:

(i) the model roughly converges in predicting the stellar content of
bright galaxies, but does not for small galaxies.  In particular, at
increasing resolution (decreasing particle mass) the stellar mass
function gets steeper at the low-mass end and, as a consequence,
lowers at the knee.

(ii) The star-formation density apparently converges at lower
redshift, but gets increasingly large at $z\ga2$.  The main reason for
the low-redshift convergence is that, to mimic the effect of
reionization (following Benson et al. 2002; see
section~\ref{section:ic}) cooling is quenched in halos with circular
velocity larger than 50 \kms.  The smallest sampled progenitor
overtakes this limit at redshifts higher than 1, 2.8 and 7.5 for the
three runs, in order of decreasing particle mass.  This means that
each box is missing star-forming halos at redshift higher than that
limit, and explains why convergence in the star formation density is
not visible in the figure.

(iii) From the stellar mass function and star-formation rate, it is
clear that the excess of small-mass galaxies noticeable for the
highest resolution run is connected to the excess of star formation at
high redshift, so that the exceeding dwarf galaxies will be very old
objects.  This unwanted feature is clearly connected to the excess of
galaxies with stellar mass $\la10^{10}$ {\msun} at $z\sim 1$ found by
Fontana et al. (2006) by comparing models (including {\gal}) to the
results of the GOODS-MUSIC survey, and to the excess of small bulges
visible in figure~\ref{fig:mfstar}.  Clearly, a source of feedback,
which would limit the formation of dwarf galaxies at $z>1$, is
missing, and this lack is more and more evident when the particle
mass decreases.

(iv) With the low-$V_c$ cutoff, convergence is reached in the
star-formation rate at $z<1$, because only the run with the largest
particle mass shows higher star formation density and a tail of
galaxies more massive than $3\times10^{12}$ \msun.  This is related to
the efficiency of the quenching of the late cooling flows by
radio-loud AGNs.  This demonstrates that our self-consistent quenching
needs a particle mass not larger than $10^9$ \msun.  In case of poor
mass resolution it is possible to apply a procedure similar to Croton
et al. (2006) and Bower et al. (2006) to force the quenching of the
cooling flow.  In this case, whenever a cooling flow is present it is
assumed that a fraction of that flow would immediately accrete onto
the BH (subject to the Eddington rate) and release energy to the hot
halo phase.  Whenever this ``fiducial'' energy is higher than that
lost by cooling, quenching is switched off.  This ``forced cooling''
procedure is clearly less physical than that described in
section~\ref{section:jets}, and for a sufficiently small particle mass
it gives a slightly stronger quenching of the cooling flows, but it
can be considered as a rough numerical trick to achieve a good level
of quenching when mass resolution is poor. This forced quenching
procedure is used in Fontanot et al. (2006a).

(v) The high-mass cutoff of the cold gas mass function is the most
sensitive prediction to mass resolution.  We don't find any hint of
convergence for this quantity at $z=0$, despite the star content of
the same galaxies is convergent.  On the other hand, the low-mass tail
of the same distribution is rather stable.

We can then conclude that the model does not really converge with the
resolution, and that convergence at large masses is obtained by using
Benson et al.'s motivated recipe.  Some relevant ingredient, able to
hamper star formation in small-mass (but relatively high circular
velocity) galaxies at high redshift, is still missing.

\bsp

\label{lastpage}

\end{document}